\documentclass[aip,pop,reprint,numerical,a4paper,superscriptaddress]{revtex4-1}
\usepackage{epsfig,graphicx,natbib}
\usepackage{amsmath,amsfonts,bm}

\newcommand{\fig}[1]{Fig.\ \ref{#1}}
\newcommand{\Eq}[1]{Eq.\ \ref{#1}}
\newcommand{\Eqs}[1]{Eqs.\ \ref{#1}}
\newcommand{\ppcode}{\textsc{PhotonPlasma }code}
\newcommand{\staggercode}{\textsc{Stagger }code}
\newcommand{\wpet}{\omega_{pe}\,t}
\newcommand{\ddt}[1]{\frac{\partial #1}{\partial t}}

\renewcommand{\d}{\partial}
\renewcommand{\div}{\nabla\cdot}
\newcommand{\laplace}{\nabla^2}
\newcommand{\divn}{\hat{\nabla}\cdot}
\newcommand{\curl}{\nabla\times}

\newcommand{\uu}{\bm{u}}

\newcommand{\uug}{\bm{u}\gamma}

\newcommand{\pp}{\bm{p}}
\newcommand{\JJ}{\bm{J}}
\newcommand{\BB}{\bm{B}}
\newcommand{\EE}{\bm{E}}

\newcommand{\xx}{\bm{x}}
\newcommand{\rr}{\bm{r}}
\newcommand{\vv}{\bm{v}}

\newcommand{\bbeta}{\bm{\beta}}
\newcommand{\apj}{The Astrophysical Journal}
\newcommand{\apjl}{Astrophysical Journal Letters}

\newcommand{\jgr}{Journal of Geophysical Research}
\newcommand{\pre}{Physical Review E}

\begin{document}
\title{Photon-Plasma: a modern high-order particle-in-cell code}
\author{Troels Haugb{\o}lle}
\email{haugboel@nbi.dk}
\affiliation{Centre for Star and Planet Formation, Natural History Museum of Denmark, University of Copenhagen,
{\O}ster Voldgade 5-7, DK-1350 Copenhagen, Denmark}
\author{Jacob Trier Frederiksen}
\affiliation{Niels Bohr Institute, University of Copenhagen, Juliane Maries Vej 30, DK-2100 Copenhagen, Denmark}
\author{{\AA}ke Nordlund}
\affiliation{Niels Bohr Institute, University of Copenhagen, Juliane Maries Vej 30, DK-2100 Copenhagen, Denmark}
\affiliation{Centre for Star and Planet Formation, Natural History Museum of Denmark, University of Copenhagen,
{\O}ster Voldgade 5-7, DK-1350 Copenhagen, Denmark}
\begin{abstract}
We present the Photon-Plasma code, a modern high order charge conserving particle-in-cell code for
simulating relativistic plasmas. The code is using a high order implicit field solver and a novel high order charge
conserving interpolation scheme for particle-to-cell interpolation and charge deposition. It includes powerful
diagnostics tools with on-the-fly particle tracking, synthetic spectra integration, 2D volume slicing, and
a new method to correctly account for radiative cooling in the simulations. A robust
technique for imposing (time-dependent) particle and field fluxes on the boundaries is also presented. Using a hybrid OpenMP and MPI
approach the code scales efficiently from 8 to more than 250.000 cores with almost linear weak scaling on a range of architectures.
The code is tested with the classical benchmarks particle heating, cold beam instability, and two-stream instability. We also
present particle-in-cell simulations of the Kelvin-Helmholtz instability, and new results on radiative collisionless shocks.
\end{abstract}

\maketitle

\section{Introduction}
Particle-In-Cell models have gained widespread use in astrophysics as a means to understand
plasma dynamics, particularly in collisionless plasmas, where non-linear instabilities can
play a crucial role for launching plasma waves and accelerating particles. The advent of of
tera- and now peta-flop computers has made it possible to study the macroscopic properties of both
relativistic and non-relativistic plasmas from first principles in large scale 2D and 3D models,
and sophisticated methods, such as the extraction of synthetic spectra is bridging the gap between
models and observations.

While Particle-In-Cell codes were some of the first codes to be developed for computers\cite{Harlow:1957,Harlow:1964},
and several classic books have been written on the subject\cite{birdsall:1985,hockney:1988},
modern numerical methods are in use today in the community, which did not exist then, and the temporal
and spatial scales of the problems have grown enormously. Furthermore, in the context of astrophysics the
modeling of relativistic plasmas has become of prime importance.

In this paper we present the relativistic particle-in-cell \ppcode{} in use at the University of Copenhagen,
and the numerical and technical methods implemented in the code. The code was initially created during the 'Thinkshop'
workshop at Stockholm University in 2005, and has since then been under continuous development. It has been
used on many architectures from SGI, IBM, and SUN shared memory machines to modern hybrid Linux GPU clusters.
Currently our main platforms of choice are Blue-Gene and Linux clusters with 8-16 cores per node and infiniband.
We have also developed a special GPU version that achieves a 20x speedup compared to a single 3GHz Nehalem
core (to be presented in a forthcoming paper). The code has excellent scaling, with more than 80\% efficiency
on Blue-Gene/P scaling weakly from 8 to 262.144 cores, and on Blue-Gene/Q from 64 to 524.288 threads. The I/O
and diagnostics is fully parallelized and on some installations we reach more than 45 GB/s read and 8 GB/s write
I/O performance.

In Section II and III we introduce the underlying equations of motion, and the numerical techniques for
solving the equations. We present our formulation of radiative cooling in Section IV, and in Section V the
various initial and boundary conditions supported by the code. Section VI presents on-the-fly diagnostics,
including the extraction of synthetic spectra. Section VII describes the binary collision modules, while Section VIII
contains test problems. In Section IX we discuss aspects of parallelization and scalability, and finally in
section X we finish with concluding remarks.

\section{Equations of motion}
The \ppcode{} is used to find an approximate solution to the relativistic Maxwell-Vlasov system
\begin{equation}\label{eq:vlasov}
\ddt{f^s} + \uu\cdot\frac{\partial f^s}{\partial \xx} +
  \frac{q^s}{m^s}(\EE + \uu\times\BB)\cdot\frac{\partial f^s}{\partial (\uug)} = \mathcal{C}
\end{equation}
\begin{align}\label{eq:gauss}
\div \EE & = \frac{\rho_c}{\epsilon_0} \\
\label{eq:divb} \div \BB & = 0 \\
\label{eq:faraday} \ddt{\BB} & = - \curl \EE \\
\label{eq:ampere} \frac{1}{c^2}\ddt{\EE} & = \curl \BB - \mu_0\JJ\,,
\end{align}
where $s$ denotes particle species in the plasma (electrons, protons, \ldots), $\gamma = {(1-(u/c)^2)^{-1/2}}$ is the Lorentz
factor, and $\mathcal{C} \equiv \partial{f^s}/\partial{t}\big|_{coll}$ denotes some collision operator.

In a completely collisionless plasma $\mathcal{C}$ is zero, but in the code we also allow for binary collisions between
particles. As shown below in the tests, discretization effects in the interpolation of fields and sources between the mesh
and the particles and the integration of the equations of motion lead to a non-zero, non-physical, collision term, which
should be minimized, especially in the case of collisionless plasmas, but also with respect to any collisional modeling
term introduced explicitly. The charge and current densities are given by taking moments of the distribution function
over momentum space
\begin{align}\label{eq:rho}
\rho_c(\xx) = \int \textrm{d}\uu \sum_s q^s f^s(\xx,\uu) \\
\JJ(\xx) = \int \textrm{d}\uu \sum_s \uu\, q^s f^s(\xx,\uu) \,.
\end{align}
To find an approximate representation for this six-dimensional system in the particle-in-cell method so-called macro
particles are introduced to sample phase space. Macro particles can either be thought of as Lagrangian markers that
measure the distribution function in momentum space at certain positions, or as ensembles of particles
that maintain the same momentum while moving through the volume. If the trajectory of a macro particle
is a solution to the Vlasov equation, given the linearity, a set of macro particles will also be a solution to the system.
Other continuum fields, which only depend on position, are sampled on a regular mesh.
Macro particles are characterized by their positions $\xx_p$ and proper
velocities $\pp_p=\uug$. They have a weight factor $w_p$, giving the number density of physical particles inside a macro
particle, and a physical shape $S$. The shape is chosen to be a symmetric, positive and separable function,
with a volume that is normalized to 1. For example in three dimensions it can be written
$S(\xx-\xx_p) = S_\textrm{1D}(x-x_p)S_\textrm{1D}(y-y_p)S_\textrm{1D}(z-z_p)$, and $\int S(\xx-\xx_p) \textrm{d}\xx = 1$.
The full distribution function of a single macro particle is then
\begin{equation}\label{eq:pseudop}
f_p(\xx,\pp) = w_p\, \delta(\pp-\pp_p)\,S(\xx-\xx_p)\,.
\end{equation}
Inserting the above in \Eq{eq:vlasov} and taking moments \cite{Lapenta:2006,hockney:1988,birdsall:1985}
we find the equations of motion for a single macro particle,
\begin{align}\label{eq:pmotion}
\frac{\textrm{d}\xx_p}{\textrm{d}t} &= \uu_p &
\frac{\textrm{d}\uu_p\gamma_p}{\textrm{d}t} &= \frac{q}{m}\left(\EE_p + \uu_p \times \BB_p \right)\,,
\end{align}
where the electromagnetic fields are sampled at the macro particle position through the shape functions
\begin{align}
\EE_p &= \EE(\xx_p) = \int \!\! d\xx \, \EE(\xx) \, S(\xx-\xx_p) \\
\BB_p &= \BB(\xx_p) = \int \!\! d\xx \, \BB(\xx) \, S(\xx-\xx_p)\,.
\end{align}
The Vlasov equation (\Eq{eq:vlasov}) is linear, and if a single macro particles obeys \Eqs{eq:pmotion} a
collection of macro particles, describing the plasma, will also obey it. Using that the shape functions are
symmetric, and assuming that the electromagnetic fields are constant inside each cell volume we find
\begin{align}\label{eq:pfields}
\EE_p &= \!\!\!\!\!\! \sum_{\xx_c = \textrm{cell vertices}} \!\!\!\!\!\! \EE(\xx_c) \, W(\xx_c-\xx_p) \\
\BB_p &= \!\!\!\!\!\! \sum_{\xx_c = \textrm{cell vertices}} \!\!\!\!\!\! \BB(\xx_c) \, W(\xx_c-\xx_p)\,,
\end{align}
where the weight function $W$ is given by integrating the shape function over the cell volume
\begin{equation}\label{eq:weightfunction}
W(\xx_c-\xx_p) = \int_{\xx_c-\frac{\Delta\xx}{2}}^{\xx_c+\frac{\Delta\xx}{2}} \!\!\!\!\!\! \textrm{d}\xx \, S(\xx - \xx_p)\,.
\end{equation}

\begin{figure*}
     \begin{center}
           \includegraphics[width=0.4\linewidth]{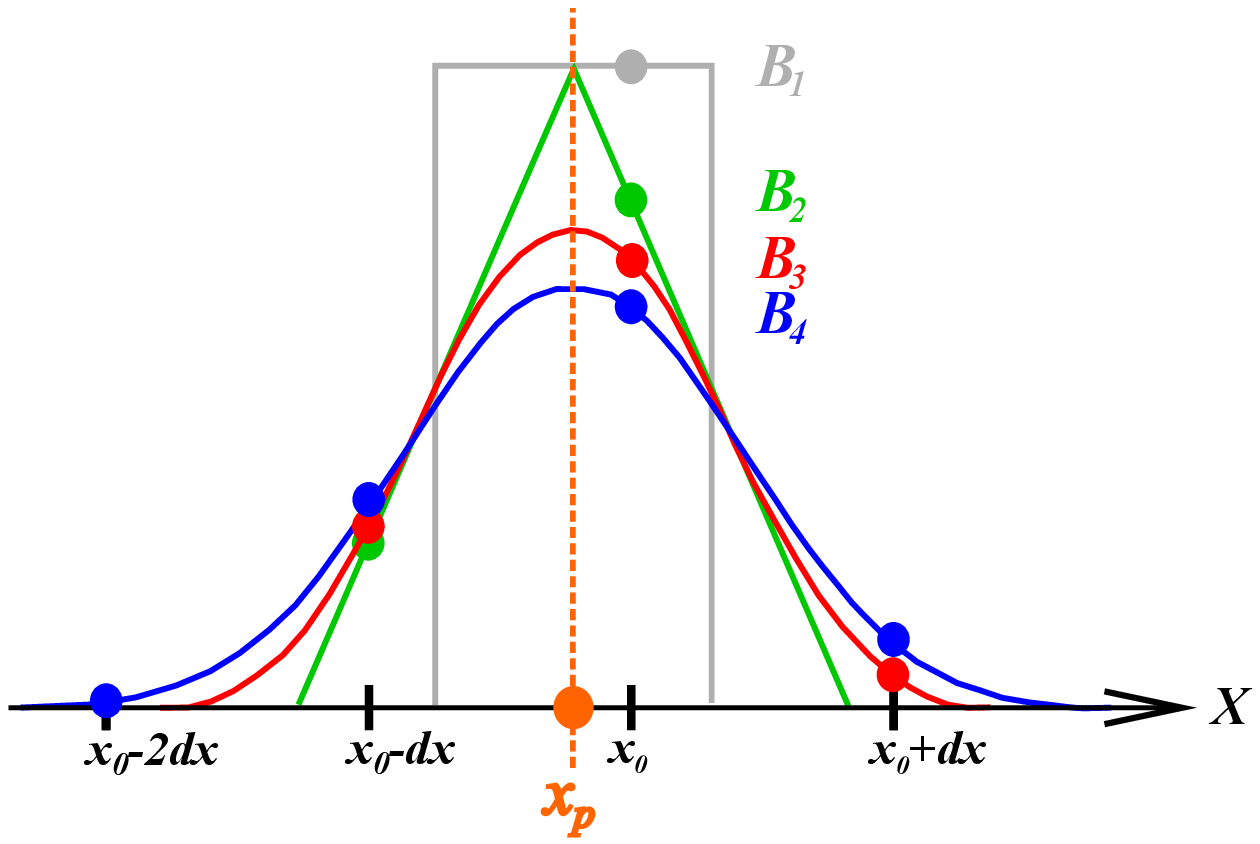}
           \includegraphics[width=0.4\linewidth]{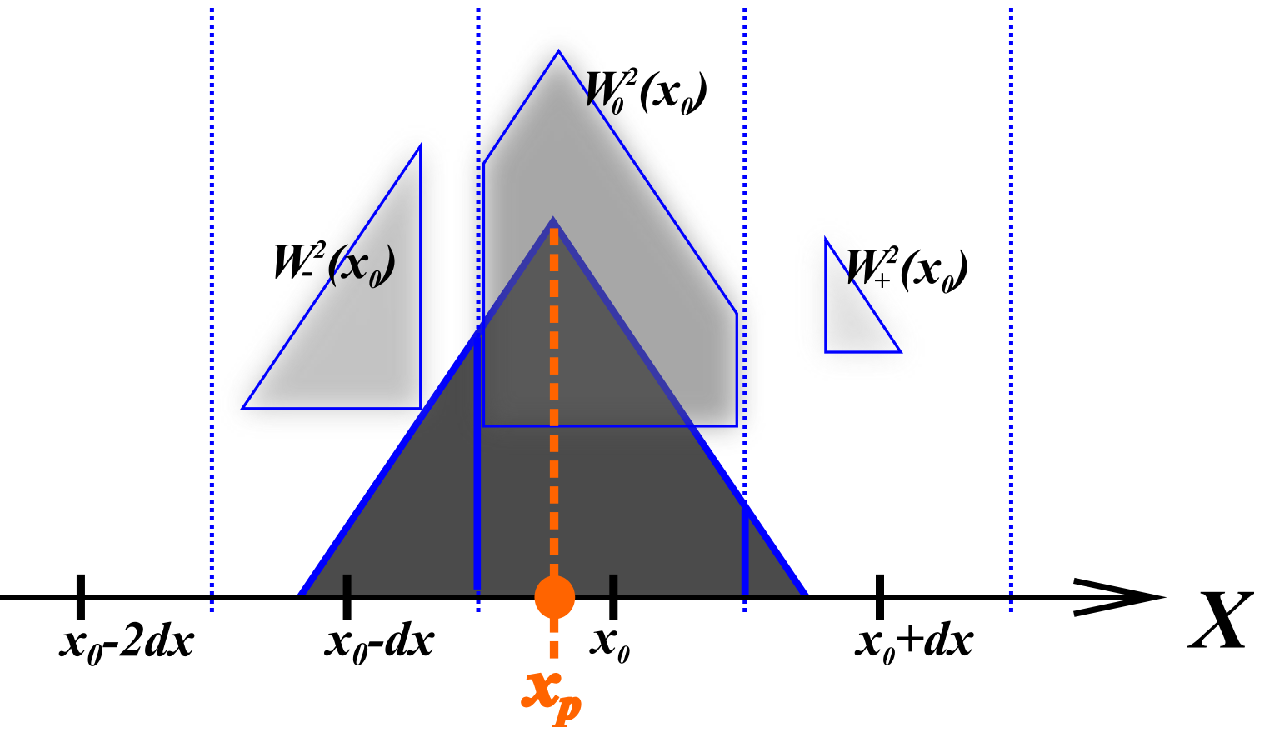}
     \end{center}
     \caption{To the left, the first four $B$-splines, $\left\{{B^1},\ldots,{B^4}\right\}$,
     which define the particles' shape functions (interpolation kernels). These four are all
     accessible in the code. To the right, the cell (area) weighting of the interpolation
     kernel for the triangular-shaped-cloud. This \textit{shape} function is the piecewise
     linear spline, whereas the piecewise quadratic spline shape produces the bell-shaped
     (cubic spline) area weighting function -- cf.\ \Eq{eq:weightfunction}.}\label{fig:bsplines}
\end{figure*}

In principle any shape function would be valid. However, in practice most PIC codes employ shape
functions belonging to a family of basis functions known as $B$-splines. They have a number of useful
properties\cite{Chaniotis2004}:
\begin{itemize}
    \item The particle interpolation function, a $B$-spline of order $\mathcal{O}$, is ${\mathcal{O}}-1$ times
              differentiable continuous, for ${\mathcal{O}}>1$
    \item Particle weight functions also become $B$-splines of order ${\mathcal{O}}+1$
    \item Their support is bounded; their width in number of mesh points is equal to their order + 1.
    \item The sum on mesh points over the support is $\Sigma^\textrm{support}_{i} B_i^\mathcal{O} \equiv 1$, for $\mathcal{O}>0$
\end{itemize}
In the \ppcode{} one can select particle shape functions from one of the four lowest order $B$-splines
(see also \fig{fig:bsplines}) giving the weight functions:
\begin{description}
\item[NGP] 'Nearest grid point',
\[
W^0(x) =
  \begin{cases}
   1 & \text{if } \left|\delta\right| \leq \frac{1}{2} \\
   0 & \text{otherwise }
  \end{cases}
\]
\item[CIC] 'Cloud-In-Cell',
\[
W^1(x) =
  \begin{cases}
   1 - \left|\delta\right| & \text{if } \left|\delta\right| < 1 \\
   0 & \text{otherwise }
  \end{cases}
\]
\item[TSC] 'Triangular Shaped Cloud',
\[
W^2(x) =
  \begin{cases}
   \frac{3}{4} - \delta^2 & \text{if } \left|\delta\right| < \frac{1}{2} \\
   \frac{1}{2} \left(\frac{3}{2} - \left|\delta\right|\right)^2 & \text{if } \frac{1}{2} \leq \left|\delta\right| < \frac{3}{2} \\
   0 & \text{otherwise }
  \end{cases}
\]
\item[PCS] 'Piecewise Cubic Spline',
\[
W^3(x) =
  \begin{cases}
   \frac{1}{6}\left(4-6\delta^2+3\left|\delta\right|^3\right) & \text{if } 0 \leq \left|\delta\right| < 1  \\
   \frac{1}{6}\left(2-\left|\delta\right|\right)^3 & \text{if } 1 \leq \left|\delta\right| < 2 \\
   0 & \text{otherwise }
  \end{cases}\,,
\]
\end{description}
where $W^i(x)$ is the one dimensional weight function, and $\delta \equiv (x_c-x_p) / \Delta x$.

Normally, there are many more particles than mesh points in a particle-in-cell model. One important benefit of introducing
a higher order particle shape function is a reduction in the aliasing effects associated with the under-sampling when
interpolating particle data on the mesh. When employing a higher order field solver, also, a higher order particle shape function
such that the effective width of the particle shape function response and the field differencing operator response
are 'not too far apart'. If the particle shape function has a very low order --- say $\mathcal{O}=0$ --- the strong
aliasing at high frequency may be visible to a high order --- say $\mathcal{O}=6$ --- differential operator, which will
introduce spurious contributions from the Maxwell source term(s). One should thus take care to keep the order of the particle
shape function 'high', if the field solver difference operators have order 'high'.
On balance, using the highest order cubic spline interpolation with 64 mesh points in three dimensions
to couple the fields and the particles has turned out to be effectively the cheapest method in most applications. The large
support leads to better noise properties, and hence a lower amount of particles can be used to reach the same quality
compared to using more particles and a lower order $B$-spline. The cost is also partially offset by the increased number
of FLOPS per memory transfer, when compared to a lower order scheme integration cycle resource consumption.

\section{Discretization and time integration}
\begin{figure}
     \begin{center}
           \includegraphics[width=0.8\linewidth]{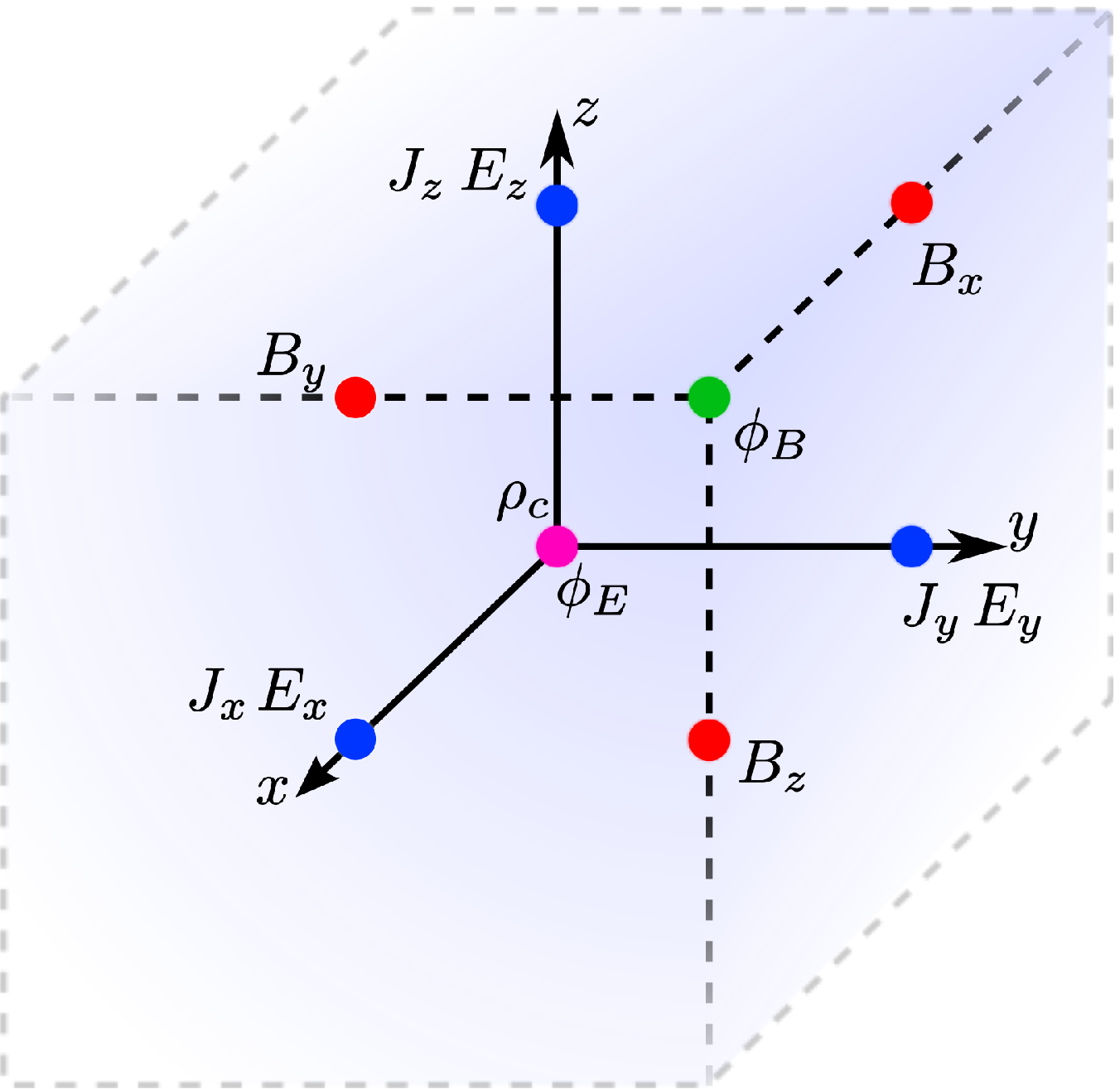}
     \end{center}
     \caption{Spatial staggering in the \ppcode.} \label{fig:yee}
\end{figure}

\begin{figure}
     \begin{center}
           \includegraphics[width=\linewidth]{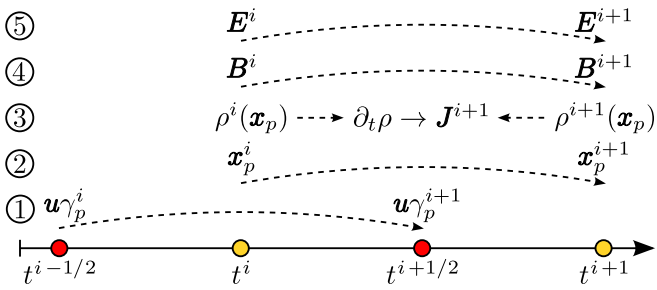}
     \end{center}
     \caption{Time staggering and integration order in the \ppcode{}. $\uug_p$ and $\JJ$ are
     staggered $\Delta t/2$ backwards in time, while everything else is time centered.} \label{fig:time}
\end{figure}

To solve the equations of motion for the coupled particle-field system (\Eq{eq:faraday}, \Eq{eq:ampere}, and
\Eq{eq:pmotion}) we need to choose both a spatial and temporal discretization, and use either explicit
or implicit spatial derivatives and time integration techniques. To optimally exploit the resolution,
and taking into account the symmetries of the Maxwell equations, we use a Yee lattice\cite{yee} to stagger the
fields. The charge density is located at cell centers. To comply with Gauss law (\Eq{eq:gauss}) the electric
fields and the current density, both entering with the same spatial distribution in Amp\`ere's law (\Eq{eq:ampere}),
are staggered upwards to cell faces, while magnetic fields, to be consistent with the curl operator in
\Eq{eq:ampere}, are placed at cell edges. With this distribution (see \fig{fig:yee}) the
derivatives in \Eqs{eq:gauss}-\ref{eq:ampere} are automatically calculated at the right spatial positions,
and no interpolations are needed. Because of the spatial staggering the numerical derivatives commute,
and the magnetic field evolution conserves divergence to round-off precision. The electric potential $\phi_E$
is located at cell centers, and the magnetic potential $\phi_B$ is at cell corners. For the time integration we stagger the
proper velocity of the macro particles $\uug_p$ and the associated current density $\JJ$ backwards time.
In the \ppcode{} the natural order to do the different updates in a time step is (see \fig{fig:time})
\begin{enumerate}

\item{Update the proper velocity $\uug_p$ of the macro particles using either the Boris or Vay particle pusher}
\item{Calculate the charge density on the mesh $\rho^i(\xx_p^i)$, update the macro particle positions to
$\xx_p^{i+1}$, and recalculate the charge density $\rho^{i+1}(\xx_p^{i+1})$}
\item{If using a charge conserving method, find the current density from the time derivative of the charge density
$\hat\partial_t \rho = [\rho^{i+1}(\xx_p^{i+1}) - \rho^i(\xx_p^i)]/\Delta t$, else calculate it directly from
the particle flux $\uu_p$ interpolated on the mesh.}
\item{Given the time staggered current density, update the magnetic field using an implicit method}
\item{Combine the old and new magnetic field values to make a time centered update of the electric
field}
\end{enumerate}
In between step 3 and 4 first the particle-, charge- and current boundary conditions are applied, together with an exchange
of particles between MPI threads. Then the particle array, local to each thread, is sorted sequentially according
to the cell position. The sorting assures optimal cache locality when computing the charge density, and step 1 to 3 can
be executed in one sweep, and using small cache friendly buffer arrays for accumulating $\rho_c$ and $\hat\partial_t\rho_c$
or $\JJ$. The boundary conditions for the magnetic and electric fields are applied while solving their time evolution
in step 4 and 5. The different steps are described in more detail below.

The time update shown above corresponds to a second order Leap Frog method.
Compared to for example Runge-Kutta methods, or other methods where all variables are time centered, the advantage
is that it requires zero extra storage for intermediate steps, and the particle update is symplectic, giving stable
particle orbits. In general, symplectic integrators use a pair of conjugate variables. In this case they are the position,
electromagnetic fields and charge density and the particle proper velocities and current densities. As detailed below,
the leap frog update is only possible because we evaluate the $\uu\times\BB$ term in the Lorentz force using an
implicit evaluation of the velocity. There exists higher order symplectic integrators with several substeps of the general
positions and momenta composing a full time update. But only for a subset of these integrators the position is always at
the mid-distance in time between the current and new momenta; a necessary condition to be able to update the particle momenta.

We have implemented a symmetric fourth order symplectic integrator\cite{Yoshida:1990,Forest:1990,Candy:1991}
in the code. Due to its symmetric nature, a single, full timestep is performed by
simply taking four leap frog substeps, making it an almost trivial change to implement,
with updates needed only in the main driver.  The method
significantly increases the long term stability of the evolution for e.g.\ streaming plasma simulations, at the price of increasing
the cost of a single time step by a factor of 4. For three dimensional simulations, where increasing the resolution
by a factor of 2 costs a factor of 16 in cpu time this can be very worth while, compared to using the second order
leap frog method with a $\sqrt{2}$ higher resolution.

\subsection{Particle motion}
To move the macro particles forward in time we have to solve \Eq{eq:pmotion}. Because of the time staggering
it is straight forward to make an $\mathcal{O}(\Delta t^2)$ precise position update
\begin{equation}
\xx_p^{i+1} = \xx_p^{i} + \Delta t\, \uu_p^{i+1}\,,
\end{equation}
where the upper index $i$ indicates the iteration number. The position $\xx_p^i$ is evaluated at $t^i$, while the
proper velocity $\uug_p^i$ is staggered backwards in time and evaluated at $t^{i-1/2}=t^{i}-\Delta t/2$.

The proper velocity update is a bit more delicate. To find a time centered Lorentz force we need a time centered
value for the velocity $\uu_p$. This gives an implicit equation for $\uug_p^{i+1}$, and traditionally the Boris particle
pusher\cite{boris} has been used. It is formulated so that first the time centered proper velocity is computed as the
average $\uug_p(t^i)=(\uug_p^i + \uug_p^{i+1})/2$, and from that the normal velocity $\uu_p(t^i)$ is derived. It is
still an option in the code, but Vay has showed\cite{vay:2008} that the Boris pusher is not Lorentz invariant, and
gives incorrect solutions in simple relativistic test cases. Instead, the Vay particle pusher calculates
$\uu_p(t^i)=(\uu_p^i + \uu_p^{i+1})/2$ directly, as the average of the two three velocities. Even though the resulting
equations for $\uug_p^{i+1}$ involve the root of a fourth order polynomial, there is an analytic solution, and
the end result is a Lorentz invariant proper velocity update. Therefore, the Vay particle pusher is now the
standard option for updating the proper velocity.  It is $\mathcal{O}(\Delta t^2)$ precise.

\subsection{Solving Maxwell's Equations}
In the \ppcode{} Maxwell's equations are solved by means of an implicit scheme for evolving the magnetic field,
$\BB^{i} \rightarrow \BB^{i+1}$, followed by an explicit update of the electric field, $\EE^{i} \rightarrow \EE^{i+1}$.
De-centering of the integrator may be employed, such that the implicit magnetic field term, $\BB^{i+1}$, is weighted
higher than the explicit term, $\BB^i$, by setting a parameter $\alpha$, at run initialization.

With this de-centering of the implicit averaging of $\BB$ taken at two consecutive time steps, the discretized forms
of Faraday's and Amp\`{e}re's Laws read
\begin{equation} \label{eq:bnplus1}
\frac{{\bf B}^{i+1}-{\bf B}^i}{\Delta
t} = - \hat\nabla^+ \times \left(  \alpha  {\bf E}^{i+1} + \beta  {\bf E}^i
\right)\,,
\end{equation}
and
\begin{equation} \label{eq:enplus1}
\frac{{\bf E}^{i+1}-{\bf E}^i}{\Delta t} = c^2 \left( \hat\nabla^- \times \left( \alpha {\bf B}^{i+1} + \beta {\bf B}^i \right)
                                           - \mu_0 {\bf J}^{i+1} \right)\,,
\end{equation}
where, $\alpha$ and $\beta=1-\alpha$ with $\alpha\geq1/2$, quantifies the forward de-centering of the implicit magnetic
field term and the high frequency wave damping strength and spectral width that accompanies the choice. The hat
and sign denotes that it is the discrete version of the differential operator and that it is applied downwards or upwards.
$i$ indicates the iteration. The current is downstaggered in time, and the $i+1$ iteration is already calculated
from the macro particle distribution, when starting to solve the Maxwell equations.

Isolating $\EE^{i+1}$ in \Eq{eq:enplus1} and taking the curl, we can insert it in to \Eq{eq:bnplus1}. Using the vector
identity $\nabla\times\nabla\times\BB=\nabla\left(\div\BB\right)-\laplace\BB$, which also holds for the staggered
discretized operators, and using $\div\BB=0$, produces an elliptic equation for $\BB^{i+1}$
\begin{align}
 (1-c^2\alpha^2\Delta t^2\hat\nabla^2)\BB^{i+1} =& \BB^i + c^2 \alpha \beta {\Delta t}^2 \hat\nabla^2 \BB^i \nonumber \\
          &      - \Delta t \hat\nabla^+ \times \EE^i  \nonumber \\ \label{eq:EllipticB}
          &      - c^2 \alpha \mu_0 {\Delta t}^2 \hat\nabla^+ \times \JJ^{i+1}\,.
\end{align}
The right hand side of \Eq{eq:EllipticB} contains only known terms (at time $t_i$
for $\EE^i$ and $\BB^i$, and $t_{i} + \Delta t/2$ for $\JJ^{i+1}$), but the operator on the left hand side
is elliptic, complicating a direct solve. We have implemented a simple iterative
solver taking $\BB^{i+1}=\BB^{i}$ as a first guess, with a solution found by successive relaxation. Elliptic
equations are non-local and our solver requires repeated updates of boundaries and ghost-zones. However,
the limitation to the parallel scalability is not serious, in that convergence is normally reached in 1-10 iterations
for most simulation setups, with tolerances on the residual error
of about $10^{-6}$. Once the relaxed solution has been found, and provided that the initial simulation setup had
$\div\BB=0$, going forward $\div\BB=0$ is guaranteed, due the constraint ($\div\BB=0$) being implicitely built into 
the derivation of \Eq{eq:EllipticB}. Having found $\BB^{i+1}$, the electric field is simply updated explicitly by 
\Eq{eq:enplus1}.

The field integrator is unconditionally stable for $1/2^+ < \alpha < 1^-$. However, the scheme tends to damp out high-frequency
waves due to the de-centered implicit nature of the scheme, and the solver is only second order accurate for values
$\alpha\approx1/2^+$. Besides providing a tunable stabilization of the field integration scheme, this parameter also 
determines how large a time step can be chosen, and the damping of high frequency waves. Empirically, with our highest 
order scheme (6$^\textrm{th}$ order fields, PCS particle assignment), for values $\alpha\geq0.525$ the scheme is 
numerically stable.

\begin{figure}
     \begin{center}
           \includegraphics[width=\linewidth]{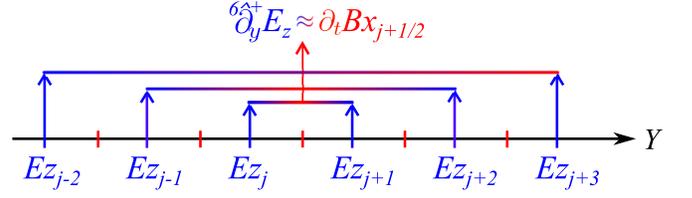}
     \end{center}
\caption{Example of the sixth order difference operation $\curl$ in 1D (along the y-axis). Due to the Yee mesh staggered
layout of variables, the central difference is computed exactly where needed. The differential operator called
\texttt{'ddyup'} (in the code) produces the derivative w.r.t.~the Y-axis, up-shifted one half mesh point on the axis. For
the example in the figure this produces the correct derivative of $B_y$ at the desired mesh point location, $y_{j+1/2}$.
In the nomenclature adopted here, this operator is denoted $^6\hat{\d}_{y}^+$. Cell centers are marked in blue, and cell
edges in red.  Further, compare this figure with the respective components in \fig{fig:yee}.} \label{fig:diffoper}
\end{figure}

Forming spatial derivatives of the field quantities is done by finite differencing on a uniform (but not necessarily
isotropic), mesh. A set of operators identical to those used in the \staggercode~ -- also
developed and maintained in Copenhagen\citep{1997LNP...489..179N} -- are implemented, with a choice between
2$^\textrm{nd}$, 4$^\textrm{th}$ and 6$^\textrm{th}$ order accuracy in space. Staggering of the variables on a Yee 
lattice~\cite{yee} leads to highly simplified computations for the difference equations. For example, for Faraday's 
Law (\Eq{eq:bnplus1}) the $x$-component, namely $\left[\d_t \BB\right]_x \equiv \d_y E_z - \d_z E_y$, along the 
$y-axis$ reduces to $\d_t B_x = \d_y E_z$. From \fig{fig:yee} we see that this computation yields the desired value 
exactly where needed, provided that we compute the central differences at the half-staggered mesh point; a single 
component of the $\curl$ operator is illustrated in \fig{fig:diffoper}.

In the \ppcode, for the 6$^\textrm{th}$ order accurate finite difference first derivative with respect to
the $y$-axis the expression reads
\begin{align}
 ^6\hat{\d}_y^+ f_{j+1/2} =       & ~ ~ ~ ~   a \left(\frac{f_{j+1}-f_{j}}{\Delta y}\right) \nonumber \\
                                  &         + b \left(\frac{f_{j+2}-f_{j-1}}{\Delta y}\right) \nonumber \\
                                  &         + c \left(\frac{f_{j+3}-f_{j-2}}{\Delta y}\right)~,
\end{align}
with coefficients $a = {25}/{21}$, $b = {-25}/{384}$ and $c = {3}/{640}$, matching those of the
\staggercode. For our example of Faraday's Law above, the corresponding expression to compute becomes
\begin{align}
 ^6\hat{\d}_y^+ E_z(y_{j+1/2}) = & ~~~~   \frac{25}{21}  \left(\frac{E_z(y_{j+1})-E_z(y_{j})  }{\Delta y}\right) \nonumber \\
                                 &      + \frac{-25}{384}\left(\frac{E_z(y_{j+2})-E_z(y_{j-1})}{\Delta y}\right) \nonumber \\
                                 &      + \frac{3}{640}  \left(\frac{E_z(y_{j+3})-E_z(y_{j-2})}{\Delta y}\right)
\end{align}
for that component along the $y$-direction.

This choice of coefficients (based on a Taylor expansion) for the higher order differential operators has enabled
us to consistently port simulation data from the \staggercode{} to the \ppcode, for coupling of MHD simulations and
PIC simulations.

\subsection{Charge conserving current density}
Assuming the charge and current density on the mesh to be volume averaged, just like the fields, and inserting the phase
space density of a set of macro particle (\Eq{eq:pseudop}) into \Eq{eq:rho} for the charge density,
we find the charge density in a cell as
\begin{equation}\label{eq:rho2}
\rho_c(\xx_c) = \!\!\!\! \sum_{\textrm{particles}} \!\!\!\! q_p \, w_p \, W(\xx_p-\xx_c)\,,
\end{equation}
where $q_p$ and $w_p$ are the charge and number density of the macro particle. Another reason for choosing volume averaging,
just like with the electromagnetic fields, is that the computation of the fields at the particle positions and the charge
density on the mesh has to use the same interpolation technique, or particles can induce
self-forces\cite{birdsall:1985,hockney:1988}.
In principle, one could use a similar definition for the current density
\begin{equation}\label{eq:jj2}
\JJ_c(\xx_c) = \!\!\!\! \sum_{\textrm{particles}} \!\!\!\! q_p \, w_p \, \uu_p W(\xx_p-\xx_c)\,,
\end{equation}
but for the discretized version of Gauss law (\Eq{eq:gauss}) to hold true the correspondingly
discretized charge conservation law,
\begin{equation}\label{eq:cc1}
\hat\partial_t \rho_c + \divn{\JJ} = 0\,,
\end{equation}
has to be satisfied. In general, with the above definitions for the charge and current densities,
this will not be the case. While it holds for particles moving inside a cell, particles that move through
a cell boundary in a single time step can violate charge conservation. Methods have been
developed\cite{eastwood:1991,villasenor:1992,esirkepov:2001,umeda:2003} for second order field solvers
that instead of using \Eq{eq:jj2} to compute $\JJ$ use different schemes to directly find $\JJ$
via \Eq{eq:cc1}. In particular Esirkepov showed\cite{esirkepov:2001} how to make a unique linear
decomposition of the change in charge density $\hat\partial_t \rho_c$ for arbitrary shape functions into different
spatial directions, and decouple \Eq{eq:cc1} into a set of differential equations, each involving
only one component of the current density
\begin{equation}\label{eq:jj3}
\hat\partial_i^- J_i = \left.\hat\partial_t \rho_c\right|_i \equiv \mathcal{D}\rho_i \,.
\end{equation}
The directional time derivatives of the charge density $\mathcal{D}\rho_i$ are computed on a macro particle basis,
as described by Esirkepov\cite{esirkepov:2001}, and for a second order differential operator,
\begin{equation}
^2\hat\partial_i^- J_i = \frac{J_i^0 - J_i^-}{\Delta x_i}\,,
\end{equation}
it is straight forward to compute $J_i$ from a single particle with a simple prefix sum, assuming
that the contribution of a particle to the current density has compact support on the mesh and far away is zero.
In the \ppcode{} the current density is found using this method when a second order field solver is in use.
Note that \emph{all} PIC codes based directly on the published charge conserving
methods\cite{eastwood:1991,villasenor:1992,esirkepov:2001,umeda:2003} work with second order field solvers \emph{only},
since the order of the discretized differential operators for Gauss law and charge conservation has to be the same.
The Esirkepov method has a very concise formulation and is well suited to generalize to
higher order field solvers, but it is not clear how to couple the methods by Eastwood\cite{eastwood:1991},
Villasenor and Buneman\cite{villasenor:1992}, or Umeda\cite{umeda:2003} to higher order field solvers.

For higher order field solvers it is more complicated to solve \Eq{eq:jj3}. For example a sixth order
differential operator involves sixth mesh points
\begin{equation} \label{eq:jj6}\footnotesize{
^6\hat\partial_i^- J_i\! = \frac{1}{\Delta x_i}\!\left[c(J_i^{++}\!\!\!\!\!- J_i^{---}) +
b(J_i^{+}\!\! - J_i^{--}) + a(J_i^0\!\! - J_i^-)\right]}
\end{equation}
and the simple prefix sum that can be used at second order turns into a linear set of equations. It is also not clear
what the boundary conditions, even for a single particle with compact support, should be. In the simplest case, for a
single particle, the matrix will look something like
\begin{equation}
\footnotesize{
\left( \begin{array}{cccccccc}
 a &  b     &  c & 0      &    &\ldots & 0 \\
-a &  a     &  b & c      & 0  &\ldots & 0 \\
-b & -a     &  a & b      & c  &\ldots & 0 \\
 0 & \vdots &    & \ddots &    &\vdots & 0 \\
 0 & \ldots & -b & -a     &  a &  b    & c \\
 0 & \ldots & -c & -b     & -a &  a    & b \\
 0 & \ldots &  0 & -c     & -b & -a    & a
\end{array}\right) \left(\begin{array}{c}
J_i^{5-} \\
J_i^{4-} \\
J_i^{3-} \\
\vdots \\
J_i^{3+} \\
J_i^{4+} \\
J_i^{5+}
\end{array}\right) = \left( \begin{array}{c}
 0 \\
 0 \\
\mathcal{D}\rho_i^{3-} \\
\vdots \\
\mathcal{D}\rho_i^{3+} \\
 0 \\
 0 \\
\end{array}\right)} \,,
\end{equation}
where we for simplicity have absorbed $\Delta x_i$ in the definition of $a$, $b$, and $c$. There are
several problems with using the above set of equations: i) it is not clear how the ``no current
far away'' condition is implemented, and because of the few points involved the cutoff will have some impact on the result
ii) the equation for a single macro particle with cubic interpolation
easily involves a 10$\times$10 matrix, and iii) as formulated above, the problem is highly unstable, because
the coefficients are alternating and of very different size ($a=25/21$, $b=-25/384$, $c=3/640$). Furthermore
to solve the above system of equations for three directions and every single macro particle is both very costly, and
will introduce noticeable round-off error on the mesh, when all contributions from all particles are summed up.
In the \ppcode{} we have developed a fast and stable alternative, similar in structure to the one published
recently\cite{londrillo:2010} for the Aladyn code. Instead of solving the current density for each macro particle,
taking advantage of the linearity of the problem, we sum up $\mathcal D\rho_i$ directly on the mesh.
Then \Eq{eq:jj6} can be solved on the whole domain. There is still the problem of the alternating coefficients,
but that can be dealt with, by first solving for the difference $\Delta J_i \equiv J_i^0 - J_i^-$, and then
using a prefix sum, just like in the second order method, to find $J_i(x_i)$. In terms of the differences the differential
operator becomes
\begin{equation} \label{eq:jj6d}
^6\hat\partial^-_i J_i\! = \!\tilde c (\Delta J_i^{++}\!+ \Delta J_i^{--}) +
                                            \tilde b (\Delta J_i^+ + \Delta J_i^-) + \tilde a\Delta J_i^0\,,
\end{equation}
where the coefficients are $\tilde a = (a+b+c)/\Delta x_i$, $\tilde b = (b+c)/\Delta x_i$, $\tilde c = c/\Delta x_i$,
and the corresponding linear system, which now has to be solved on all of the mesh, is a symmetric penta-diagonal system
\begin{equation}\label{eq:penta}
\footnotesize{
\left( \begin{array}{cccccccc}
\tilde a & \tilde b & \tilde c & 0        &          & \ldots    & 0 \\
\tilde b & \tilde a & \tilde b & \tilde c & 0        & \ldots    & 0 \\
\tilde c & \tilde b & \tilde a & \tilde b & \tilde c & \ldots    & 0 \\
0        & \vdots   &          & \ddots   &          & \vdots    & 0 \\
0        & \ldots   & \tilde c & \tilde b & \tilde a & \tilde b  & \tilde c \\
0        & \ldots   & 0        & \tilde c & \tilde b & \tilde a  & \tilde b \\
0        & \ldots   &          & 0        & \tilde c & \tilde b  & \tilde a \\
\end{array}\right) \left(\begin{array}{c}
\Delta J_i^0 \\
\Delta J_i^1 \\
\Delta J_i^2 \\
\vdots \\
\Delta J_i^{n-2} \\
\Delta J_i^{n-1} \\
\Delta J_i^n
\end{array}\right) = \left( \begin{array}{c}
\mathcal{D}\rho_i^{0} \\
\mathcal{D}\rho_i^{1} \\
\mathcal{D}\rho_i^{2} \\
\vdots \\
\mathcal{D}\rho_i^{n-2} \\
\mathcal{D}\rho_i^{n-1} \\
\mathcal{D}\rho_i^{n}
\end{array}\right) \,.
}
\end{equation}
Pentadiagonal systems have stable and efficient solvers\cite{penta:1996}, and it costs a negligible amount of cpu time
compared to the rest of the code to find the current density on the mesh, $\JJ$, from the decomposed time derivative in the
charge density, $\bm{\mathcal{D}\rho}$. Furthermore, because the evaluation is done directly on the mesh, the accumulated errors
in the charge conservation are smaller than if the current density is computed on a macro particle basis. To avoid having to
specify boundary conditions for $\bm{\mathcal{D}\rho}$ and introduce new terms into the matrix in \Eq{eq:penta} we use
an extra virtual cell layer. Before a timestep, by definition, the charge density will always be zero in the virtual
cell layer, and the
current density can then be computed correctly with a ``zero-current'' ansatz for the current density on the lower boundary.
Afterwards the normal boundary conditions for the current
density can be applied, just like when using \Eq{eq:jj2} for computing the current density. This method
also completely decouples the solution, when running the code with multiple MPI threads. If instead
of a sixth order field solver a fourth order field solver is used the system \Eq{eq:penta} becomes tridiagonal.

In the original version of the \ppcode{} \Eq{eq:jj2} is used to compute the current density. The error
introduced into Gauss law (\Eq{eq:gauss}) is mostly on the Nyquist scale, and a simple iterative Gauss-Seidel
filtering technique is used to correct $\EE$.
The module also computes the error, and has been used to validate the charge conserving methods.
In most applications, the relative error with the old method can be kept on a $10^{-4}$-$10^{-5}$ level by running the filter
5 to 10 times per iteration, but there is no unique way to correct the electric field, and the divergence cleaning
introduces tiny electric fluctuations, which couple back to the particles through the Lorentz force. Apart from the
higher cost of the elliptic filter when running on many cores, the method is worse at conserving energy and has a
larger numerical heating rate for cold plasma beams than a charge
conserving method, and without a current filter it can be unstable.
When running the code with charge conservation and in single precision at high resolution (e.g.~$\sim\!\!1000^3$ cells
and 10-100 billion particles), over time the numerical round-off noise will eventually build up errors both in Gauss' law and
in the solenoidal nature of the magnetic field (\Eq{eq:divb}). A single filtering step on the electric and magnetic fields every
$\sim$5$^\textrm{th}$ iteration is enough to keep the relative errors at the $10^{-5}$ level.
When running the code in double precision we have not seen any need to use divergence cleaning, and if the initial and
boundary conditions obey \Eqs{eq:gauss} and \ref{eq:divb}, then the relative error typically stays below
$\sim\!\!10^{-7}$.

\section{Radiative cooling}
Very high energy electrons loose momentum by emitting radiation.
The emission in itself is a very valuable diagnostic and the extraction of the spectrum is
treated below, but the energy loss for the high energy particles is normally not computed in
particle-in-cell codes. Building on the work on radiative losses done by
Hededal\cite{Hededal:2005b} in our old PIC code, we have developed a numerically stable method
for correctly calculating radiative losses in the \ppcode{}. The radiated power from a single
particle $P_\textrm{rad}$ can be written as\cite{Hededal:2005b}
\begin{equation}\label{eq:pradv}
P_\textrm{rad} = \frac{\mu_0 q^2}{6\pi c} \left[ \gamma^4 \dot\uu^2 +
\frac{\gamma^6}{c^2} (\dot\uu\cdot\uu)^2\right]\,,
\end{equation}
where dot denotes the time derivative.

Denoting the proper velocity $\pp=\uug$, and using the identities
$\pp\cdot\dot\pp = \gamma^4 \uu\cdot\dot\uu$, and
$\dot\pp = \gamma \dot\uu + c^{-2}\gamma^3 \uu (\uu\cdot\dot\uu)$, we
can rewrite \Eq{eq:pradv} to
\begin{align}\nonumber
P_\textrm{rad} & = \frac{\mu_0 q^2}{6\pi c} \left[ \gamma^2 \dot\pp^2 -
\frac{1}{c^2} (\dot\pp\cdot\pp)^2\right] \\ \label{eq:prad}
& = \frac{\mu_0 q^2}{6\pi c} \left[ \dot\pp^2 + \frac{1}{c^2}
                                   (\dot\pp\times\pp)^2\right] \,.
\end{align}
Notice how \Eq{eq:prad} only involves proper velocities, and is numerically stable at
both high and low Lorentz factors, while the first version relies on a cancellation between
$\gamma^2$ and $\pp^2$, and \Eq{eq:pradv} uses three velocities, prone to numerical errors.

The radiative cooling always acts in the opposite direction to the proper velocity vector, and
therefore the change in the length of the proper velocity vector is directly related to the change
in the kinetic energy and $P_{rad}$:
\begin{align}\nonumber
\dot\pp_\textrm{rad} &= \frac{\pp\cdot\dot\pp_\textrm{rad}}{\pp^2} \pp \\ \label{eq:udot}
  &= - \frac{\mu_0 q^2}{6 \pi m c} \frac{\gamma}{\pp^2}
   \left[ \dot\pp^2 + \frac{1}{c^2} (\dot\pp\times\pp)^2\right] \pp\,,
\end{align}
where we have used that $\pp\cdot\dot\pp = c^2 \gamma\,\dot\gamma$, and
 $m c^2 (\gamma - 1\dot{)\,} = -P_\textrm{rad}$.

In the \ppcode{} the Boris or the Vay pusher is used to advance the particles.
To integrate the effect of radiative cooling, for simplicity and to keep the scheme
explicit, we assume that the cooling in a single timestep only changes the energy with
a minor amount. The particle pushers advances the four velocities from time step
$t - \Delta t/2$ to $t + \Delta t/2$, while accelerations are computed time centered
at $t$. Below we denote the three times with $-$, $+$, and $0$ respectively.

To get a time centered cooling rate $\pp_0$, $\dot\pp_0$, and $\gamma_0$ are needed.
The proper acceleration $\dot\pp_0$ is already naturally time centered, but with the Vay
Pusher, for consistency one should use the time centered three velocity to compute $ww_0$
and $\gamma_0$, which however is numerically imprecise when using single precision.
To get a more numerically precise, albeit ever so slightly inconsistent, measure for
$\dot\pp_0$ and $\gamma_0$ we use the time averaged proper velocities
\begin{align}
\pp_0 & = \frac{\pp_+ + \pp_-}{2} &
\gamma_0 & = \sqrt{1 + \pp_0^2 c^{-2}} \,.
\end{align}
These values can then be plugged into \Eqs{eq:prad} and \ref{eq:udot} to find the change in
momentum due to radiative cooling as
\begin{align}\nonumber
\pp_+ & = \pp_- +  \dot\pp_0 \Delta t \\ \nonumber
& = \pp_- + (\dot\pp_\textrm{EM} + \dot\pp_\textrm{rad})\Delta t \\
       & = \pp_+^\textrm{EM} + \dot\pp_\textrm{rad} \Delta t\,.
\end{align}
In principle one could find the converged solution to the above non-linear equation system, under
the assumption that cooling is a small correction, by iterating the radiative cooling computation a
couple of times, while updating $\pp_+$ and $\dot\pp_0$. But in practice we have
found that the direct explicit calculation of the cooling rate done by making no iterations is acceptable.

\section{Initial and boundary conditions}
Setting up consistent boundary and initial conditions for a particle-in-cell code can be non-trivial, due to the
mixed particle and cell nature and the staggering in space and time. In the \ppcode{} the initial and boundary
conditions are supported through the loading of different boundary and initial condition modules, and over time
a number of modules have been developed.

\subsection{Particle Injection}
Macro particles in a particle-in-cell code sample phase space and are by construction placed using a
random generator. Any call to a random generator is related to the position in the mesh, where the pseudo-random
number is needed. To make the random number generation scalable, but independent of the parallelization technique,
we have implemented two types of random generators: i) We have developed a multi stream variant
of the Mersenne Twister random generator\cite{mersenne} to generate high quality random numbers, and setup one
stream per $xy$-slice of cells.
This generator is used in e.g.~shock simulations where the particle average density and velocity is a function of a single
coordinate. ii) In the case of more complicated setups, e.g. when using snapshots from MHD simulations as described below,
a simpler random generator is used, where the state is contained in a single 32-bit integer, but each mesh point has a its own
state. The Mersenne Twister generator is used to generate the initial seed in each cell for the simple random generator.
To sample the velocity phase-space we have implemented cumulative 2D and 3D relativistic Maxwell
distributions\cite{Dunkel:2009} using an inverted lookup table\cite{birdsall:1985}. It is important to use the correct
dimensionality when initializing the velocities, or the corresponding 2D or 3D temperature will be incorrect.
The particle positions are correspondingly injected either uniformly in an $xy$-slice or in a single cell. Notice that using
an injection method in a full $xy$-slice allow for larger density fluctuations inside the cells, while if using a cell-by-cell
injection method there will only be fluctuations on the sub-cell level. Depending on the physical model on hand one or the
other method may be more desirable. We typically use a standard technique to inject particles in pairs of different
charge type to avoid having free charges initially. But the code is flexible, and has a built in module to correct Gauss law.
Therefore it is also possible to inject particles completely at random and correct the electric field to include the
non trivial effect of electrostatic fluctuations from the initial non-neutral charge distribution.

\subsection{Boundary conditions}
We handle periodic boundaries trivially by padding the domain with ghostzones, a technique also used at the edge of
MPI domains, and copying fields from the top of the domain to the bottom and vice versa while updating boundaries.
The particles, on the other hand, are simply allowed to stream freely and the position is in all but a few cases calculated
modulo the domain size. To maintain uniform numerical precision even for very large domains the position is
decomposed internally as an integer cell number, and a floating point number giving the fractional position inside the cell.

Reflective boundaries are implemented using ``virtual particles'' on the other side of the reflecting boundary,
taking in to account the symmetries of the Maxwell equations and the staggering of the mesh. By convention the
boundary is placed at the center of the cell (i.e. the charge density is \emph{on} the boundary), below
for simplicity taken to be an upper boundary. When the charge and current densities on the mesh are calculated
for each particle inside the boundary a corresponding virtual ``ghost particle'' outside the boundary has to be
accounted for. Particles close to the reflecting boundary will contribute to the charge and current densities outside
the boundary, while the virtual particles will contribute a corresponding charge and current density inside the
boundary. This can most efficiently be calculated by disregarding the boundary at first after a particle update,
and calculate the charge and current density as if the particles were streaming freely. The contribution to the
charge and current densities inside the boundary from the virtual particles can then be calculated by taking into
account the symmetry. It is easily seen that this corresponds, up to a sign, to the contribution of the normal
particles outside the boundary
\begin{align}
   \rho_c(x_\textrm{b} - \delta x) &=    \rho_c(x_\textrm{b} - \delta x) +    \rho_c(x_\textrm{b} + \delta x) \\
\JJ_\parallel(x_\textrm{b} - \delta x) &= \JJ_\parallel(x_\textrm{b} - \delta x) + \JJ_\parallel(x_\textrm{b} + \delta x) \\
 J_\perp(x_\textrm{b} - \delta x) &=  J_\perp(x_\textrm{b} - \delta x) -  J_\perp(x_\textrm{b} + \delta x)\,,
\end{align}
where $x_\textrm{b}$ is the position of the boundary, $\delta x$ is the distance; i.e.~integer $\Delta x$ for centered
quantities, and half integer for staggered, $\parallel$ indicates components parallel and $\perp$ perpendicular to the boundary.
The perpendicular component of the current density is staggered and anti-symmetric across the boundary.
Only after calculating the current density on the mesh are all particles outside the boundary reflected according to
\begin{align}
x &\to 2\, x_\textrm{b} - x & v\gamma_\parallel &\to v\gamma_\parallel & v\gamma_\perp &\to - v\gamma_\perp
\,.
\end{align}
When the charge and current densities are correctly calculated inside and on the boundary we can start
considering the values outside. The staggered fields, i.e.~the perpendicular current density and electric field,
the parallel magnetic field, and the magnetic potential, may be shown to be antisymmetric across the
boundary
\begin{align}
   J_\perp(x_\textrm{b} + \delta x) &=    - J_\perp(x_\textrm{b} - \delta x) \\
   E_\perp(x_\textrm{b} + \delta x) &=    - E_\perp(x_\textrm{b} - \delta x) \\
\BB_\parallel(x_\textrm{b} + \delta x) &= - \BB_\parallel(x_\textrm{b} - \delta x) \\
   \phi_B(x_\textrm{b} + \delta x) &=    - \phi_B(x_\textrm{b} - \delta x) \,.
\end{align}
Conversely, the centered fields (in the direction of the boundary), charge density, parallel current
densities, electric fields, the electric potential, and the perpendicular component of the magnetic
field are symmetric across the boundary
\begin{align}
   \rho_c(x_\textrm{b} + \delta x) &=    \rho_c(x_\textrm{b} - \delta x) \\
\JJ_\parallel(x_\textrm{b} + \delta x) &= \JJ_\parallel(x_\textrm{b} - \delta x) \\
   \phi_E(x_\textrm{b} + \delta x) &=    \phi_E(x_\textrm{b} - \delta x) \\
\EE_\parallel(x_\textrm{b} + \delta x) &= \EE_\parallel(x_\textrm{b} - \delta x) \\
   B_\perp(x_\textrm{b} + \delta x) &=    B_\perp(x_\textrm{b} - \delta x) \,,
\end{align}
and we only need to determine the values of the centered fields at the boundary. The charge and current
densities are derived from the particle distribution and are therefore already given at all points, including the boundary.
$B_\perp(x_\textrm{b})$ is the only unknown component of the magnetic field, and it can be computed
from the solenoidal constraint $\hat\nabla^+\cdot \BB=0$ calculated at the boundary.
Given that the magnetic field is the first to be evolved forward in time it is then possible to self consistently
calculate the parallel electric field on the boundary $\EE_\parallel(x_\textrm{b})$, directly from the evolution equation.

Outflow boundaries are less constrained than reflecting boundaries, given the extrapolating nature, and various
types of damping layers and extrapolations have been considered\cite{PML2D,PML3D,umeda:2001}. We use a damping
layer of mesh points --- typically 20 --- to damp all perpendicular components of the electromagnetic fields, effectively
absorbing reflected waves, and combine it with an extrapolating boundary condition (i.e.~symmetric first derivative).
We allow particles to still generate current and charge densities on the mesh inside the box until they are well outside
the boundary. As long as the disturbances in the outgoing flow are small this works well and is stable for extended run times.

\subsection{Sliding window and injection of particles}
In highly relativistic flows it can be advantageous to simulate the plasma in a frame where the region of interest is
relativistic. This constrains the time such a region can be followed, because the computational domain has to be
continuously expanding or has to have an enormous aspect ratio in the flow direction. This is the case for example
for relativistic collisionless shocks. An alternative to this is to use a sliding window as the computational domain centered
on the region of interest and moving with the same velocity maintaining it at the center of the box\cite{movingframe}.
We have implemented this technique in the \ppcode{}, together with a moveable open boundary and particle injector. If
the window is moving with a velocity $v$ relative to the lab frame then in each iteration we check if
$v (t_\textrm{i} - t_{\textrm{old}}) > \Delta x$ and in that case we move the box a full mesh point, if necessary.
I.e.~if $v = 0.5 c$ and $\Delta t = 0.5 \Delta x / c$ the code will roughly move the window one point in every 4 iterations.
The move is implemented by removing one cell at one end of the box, translating everything one point, and injecting an
extra layer of inflow particles at the other end. This technique was used to successfully capture the long term
evolution of an 3D ion-electron collisionless shock\cite{haugboelle:2011}.

\subsection{Embedded particle-in-cell models}\label{sec:embed}
MHD models have successfully been applied for many years to study the large scale structure of plasmas from laboratory
length scales to the largest scales in the universe. MHD can reach over such enormous scales, because
a statistical description of the plasma is employed, where the microscopic state is captured using statistical quantities like the
temperature, viscosity and resistivity. On the other hand, the kinetic description of a plasma used in a particle-in-cell code
is ideally suited to investigate non-thermal processes, such as particle acceleration and particle-wave interactions, and
can be used to model and understand a much broader range of plasma instabilities than MHD. The drawback is that exactly
because the plasma is described in kinetic terms, explicit particle-in-cell codes, like the \ppcode{}, have to respect
microscopic constraints and resolve the Debye length, the plasma skin depth, and the light crossing time of a single cell.
Recently, we have developed a technique to couple the two approaches using the results from MHD simulations to
supply initial and boundary conditions for our PIC code. This has enabled us to for the first time make a realistic
particle-in-cell description of active coronal regions\cite{baumann:2012a,baumann:2012b}, and investigate the mechanism
that accelerates particles in the solar corona. The MHD code is used to evolve the global plasma over
several solar hours, and a snapshot just before e.g.~a major reconnection event is used to study a small time
sequence, of the order of tens of solar seconds, using the PIC code.

Given an MHD snapshot, typically a smaller cutout from a larger simulation of the region of interest,
we first interpolate to the resolution that is to be used in the PIC code. The interpolated magnetic fields are
corrected with a divergence cleaner that uses the same numerical derivative operator that are used in the PIC code,
to assure that the initial magnetic fields are solenoidal to roundoff precision. In a given cell the density in
the MHD snapshot is used to set the weight of individual particles. The particles are placed randomly inside the
cell, but in pairs, so that initially there are no free charges. The velocity of the particles have three
contributions,
\begin{equation}
v\gamma = v\gamma_{\textrm{bulk}} + v\gamma_{\textrm{thermal}} + v\gamma_{\textrm{current}}\,.
\end{equation}
The bulk momentum is taken from the MHD snapshot. The thermal velocity is sampled from a Maxwell distribution
using the MHD temperature, and finally the current speed is found from the ideal MHD current
$\mu_0\JJ = \mu_0 \sum_i q^i n^i v^i_\textrm{current} = \nabla \times \BB$. The average momentum has to
correspond to the bulk momentum in the MHD snapshot. Taking into account the mass ratio, then for example in a
neutral two component proton-electron plasma, with $n = \rho_{\textrm{MHD}} / (m_e + m_p)$, the weighting is
\begin{align}
v^e_\textrm{current} &= - \frac{m_p}{\mu_0 |q| \rho } \nabla \times \BB &
v^p_\textrm{current} &=   \frac{m_e}{\mu_0 |q| \rho } \nabla \times \BB\,,
\end{align}
and in general the correct way is to use harmonic weighting.
Finally, an initial condition has to be specified for the electric field $\EE$. One possibility is $\EE=0$.
It satisfies Gauss law -- the plasma is neutral initially -- but is inconsistent with the EMF from the MHD equations.
If using this initial condition, in the beginning of the run a powerful small scale electromagnetic wave is
launched throughout the box, when the $\partial_t \EE$ term in Amp\`ere's law adjusts the electric field on a
plasma oscillation
time scale. Another possibility is to set $\EE = - \uu \times \BB$, in accordance with the ideal MHD equations.
Then there is no guarantee that Gauss law is satisfied. In the code the second choice is used, but small scale
features in the electric field are corrected by running the build-in Gauss law divergence cleaner for a few iterations. The
remaining difference is adjusted by changing slightly the ion- and electron-density, making the plasma charged.
Typically this only leads to small scale changes. The resulting initial electric field then both satisfies Gauss law, and
is almost in accordance with the MHD EMF. Apart from using the MHD EMF, we have also options to add the Hall
and Battery effect terms.

To evolve the model, boundary conditions on all six boundaries are needed. For the plasma they are constructed
exactly like the initial conditions. They can easily be made time dependent, by loading several MHD snapshots and
interpolating in time. In every time step, when applying the boundary conditions, first all particles in the boundary
zones are removed -- also particles that have crossed the boundary from the interior of the box -- and are then
replaced with a fresh plasma, according to the MHD snapshot. The boundary is typically 3 zones broad;
enough to allow for the sixth order differential operators on the interior of the mesh, and enough to make a well
defined charge and current density with the cubic spline interpolation. This plasma is retained when evolving, and
particles from the boundary zones are allowed to cross into the interior of the computational domain. By maintaining a
correct thermal distribution in the boundary zones, and simply letting the dynamics decide which particles stream
into the box, the inflow maintains a perfect Maxwell distribution, and the resulting plasma is practically identical
to what would have been obtained with the open boundary method of Birdsall et al\cite{birdsall:1985},
but is much simpler to implement, and correctly accounts for bulk velocities and currents in and out of
the box. If there is a differential between the charge or electric current in the boundary. Inside the domain the
$\partial_t \EE$ term adjusts the plasma almost instantaneously, and the balance is
maintained. This boundary condition is very similar to a perfect thermal bath, but with in- and out-going bulk
velocities and currents.

We do not keep the fields fixed at the boundary condition, but instead let them evolve freely, only subject to
reasonable symmetry conditions at the boundary, which keep them consistent with the Maxwell equations.
To respect the symmetry of the equations we let
\begin{itemize}
\item{$E_\perp$ symmetric, $\EE_\parallel$ antisymmetric}
\item{$\partial_{\perp} B_\perp$ symmetric, $\BB_\parallel$ symmetric}
\item{$\rho_c$ and $\JJ$ specified according to MHD snapshot}
\item{$\phi_E$ symmetric, $\phi_B$ antisymmetric}\,,
\end{itemize}
where $\parallel$ are the components parallel with the boundary and $\perp$ the component perpendicular to
the boundary. For the relatively short times that we have evolved imbedded PIC
simulations\cite{baumann:2012a,baumann:2012b} these boundaries are stable.

A severe limitation for coupling PIC and MHD codes is that in many situations the Debye length, plasma frequency
and other microphysical length and time scales are many orders of magnitude smaller than the scales of interest.
For example, in the solar corona the Debye length is measured in millimeters, while interesting macroscopic
scales are measured in
megameters. If we were to simulate the true system using an explicit PIC code we would need roughly $(10^8)^3$ cells,
which is computationally unfeasible in the foreseeable future. To circumvent this problem we have developed a novel
method, in which we rescale the physical units while maintaining the hierarchy of time, length and velocity scales.
The \ppcode{} is flexible and can be employed with a range of different unit systems. Furthermore all natural
constants are maintained in the code. The rescaling technique is discussed in detail in Baumann et al\cite{baumann:2012a}.

\section{Diagnostics}
\subsection{Particle tracking and field slicing}
Particle-in-Cell simulations are routinely run with billions of particles, with each particle taking up $\sim$50 bytes, and for
$10^3$-$10^6$ iterations. To store the full data set from every single iteration would take up petabytes of storage,
and is impracticable. Instead, a standard practice when running PIC simulations is to only store every n$^\textrm{th}$
particle in a snapshot, and dump snapshots with a reduced frequency, decreasing the data volume dramatically. But
to understand the underlying physics of for example particle acceleration in detail a more fine-grained approach is
warranted. To that end we have implemented dumping of field slices and tracking of individual
particles. Any particle in the code can be tagged for particle tracking, according to a number of criteria. For example
based on its energy, at random, or according to the specific ID of the particle, which is reproducible between runs.
The tagged particles are harvested by each MPI thread individually, and together with the position and momentum the
local values of the current, density, electric and magnetic field are recorded, by interpolation from the mesh to the particle
position. Everything is arranged in a single array that is sent to the master thread. The master thread then dumps the particle
records to a single file, appended to in each iteration. On x86 clusters we can sustain tracing a million particles without significant
performance degradation, while on clusters with weaker CPUs, such as BlueGene/P, we are limited to $\sim10^5$ particles.
To put the particle tracks into context, data on of the field evolution is also needed. To save time-resolved field data
we have made a field slicing module, where a large selection of fields (e.g.~the electric field $\EE$, $\EE\cdot\BB$,
$\JJ\times\BB$ etc) may be stored as 2D slices. The extent of a slice, and the number of field layers in the perpendicular
direction to the slice, used for averaging, is user selectable. These two techniques have been used in concert,
to understand the mechanism behind particle acceleration in reconnection events in the solar corona
\cite{baumann:2012a,baumann:2012b}, and in collisionless shocks \cite{haugboelle:2011}. Both diagnostics a interactively
steerable: parameters can be changed, and diagnostics can be turned on and off while a simulation is running.

\subsection{Synthetic Spectra}
The radiation signature from an large number of of accelerated charges is not easily computed analytically from first
principle, for plasmas with complex fields topologies, rich phase space structure, and temporal
evolution. Application examples include relativistic outflows and collisionless shocks; more specifically,
for example, gamma-ray  bursts, where magneto-bremsstrahlung is very likely to constitute a major part
of the observational signal.

However, since particle-in-cell codes automatically provide all variables needed for producing a radiation
spectrum, namely $\rr$ (position), $\bbeta$ ($\vv/c$, velocity), and $\dot{\bbeta}$ ($\dot{\vv}/c$,
acceleration), a radiation spectral synthesizer has been integrated into the \ppcode. We need only
designate observer position(s) and match the frequency range to the plasma conditions to complete the
setup for the computing the radiation integral (\Eq{eq:radpower} below). During run-time the synthesizer computes the
radiation signature for an ensemble of charged particles (in most cases electrons) in the simulation volume,;
the formula for the spectrum is given by
\begin{align}\nonumber
    \frac{d^2W}{d\Omega d\omega} & = \frac{\mu_0cq_e^2}{16\pi^3} \times \\ \label{eq:radpower}
    & \!\!\!\! \left| \int^{\infty}_{-\infty} \frac{\textbf{n} \times \big(
    (\textbf{n}-\bbeta) \times
    \dot{\bbeta}\big)}{(1-\textbf{n}\cdot\bbeta)^2}e^{i\omega(t'-\textbf{n}\cdot\rr_0(t')/c)}dt'\right|^2
\end{align}
with $t'$ the retarded time and $\textbf{n}$ the direction of the observer. A first comprehensive and thorough study
of the spectral synthesis method is given by Hededal\cite{Hededal:2005b}, which also covers a range of test
examples. While in that study, the spectral synthesis was done as post-production, in the \ppcode{} all parts
of the integration are done at run-time, with very little overhead, even for large numbers of particle
traces\cite{Trier:2010,Medvedev:2011}.

The discretization of \Eq{eq:radpower} is done in four parts:
\begin{enumerate}
    \item \emph{Frequency range} is specified as an interval and is discretized into $N_{\omega}$ bins, typically
    of order $10^3$, either with linear or (more often in practice) with logarithmic binning.
    \item \emph{Observer positions}, often more than one ($N_{obs}>1$), are specified at run initiation time
    (input), typically with directionality perpendicular to a sphere centered on the simulation volume, or
    any important direction.
    \item \emph{Time subsampling} may be chosen; this partitions the integration for every simulation time
    step into a number of subcycled integration intervals: Subcycling is employed on particles selected for
    synthetic tracing since, for highly relativistic situations, the retarded
    electric field can be extremely compressed in spikes (for example in the case of synchrotron motion with
    $\gamma(v)\ll1$). In such situations the subcycling provides a much cheaper alternative than to
    restrict the Courant condition for the entire simulation.
    \item \emph{Radiative regions} are defined in a uniform meshing (independently of MPI and simulation
    mesh geometries), which gives the advantage of offering the possibility to sample very local volumes
    of the plasma. This may be of
    interest in simulations with --- at the same time --- subvolumes of very high and low anisotropy,
    such as is the case in fully resolved shock simulations, and relativistic streams.
\end{enumerate}

A seamless integration into the \ppcode{} has made the synthesis module computationally
efficient, and due to the embarrassingly parallel nature of the spectra collection procedure plasma simulations
have been run with millions of particles used for sampling the synthetic spectra. Particles are
chosen for spectral integration before or during a run by tagging for synthetic sampling. The detailed sampling
is important in highly relativistic cases with bulk flows -- for example when investigating
the radiation signature from relativistic collisionless shocks, and streaming instabilities, more
generally~\cite{Hededal:2005b,Trier:2010,Medvedev:2011}.

\section{Binary collision operators}
The classic particle-in-cell framework does not take into account physical collision processes,
and all particle interactions are mediated through the electromagnetic fields on the mesh. Low energy
electromagnetic waves, and large impact parameter electrostatic scatterings between
charged particles can be resolved directly on the grid through particle-wave interaction,
but the photon energy is limited by the grid resolution and binary electro static scattering
is not correctly represented. To allow for binary interactions, high energy photons, and
in general interactions of neutral and charged particles and gas drag forces, we have to
model them explicitly in cases where they are of importance, such as in high energy density
plasmas, and partially ionized mediums. The \ppcode{} supports the inclusion of particle-particle
interactions, decay of particles, and allow for neutral particles, in particular photons, in the
model.
The first implementation of binary interactions---in particular Compton scattering---together with
tests of the method was given by Haugb{\o}lle\cite{Haugboelle:2005} and Hededal\cite{Hededal:2005b}.
This first implementation motivated the name for the code, the \ppcode{}. The Compton scattering
module was used to model the interaction of a gamma-ray burst with a  circumstellar
medium\cite{Trier:2008.2,Trier:2008.3,Trier:2008.1}. Coulomb collisions have later been
incorporated into the framework, to study particle acceleration in solar active regions \cite{Baumann:2012c}.

\subsection{Compton Scattering and splitting of particles}
The classic Monte Carlo approach to scattering is based on a cut-off probability: first a probability
for the process is computed and then it is compared with a random number. If the random number
is lower than the threshold the scattering for the full macro particle pair is carried through, and otherwise
nothing happens. This probabilistic approach is straight forward both numerically and conceptually, but
it can be noisy, in particular when interaction effects are strong but have low probability.

In the code the natural domain to consider is a single cell, partly because that is by definition
the volume of a single macro particle, partly because some interactions (e.g.~electro static
interactions) are mediated by the grid at larger scales. In a PIC simulation typical numbers
are $10-10^3$ particles per species per cell, and a probabilistic approach would result
in an unacceptable level of noise. Consider a beam incident on a thermal population: The
first generation of scattered particles may be computed relatively precise, but the spectra
of later generations will require an excessive amount of macro particles, if they all represent
an equal amount of physical particles, given the exponentially lower number density of later
generations. Another well known consequence is that the precision scales
inversely proportional to \emph{the square root} of the number of particles. This is a problematic
limitation, when the higher order generations are important ingredients of the physics, and the
scattering process is not just a means to thermalizing or equilibrating the phase space distribution.
\begin{figure}
\begin{center}
    \centering
        \includegraphics[width=0.48\textwidth]{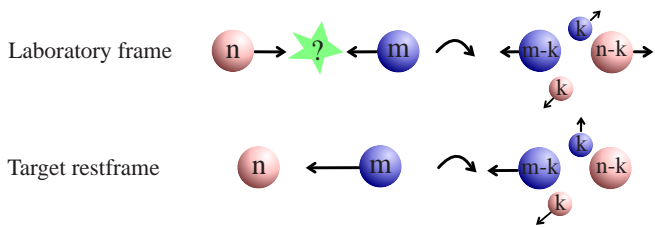}
\caption{Sketch of the scattering mechanism in the code of an incident macro particle (shown as blue/dark
gray) on a target macro particle (shown as red/light gray) resulting in the creation of a scattered
macro particle pair.}
\label{fig:detailedcollisions}
\end{center}
\end{figure}
To circumvent these problems, the Compton scattering module is instead based on an explicit splitting
approach for particles using the calculation of cross sections and allowing for individual weights for each
macro particle. To implement the scattering of two macro particles we transform to the rest frame of the
target particle and compute the probability $P(n)$ that a single incident particle during a single timestep
is scattered on the $n$ target particles. If the incident macro particle has weight $m$, then $k=P(n) m$ particles
will interact and two new macro particles representing the fraction of scattered particles are created
(see \fig{fig:detailedcollisions}). In the case of Compton scattering, the target will always be the charged
particle, while the incident is a photon, and the scattering amplitude is calculated with the Klein-Nishina
formula, which covers the full energy range from Thompson to high energy Compton scattering.
To make the process computationally more efficient prior probabilities can be applied. If instead of selecting
all pairs in a given computational cell, one only selects pairs with a prior probability $Q$, then the weight of
the scattered macro particles has to be changed to $k/Q$. By cleverly selecting the prior probability such that
for example the Thompson regime is avoided, the computational load can be greatly decreased.

Each scattering leads to the creation of a new macro particle pair, and if untamed, the number of
macro particles will grow exponentially. To keep the number of particles under control, we use particle
merging in cells where the number of particles is above a certain threshold.
The algorithm is described in detail by Haugb{\o}lle\cite{Haugboelle:2005} and Hededal\cite{Hededal:2005b}.

The Compton scattering module of the \ppcode{} has allowed us to approach exciting new topics in high energy
plasma astrophysics, where plasmas are excited and populations modified by photons, and the back-reaction
of the plasma on the (kinetic) photons produces interesting and detailed descriptions of for example the
production of inverse Compton components\cite{Trier:2008.2}, and photon beam induced plasma
filamentation\cite{Trier:2008.1}.

\subsection{Coulomb scattering}
Coulomb scattering of charged particles in an astrophysical context is for example important in order to understand the solar
chromosphere and the lower parts of the corona, where the mean free path is similar to the dynamical length scales in the system.
In the \ppcode{} we have integrated a collision model, where all macro particle pairs in a cell are considered for scattering.
We calculate the scattering process in the elastic Rutherford regime. The collision process is implemented using a physical description
for each macro particle pair, starting by calculating the time of closest approach. Only if that time is less than $\Delta t/2$ from
the current time, ie.~inside the current time interval, is the scattering carried through. In the limit of very small time steps
this makes the algorithm independent of the size of the time step. When calculating the impact parameter between two
macro particles we have to take in to account that each particle represents a large number of physical particles, therefore
the impact parameter is rescaled with the typical distance between each physical particle $n^{-1/3}$, where $n$ is the number
density. Because macro particles carry a variable weight, we use the geometric mean of the number density of the particle
pairs to calculate the effective number density $n = (n_1 n_2)^{1/2}$.

We consider three different regimes, based on the impact parameter: i) If the impact parameter is larger than the local Debye length
we assume that Debye screening between the two particles is so effective that no scattering happens. ii) If the impact
parameter is so large that the effective scattering angle is less than $\theta_c$ radians (normally taken to be 0.2
in the code) we
use a statistical approach: At small angles the scattering angle is inversely proportional to the impact parameter, and we can replace
the many small angle scattering by fewer large angle scatterings, comparing the ratio of the cut-off to the impact parameter
$b_c / b$ with a random number. If it is lower the scattering is carried through, but using the cut-off impact parameter $b_c$
for greater computational efficiency. iii) If the impact parameter is smaller than the cut-off parameter $b_c$ then we make a
detailed computation of the scattering. In both of the two last cases the scattering angle is calculated in the center-of-mass
frame as an inelastic Rutherford scattering, which conservers the energy of each macro particle.

The explicitly physical implementation at the macro particle level of Coulomb scattering is conceptually completely different
from the Compton scattering module, in particular because the main consequence of Coulomb scattering is the thermalization
and isotropization of the plasma.

\section{Test Problems}
Testing the accuracy and precision of a particle-in-cell code is particularly difficult, because of the non linear nature of plasma
dynamics, Monte-Carlo particle sampling, and the few examples of realistic test problems with analytic counterparts. To facilitate
cross comparison with other codes, below we apply the \ppcode{} to a set of classic test problems, and in some cases compare
different order splines and differential operators to highlight the impact of using high order methods. We also give an example of
of the more non-standard feature of radiative cooling. Further tests have been published in the case of Compton
scattering\cite{Haugboelle:2005,Hededal:2005b} and synthetic spectra\cite{Hededal:2005b}.

\subsection{Numerical heating and collision artifacts}
Numerical heating is a well studied feature of all PIC codes\cite{birdsall:1985}. It happens due to the interaction
of particles with the mesh; the so-called grid-collisions. If particle populations with different energy distributions exist
in the plasma, the interaction through the mesh will tend to equilibrate the kinetic energy of each particle species.
The equilibration of temperatures happens in a laboratory plasma too, albeit normally at a much slower rate, and
the difference between a laboratory and the computational plasma is the much smaller number of macro-particles
used to represent the plasma in the latter case. It is also worth pointing out that the numerical mesh heating is an equilibration of
kinetic energies, and as such much more severe for ion-electron plasmas. It is important to keep this energy
equilibration in mind when simulating plasmas with greatly varying temperatures, or when analyzing heating
rates due to real physics. If the numerical heating rate is close to the physical rate in question, the results cannot
be trusted. It is interesting to note that the heating rate is practically invariant with respect to the numerical
technique used, and instead mainly depends on the number of time steps taken.

The test is done in two dimensions, with three velocity components. There is no bulk velocity, and the kinetic energy
corresponds to a thermal velocity of both ions and electrons of $v_{th,e} = v_{th,i} = 0.1\,c$ per component, or
$E_\textrm{kin}^\textrm{s} = 0.015\,m_\textrm{s}\, c^2$. The size of the box is $12.8^2$ electron skin depths with
10 cells per skin depth for a $128^2$ resolution. We perform two tests with 5 and 50 particles per cell. The mass
ratio is $m_i / m_e = 16$. We used TSC or cubic spline interpolation, 2$^\textrm{nd}$, 4$^\textrm{th}$, or
6$^\textrm{th}$ order field solver, and in the case of the 6$^\textrm{th}$ order field solver with cubic interpolation
we also use the charge conserving (CC) method for the current density with second or fourth order time stepping.

\begin{figure*}
     \begin{center}
           \includegraphics[width=0.23\linewidth]{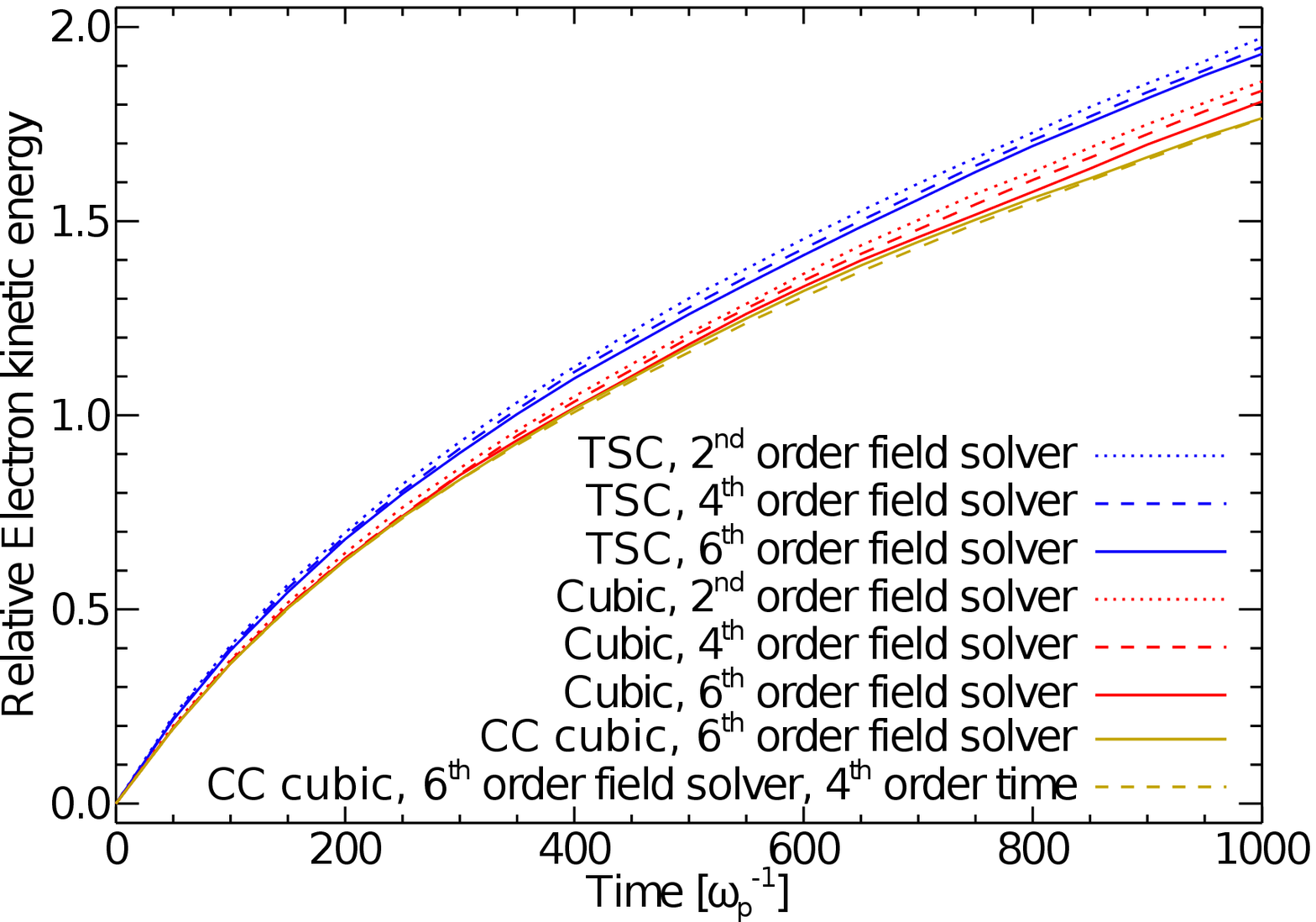}
           \includegraphics[width=0.23\linewidth]{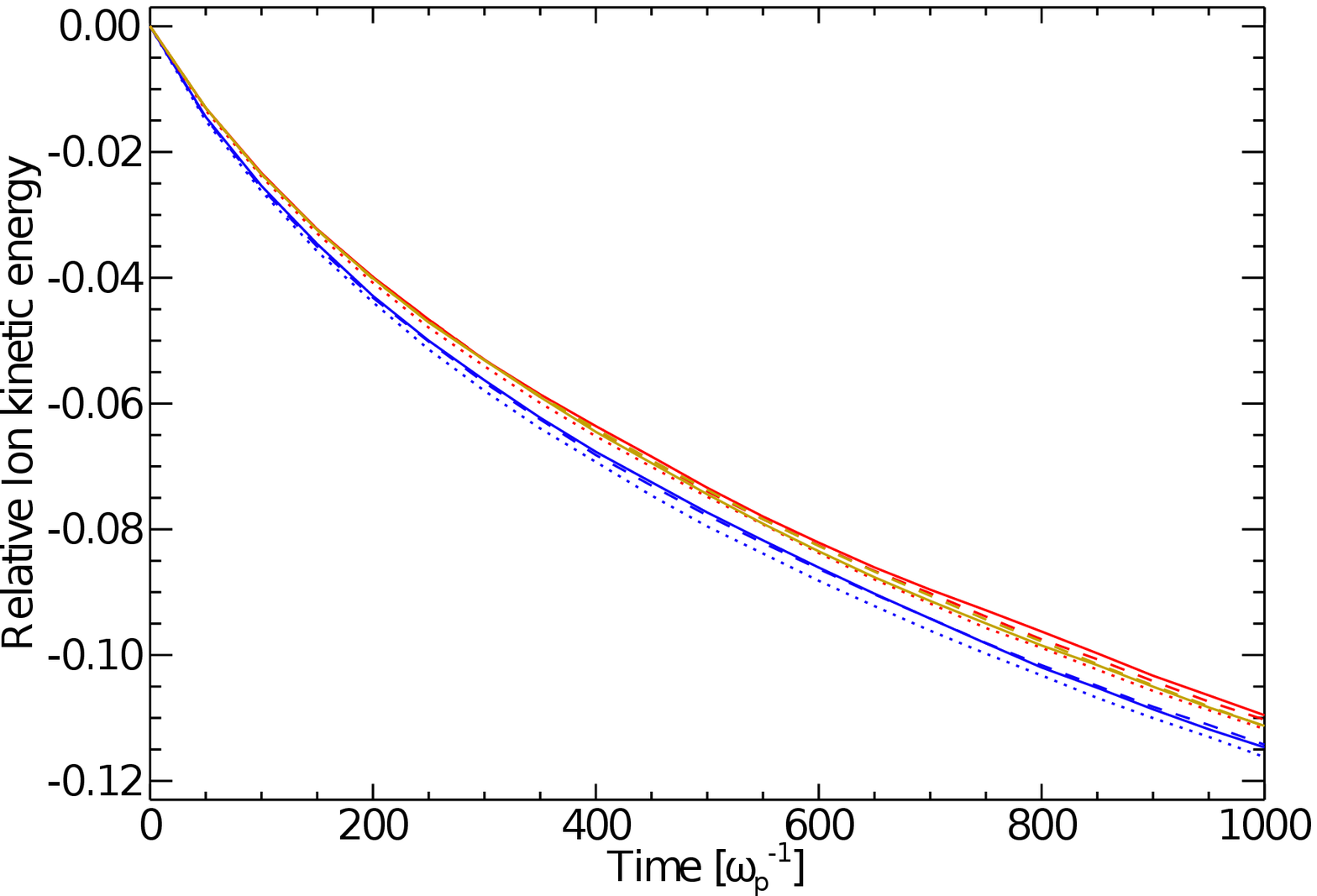}
           \includegraphics[width=0.23\linewidth]{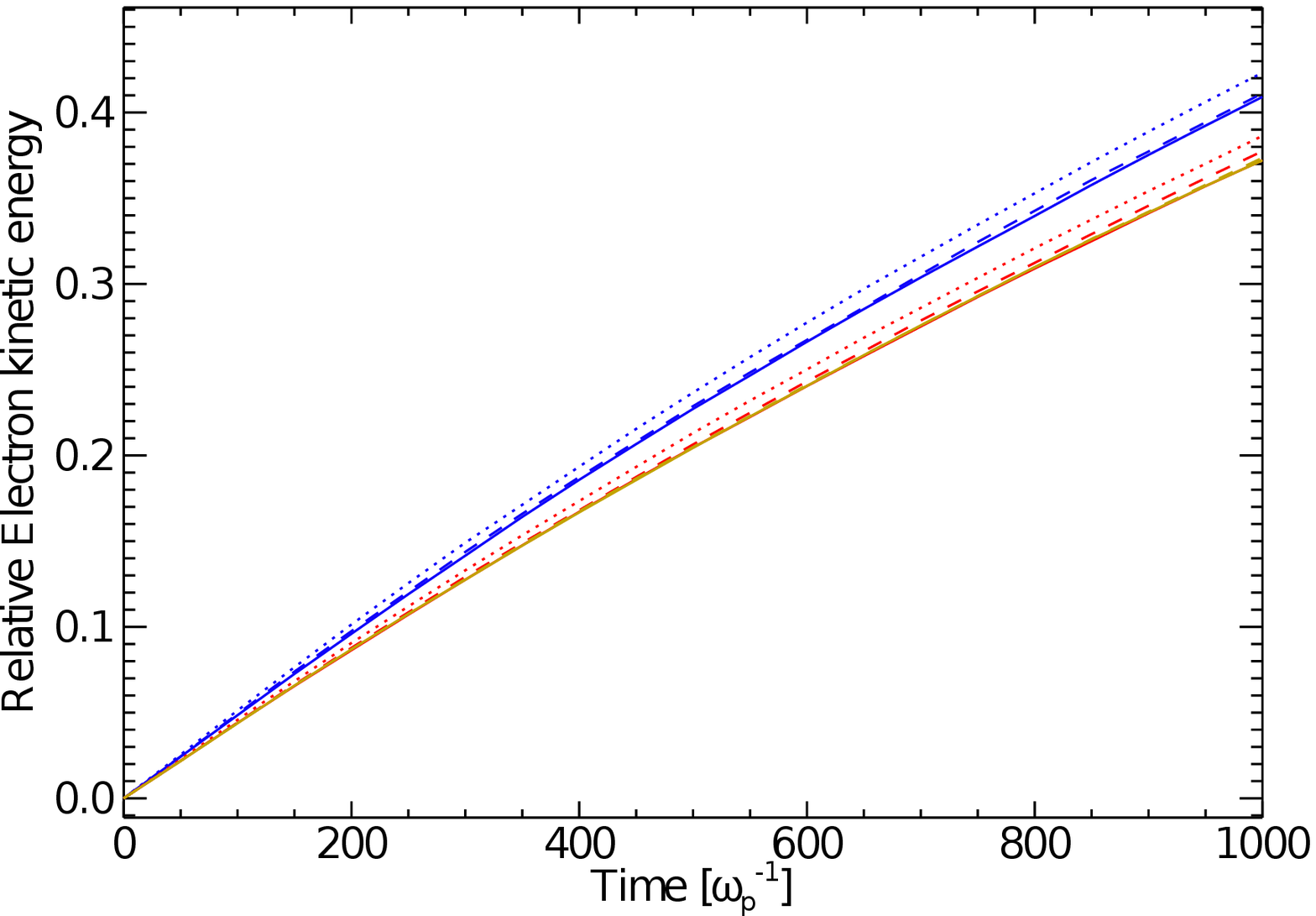}
           \includegraphics[width=0.23\linewidth]{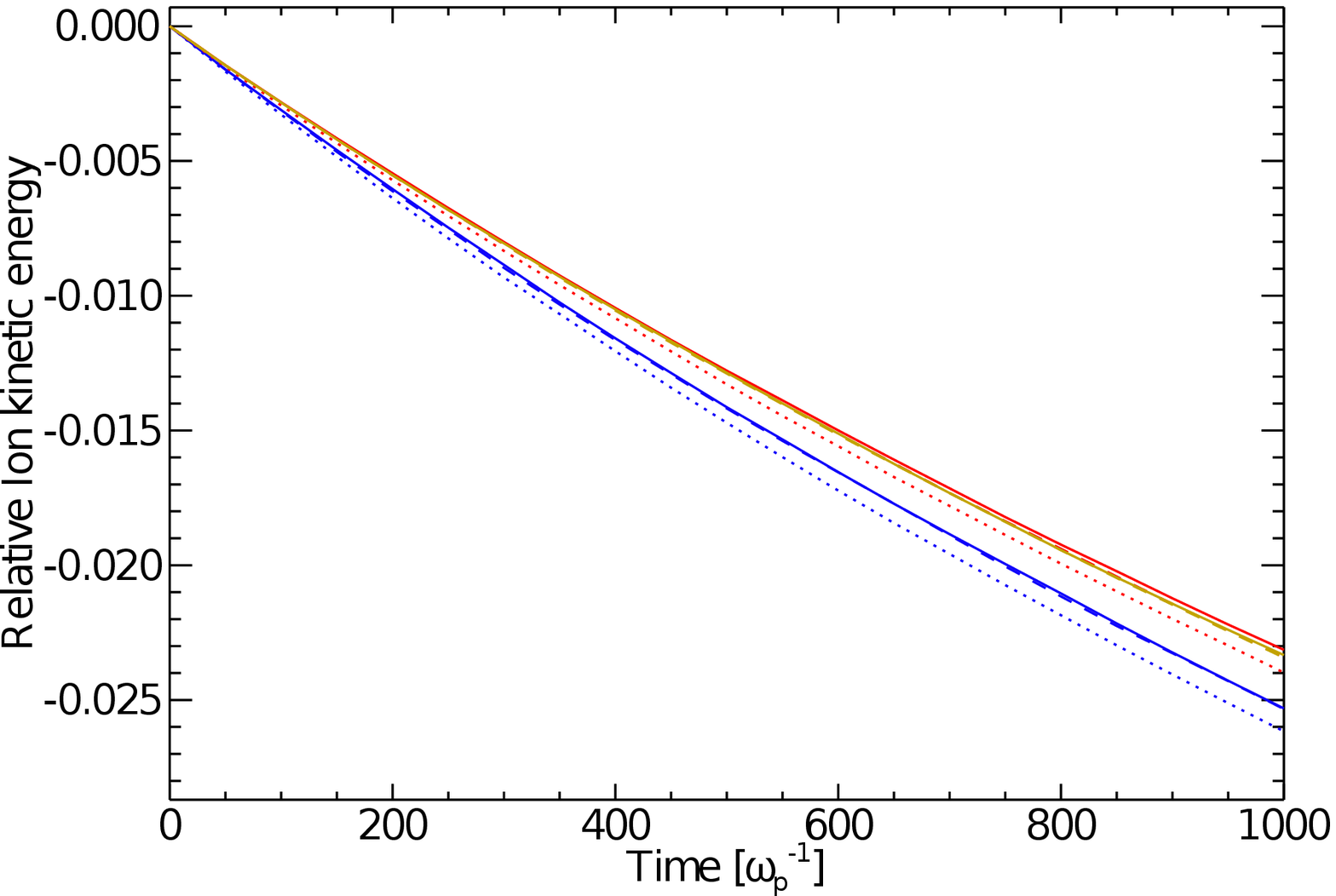}
     \end{center}
     \caption{Changes in ion and electron temperatures in the numerical heating experiment as a function of time
     for different choices of solvers. The two left (right) panels show the heating rate for 5 (50) particles per
     species per cell. Note that the heating rate is virtually independent of the type of solver, but is
     strongly dependent on the number of particles per cell.} \label{fig:numheating}
\end{figure*}

\subsection{A relativistic cold beam}\label{sec:coldbeam}
A particle-in-cell code is not Galilean invariant. When a cold plasma beam travels through the box at constant
velocity the electrostatic fluctuations inside the Debye sphere or, if it is less than the mesh spacing, inside a
single cell, will have resonant modes with the mesh spacing. This leads to an effective drag and redistribution of kinetic
energy from the stream direction to the parallel direction, and general warm up of the beam. When the temperature
reaches a critical level the instability is quenched. For relativistic beams there is the additional complication
that electromagnetic waves are represented on the mesh. The solver has an effective dispersion relation,
and short wavelength waves travel below the speed of light. On the other hand, particles
are Lagrangian, and if relativistic they can effectively travel above the speed of short wavelength waves,
giving rise to numerical Cherenkov radiation.

A classic method to limit the impact the of the cold beam instability
is applying filters to either the current density or the electric field. This may to some extent filter out the effects, but
will also filter out some of the physics. Alternatively, higher order field solvers, interpolation techniques, and time stepping
can mitigate the effects.
\begin{figure*}
     \begin{center}
           \includegraphics[width=0.46\linewidth]{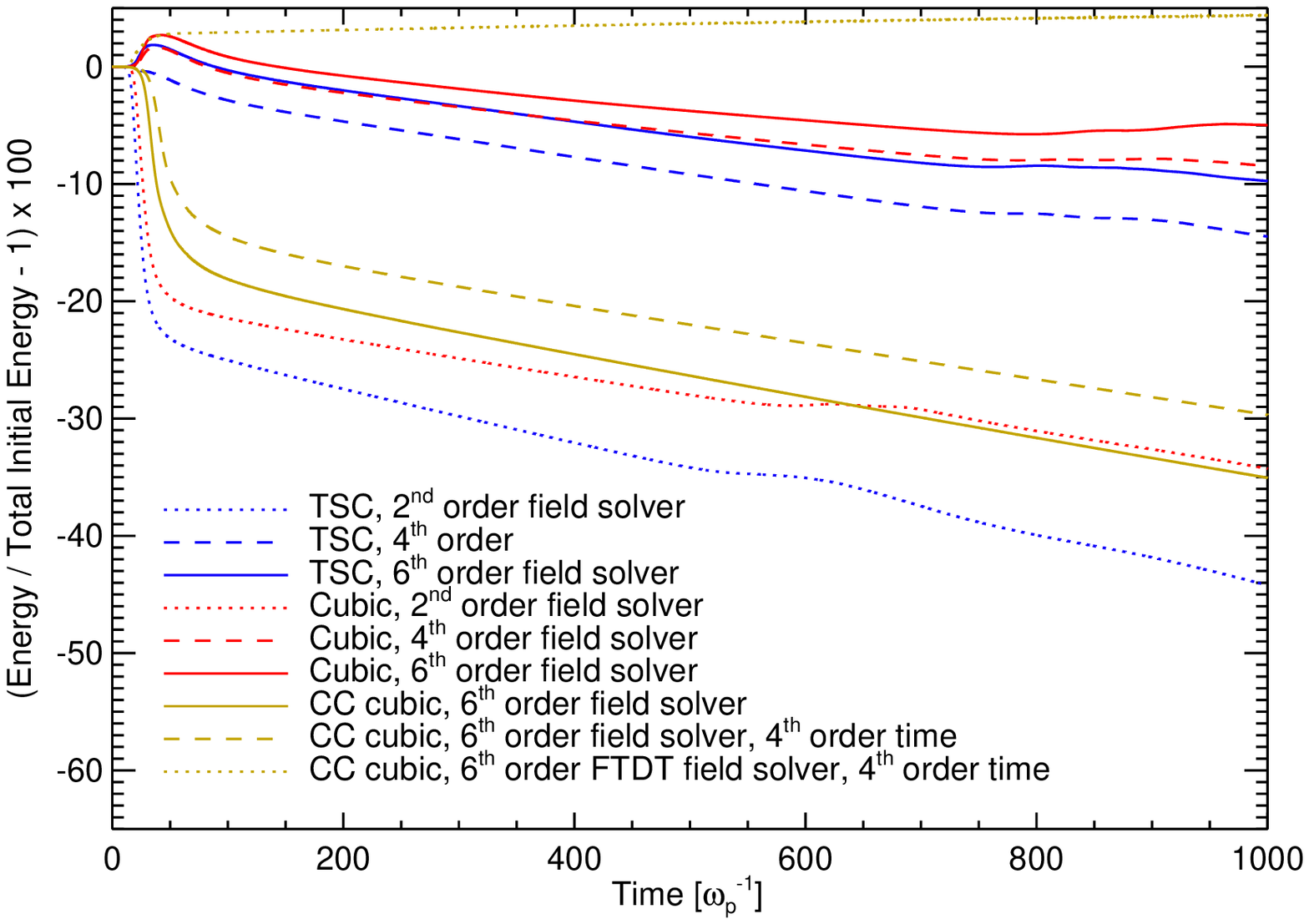}
           \includegraphics[width=0.46\linewidth]{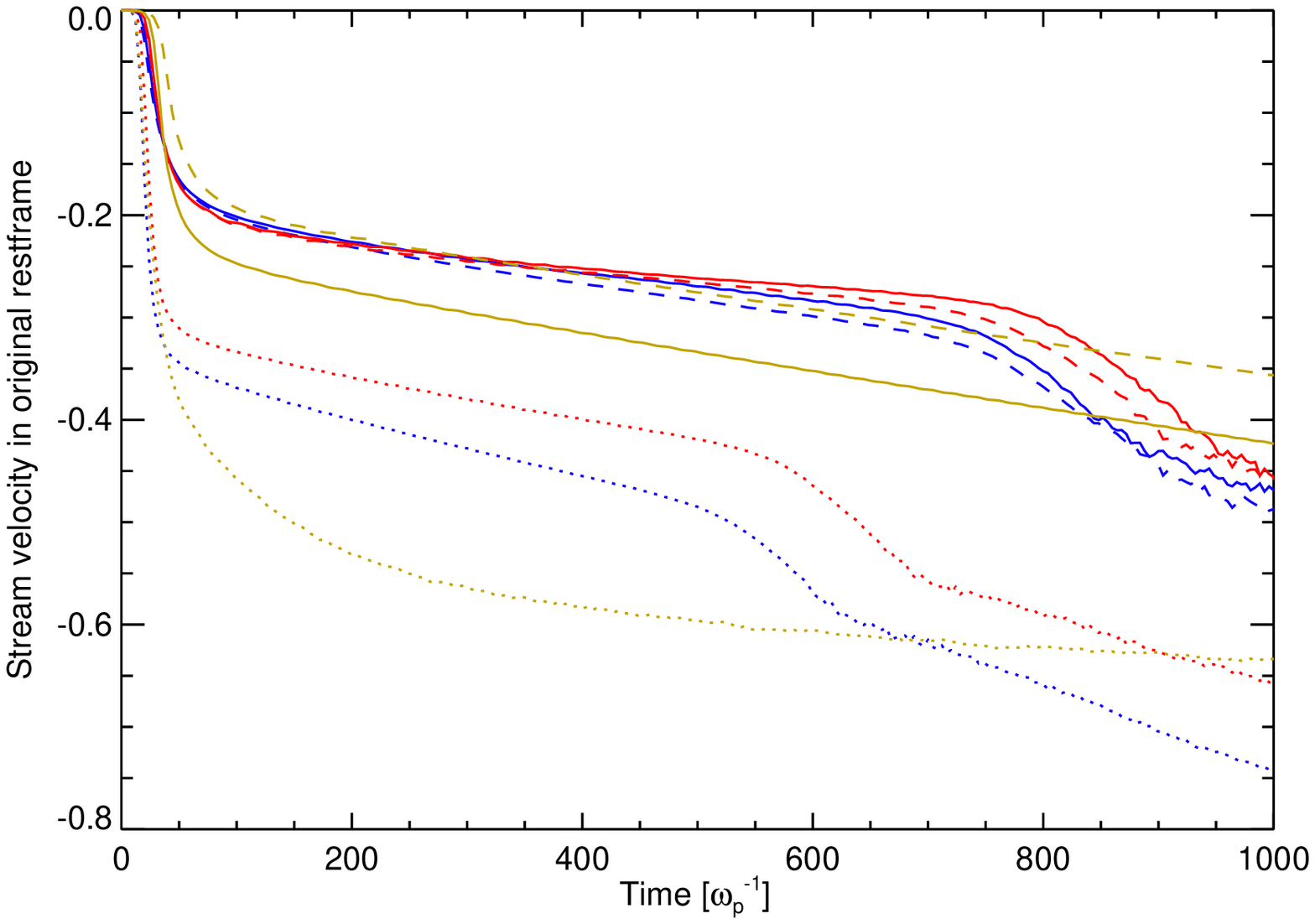} \\
           \includegraphics[width=0.46\linewidth]{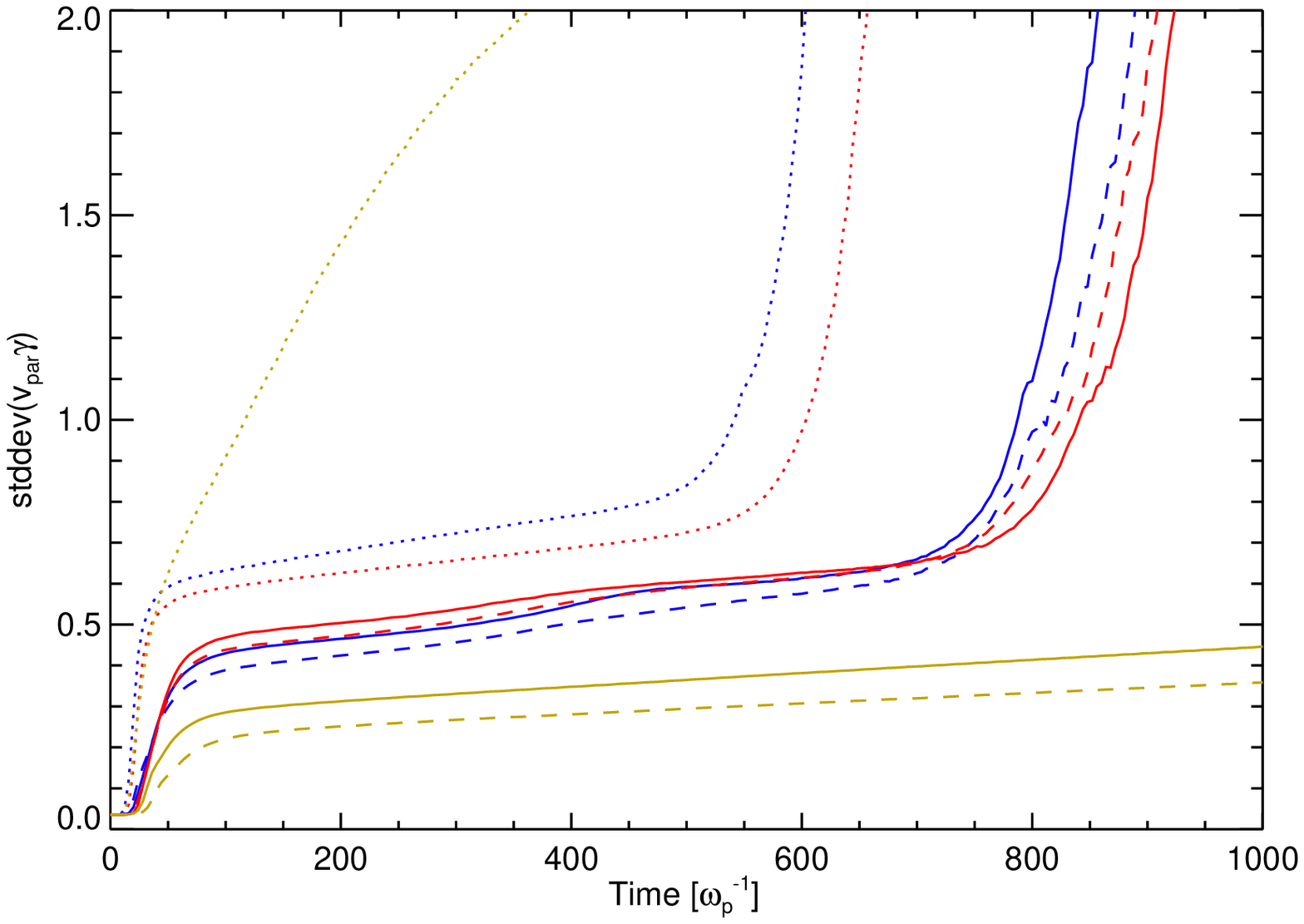}
           \includegraphics[width=0.46\linewidth]{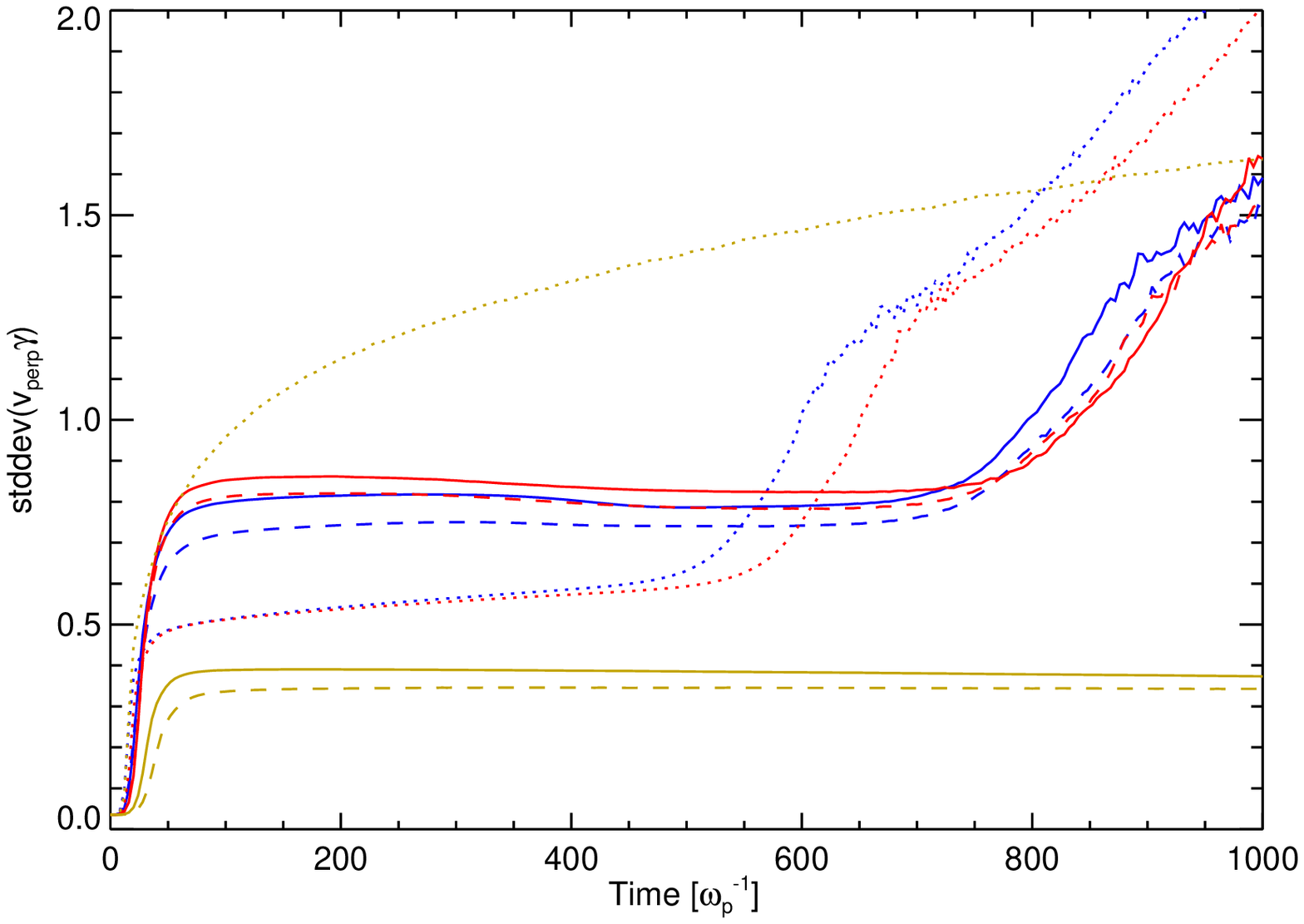}
     \end{center}
     \caption{The top row shows the energy conservation and stream velocity in the original rest frame.
     The bottom row shows the evolution in the temperatures parallel and perpendicular (i.e. $T_\parallel$ and $T_\perp$) to
     the beam direction. For reference, results are also included using a simple FDTD explicit solver for the electromagnetic
     fields. Notice that the runs have been done without applying any kind of filtering to the current density.} \label{fig:coldbeam}
\end{figure*}
To test the numerical methods used in the \ppcode{}, we have made a cold beam test with 9 different
versions of the code. We do not apply any filters to the current density, to show the actual performance of the different
code versions. Apart from the 8 methods used for
the numerical heating test there is also a version where instead of the implicit field solver a simple
(but 6$^\textrm{th}$ order) FDTD explicit solver is used. Notice how the explicit solver has numerical heating,
while implicit solvers have numerical cooling\cite{birdsall:1985}, and how the stability of the beam is greatly enhanced by the
implicit solver, compared to using an explicit solver. But only the combination of the implicit solver with a charge
conserving current deposition gives stable beams, with the smallest heating (see figure \ref{fig:coldbeam}).
It is also interesting to notice that the heating is non-isotropic. It is therefore not the same to initialize a
plasma with a low but stable temperature, as to use a very low temperature and let the cold beam stability
warm up the beam.

The test is done in 2D2V with a $\Gamma=10$ streaming pair plasma through a 256 $\times$ 256 cell domain
with 10 particles per species per cell and a relativistic skin depth $\delta = [(m c^2 \Gamma)/(4\pi n q^2)]^{1/2}$
of ten cells. The initial temperature is measured in the rest frame of the plasma and has a root-mean-square
per velocity component of $0.025c$, so that $\sum \textrm{rms}(v_\textrm{th}) = 0.05c$.

By rerunning at higher resolutions we have investigated at what resolution other methods give comparable
results to the charge conserving method with fourth order time stepping, by comparing stream velocities,
and parallel and perpendicular temperatures at $\omega_p t=1000$ (see table \ref{tab:coldbeam}). It is only
at higher resolutions that the non-charge conserving methods do not suffer from the catastrophic instability
seen in figure \ref{fig:coldbeam}, and cost-wise the sixth order charge conserving method is marginally the cheapest.
Had the test been in 3D, where the cost goes like resolution to the fourth power, the fourth order
time integration method would have been the cheapest for this particular beam test.
At high enough resolution the cold beam instability is quenched. For the fourth order charge conserving method this
happens at a resolution of approximately 3072$^2$, where instead of a large heating rate, and then a
new stable temperature, we observe a gradual heating over time. A resolution of 3072$^2$ corresponds to
resolving the Debye length, $\lambda_D = v_\textrm{th} / (\Gamma c) \delta_e$, with 1.9 cells. Taking into account
that we do not apply any damping, this is in good agreement with the common wisdom that the Debye length
has to be resolved by roughly one cell.

\begin{table}
    \centering
    \caption{Resolution study for the cold beam instability\label{tab:coldbeam}}
        \begin{tabular}{l c c c}
            \colrule\colrule
 Method & Res & Cost & $\mu$s/part\\
            \colrule
CCo 6$^\textrm{th}\!$ order field; 4$^\textrm{th}\!$ order time & 256$^2$ & 1 & 6.60 \\
CCo 6$^\textrm{th}\!$ order field & 380$^2$ & 0.9 & 1.72 \\
Cubic interpolation; 6$^\textrm{th}\!$ order field & 512$^2$ & 1.3 & 1.05\\
TSC interpolation; 2$^\textrm{nd}\!$ order field & 950$^2$ & 4.9 & 0.63 \\
             \colrule
        \end{tabular}
\end{table}

\subsection{Relativistic two-stream instability}
Relativistically counter-streaming plasmas have previously been established to be subject to a general instability
class; the oblique (or mixed-mode) two-stream-filamentation instability (MMI), which mixes the
two-stream (TSI) and filamentation (FI) instabilities. A thorough and exhaustive analysis of the MMI was
given by Bret et al\cite{Bret:2004,Bret:2007}, and has been investigated numerically by several groups (see
e.g.~Tzoufras et al\cite{Tzoufras:2006} and Dieckmann et al\cite{Dieckmann:2006}). Due to its mixed nature,
the MMI contains both an electrostatic and an electromagnetic wave component\cite{Bret:2004}.

Potentially, both electrostatic and electromagnetic turbulence (wavemode coupling leading to cascades/inverse
cascades in k-space) is possible in such systems. This potential for producing very broadband plasma
turbulence (in both E and B fields) is highly relevant to inertial confinement fusion experiments. Other
examples where electromagnetic wave turbulence\cite{Frederiksen:2008} and the general MMI are important
are astrophysical jets and shocks from gamma-ray bursts and active galactic nuclei, where ambient plasma
streams through a shock interface moving at relativistic speeds.

To test physical scenarios responsible for observational signatures from these astrophysical sources, and
from plasma experiments, we must construct plausible shocked outflow
conditions\cite{bib:Medvedev1999,Frederiksen:2004,haugboelle:2011} and then subsequently synthesize
radiation signatures\cite{Hededal:2005b,Trier:2010,Medvedev:2011} to test the assumed physical conditions
against observational evidence. Studying the MMI, in both its linear and non-linear evolution is therefore
well motivated.

Growth rates of the (general) MMI, the special case of FI and the TSI, respectively, are calculated as in\cite{Bret:2004}:
\begin{align}
    \gamma_{MMI} & = & 2^{-4/3}~ \sqrt{3}~ \alpha^{1/3}~ \Gamma(v_b)^{-1/3} \label{eq:mmigrowth}\\
    \gamma_{FI}  & = & \beta~ \alpha^{1/2}~ \Gamma(v_b)^{-1/2}~             \label{eq:figrowth}\\
    \gamma_{TSI} & = & 2^{-4/3}~ \sqrt{3}~ \alpha^{1/3}~ \Gamma(v_b)^{-1}   \label{eq:tsigrowth}
\end{align}
where $\beta \equiv v_b/c$ is the beam velocity, $\alpha \equiv n_b/n_p$ is
the beam-to-background density ratio, and $\Gamma(v_b)$ is the beam bulk flow Lorentz factor.
For thin, high-Lorentz factor beams, the MMI is the fastest growing mode, dominating over both the FI and
the TSI. In the test we have chosen the MMI is dominant, with subdominant FI and TSI components.

We perform six runs with identical initial conditions and physical scaling, using combinations of finite
difference operators and particle shape functions as given in Table\ref{tab:orders} below. This way, we
test the \ppcode{} for differences/similarities between interpolation schemes.

To capture the MMI as the fastest growing mode, we initialize a simulation volume with a cold thin neutral
beam (electrons + ions) through a warm thick neutral background (electrons + ions), with no fields initially.
The beam and background densities are $n_b=0.1$ and $n_p=0.9$, respectively. The beam velocity is chosen
to have $\Gamma(v_b)=4$. Temperatures of the beam and background are
$T_b=0.01$ and $T_p=0.1$, respectively. With these choices of physical properties, the growth rates of the fastest
growing MMI, FI, and TSI modes become $\gamma_{MMI}=0.201$, $\gamma_{FI}=0.153$, $\gamma_{TSI}=0.080$.

The computational domain is $\{L_x,L_y,L_z\}=\{12.8\delta_e,12.8\delta_e,12.8\delta_e\}$, with
$\{N_x,N_y,N_z\}=\{128,128,128\}$ cells. We use
20 particles/cell/species, or a total of 80 particles/cell (beam+background). The physical constants are scaled
as $c=1$, $q_e=1$, $m_e=1$ and $m_i/m_e\equiv1836$.

\begin{table}[!h]
    \centering
  \caption{Schemes order variation in the relativistic mixed-mode two-stream instability test case, for:
  finite difference operators (fields/sources), shape function (particle-mesh/mesh-particle interpolation),
  charge-conservation, time integration order.}\label{tab:orders}
    \begin{tabular}{cccccc}
    \colrule\colrule
        Run & Fields & Particles & charge-conservation & Time order \\
    \colrule
     1                & 2$^\textrm{nd}$ & tsc   & no  & 2   \\
     2                & 2$^\textrm{nd}$ & cubic & no  & 2   \\
     3                & 6$^\textrm{th}$ & tsc   & no  & 2   \\
     4                & 6$^\textrm{th}$ & cubic & no  & 2   \\
     5                & 6$^\textrm{th}$ & cubic & yes & 2  \\
     6                & 6$^\textrm{th}$ & cubic & yes & 4 \\
    \colrule
  \end{tabular}
\end{table}

For our choice of run parameters a MMI mode develops with a propagtion wave vector,
${\bf{k}}_{MMI}$, that is oblique with respect to the streaming direction, at
an angle given by $\theta_{MMI}=\angle({\bf{k}}_{MMI},{\bf k}_\textrm{beam})=
\textrm{arctan}\left(\sqrt{v_b/v_p}\right)\approx74^\circ$. The corresponding electric
field component is almost parallel to the direction of propagation\cite{Bret:2004}.

\begin{figure}[!h]
    \centering
        \includegraphics[width=.48\textwidth]{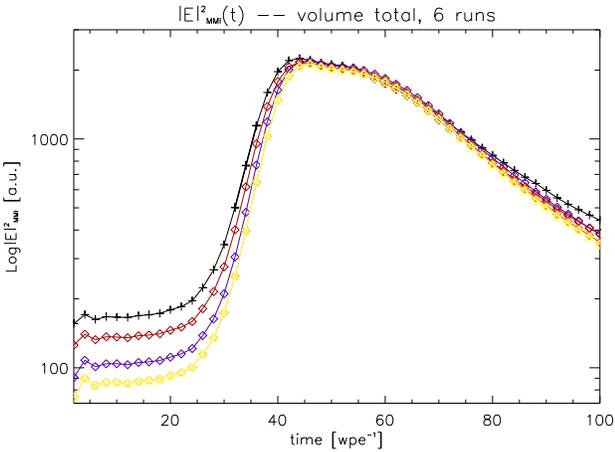}
    \caption{Growth of the volume-totaled electrostatic energy, for the electric field projected onto the direction of
    propagation of the fastest growing mode, $\EE \cdot {\bf{k}}_{MMI}$. Runs 1-6 are compared in the plot. Thickened
    line segment Run1 (black) gives the fitting interval. Runs have decreasing initial energy for increasing run
    number designation (Table~\ref{tab:orders}). Runs 5 (orange) and 6 (yellow) are completely coinciding.}
    \label{fig:mmi_growth}
\end{figure}

To find growth rates of the MMI we calculate the volume integrated electrostatic energy as a function of time
$$ E_{E,tot}(\theta_{MMI}) = \int_{V} |\EE_\perp| sin(\theta_{MMI}) + E_\parallel cos(\theta_{MMI})~dV$$
of the electric field projected on the propagation direction, $\EE(\rr)\cdot{\bf{k}}_{MMI}$.
Similarly, we also measure the TSI mode growth rate (\Eq{eq:tsigrowth})
by the same calculation, but for the TSI we have $\theta_{TSI} = \angle({\bf{k}}_{TSI},\hat{\bf{z}}=0)$. Our results
are summarized in Table~\ref{tab:growth}.
\begin{table}[!h]
    \centering
  \caption{Growth rates measured for the six runs listed in ~\ref{tab:orders} for the two cases of the MMI and
  TSI.}\label{tab:growth}

    \begin{tabular}{ccccccccc}
    \colrule\colrule
     Run              & 1 & 2 & 3 & 4 & 5 & 6 & theory \\
    \colrule
     $\gamma_{MMI}$   & 0.185 & 0.190 & 0.203 & 0.206 & 0.205 & 0.205 & \textbf{0.201} \\
     $\gamma_{TSI}$   & 0.060 & 0.079 & 0.065 & 0.082 & 0.088 & 0.088  & \textbf{0.080} \\
    \colrule
  \end{tabular}
\end{table}
From figure~\ref{fig:mmi_growth} we see that the initial noise build-up prior to instability dominance is strongest
(as expected) for Run1. This results in lower growth rates for lower order runs, since the noise tends to 'flatten«
the total energy history, with a lower $\gamma_{MMI}$ as a result. This error in the measurement decreases
with increasing run number (Run1$\rightarrow$Run2$\rightarrow$\ldots), with higher order runs' growth rates less susceptible
to noise distortion. This is similar to what was seen in the cold beam tests: in a lower order integration scheme more energy
is lost to artificial heating and Cerenkov radiation, making less energy available for the physical instabilities, and
changing the parameters (for example temperatures) of the plasma components, modifying the overall setup.

Because of the lower level of noise, the onset of instability is delayed for runs of increasing order, while the
peak energies and peak times
coincide roughly for all runs; this effect correlates with the increased growth rates of the higher order runs, relatively.
Both of the effects mentioned above are caused by the difference in dynamic response of the various schemes. A higher order scheme
is desirable since the noise levels, heating, numerical stopping power and dynamical friction are all reduced considerably.
The differences between using either TSC or cubic interpolation, second or sixth order field integration, and (no) charge
conservation are all clearly reflected in the growth rates of the MMI and TSI.

Nonetheless, growth rates are seen to converge with different methods to a value which deviates from the
theoretical prediction by about 2\% for the MMI component, and $\sim$10\% for the TSI component, for the highest
order schemes,  and the overall development is in qualitative agreement for all test cases.
Concluding this test, we have verified the growth rate of the relativistic two-stream (or oblique or mixed-mode) instability, and
found very good agreement with growth rates also when selecting the TSI branch. The slight excesses in values of
$\gamma_{TSI}([Run1,...Run6])$ is likely caused by the fact that the relativistic beam is perfectly grid aligned, thus subjected
to the finite grid instability, which introduces additional electric field energy in the beam direction, while for lower orders, this
is more than compensated by the overall dissipation to all electromagnetic and particle components.

\subsection{Radiatively cooled collisionless shocks}
\begin{figure}
     \begin{center}
       \includegraphics[width=\linewidth]{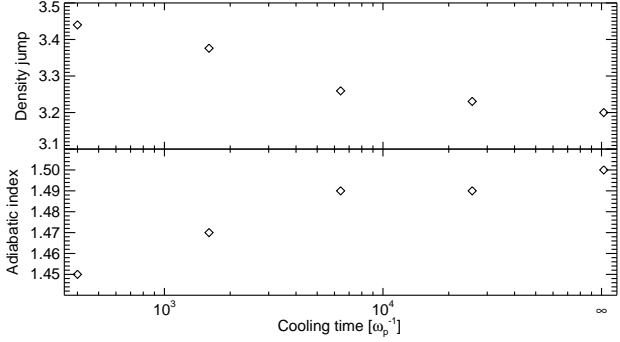}
     \end{center}
\caption{Top panel: Density ratio between up and down stream at $\wpet = 1600$, as a function
of cooling time. Bottom panel: Effective adiabatic index, derived from the shock velocity.
The right most point is for a run without radiative cooling.}\label{fig:jump}
\end{figure}

\begin{figure}
     \begin{center}
       \includegraphics[width=\linewidth]{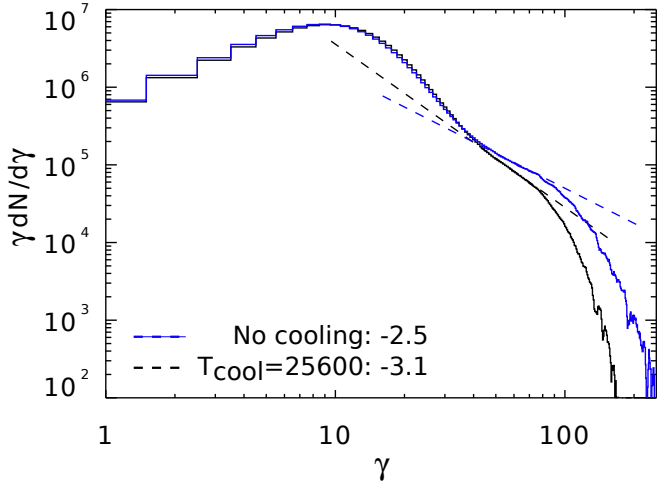}
     \end{center}
\caption{The particle distribution function sampled in the downstream region for
the case of no cooling and $T_\textrm{cool}=25600$, firmly in the weakly radiative
regime. The power law index is indicated with the dashed line, and given in the legend.}\label{fig:pdf}
\end{figure}

The radiative cooling currently implemented is inspired by the early work of Hededal; he validated it
for a simple test case of a radiatively cooling charged particle in a homogeneous magnetic field\cite{Hededal:2005b}.
For a more non trivial application we here for the first time present a series of simulations of radiatively
cooled initially non-magnetized collisionless shocks. If we assume that the acceleration of a particle in
a collisionless shock is mostly due to a homogeneous magnetic field $B$, we can estimate the synchrotron cooling
time for an initial Lorentz factor $\gamma_0$ as
\begin{equation}
t^\textrm{syn}_\textrm{cool} = \frac{6 \pi \epsilon_0 m^3 c^3}{q^4 B^2 \gamma_0}
\end{equation}
Consider a relativistic collisionless shock in the contact discontinuity frame, with the upstream moving
at a Lorentz factor $\Gamma$ and having a number density $n$. The kinetic energy density in the upstream is
\begin{equation}
E_{kin} = (\Gamma - 1) n \sum m c^2 = (\Gamma - 1) M c^2 n\,.
\end{equation}
We assume that the magnetic energy density at the shock interface $E_B = B^2 / 2 \mu_0$ is
related through some efficiency factor $\alpha\simeq0.1$ to the upstream kinetic energy density.
The relativistic plasma frequency in the upstream medium is $\omega_{pe}^2 = q^2 n / (\epsilon_0 m_e \Gamma)$.
Putting it all together we can find the synchrotron cooling time at the shock interface, for a particle
with a gamma factor $\gamma_0$, to be
\begin{equation}
T_\textrm{cool} = \omega_{pe} t^\textrm{syn}_\textrm{cool} =
  \frac{3 \pi^{1/2} \epsilon_0^{3/2} m_e^{5/2} c^3}{\alpha q^3 n^{1/2} \Gamma^{1/2} (\Gamma - 1) M \gamma_0} \,,
\end{equation}
For a pair plasma $M=2m_e$, and if we consider upstream particles $\gamma_0=\Gamma$ it reduces to
\begin{equation}
T_\textrm{cool} =
  \frac{3 \pi^{1/2} \epsilon_0^{3/2} m_e^{3/2} c^3}{2 \alpha q^3 (\Gamma - 1) \Gamma^{3/2} n^{1/2}}\,,
\end{equation}
We use this definition to label the different runs. Our definition of the cooling time is similar to
the one given in Medvedev and Spitkovsky\cite{Medvedev:2009}, but differs because they considered the downstream
skin depth: $T^\textrm{syn}_\textrm{cool,our} = 9/4 \Gamma^{1/2} T^\textrm{syn}_\textrm{cool,M-S}$.
\begin{figure*}
     \begin{center}
       \includegraphics[width=0.48\linewidth]{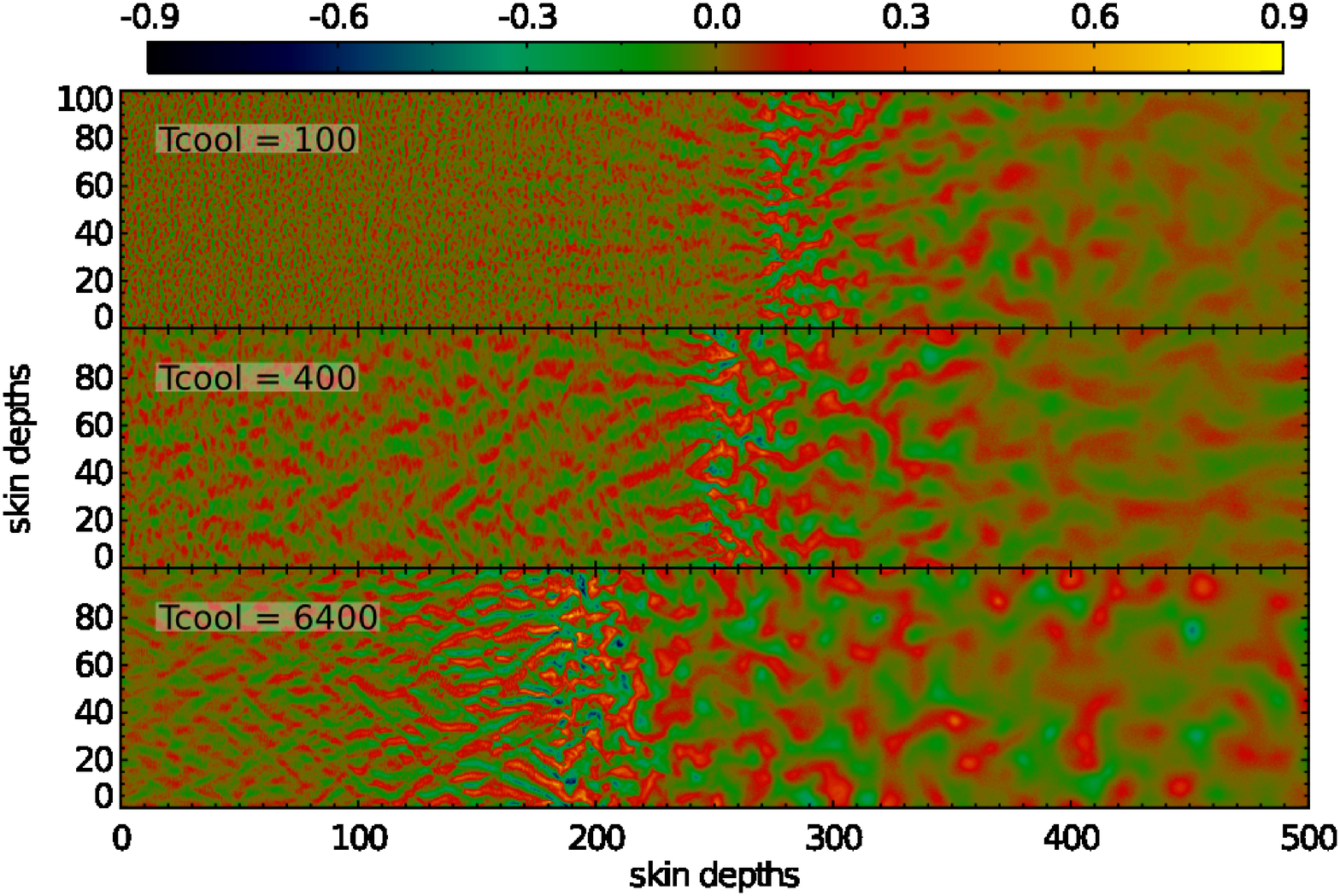}
       \includegraphics[width=0.42\linewidth]{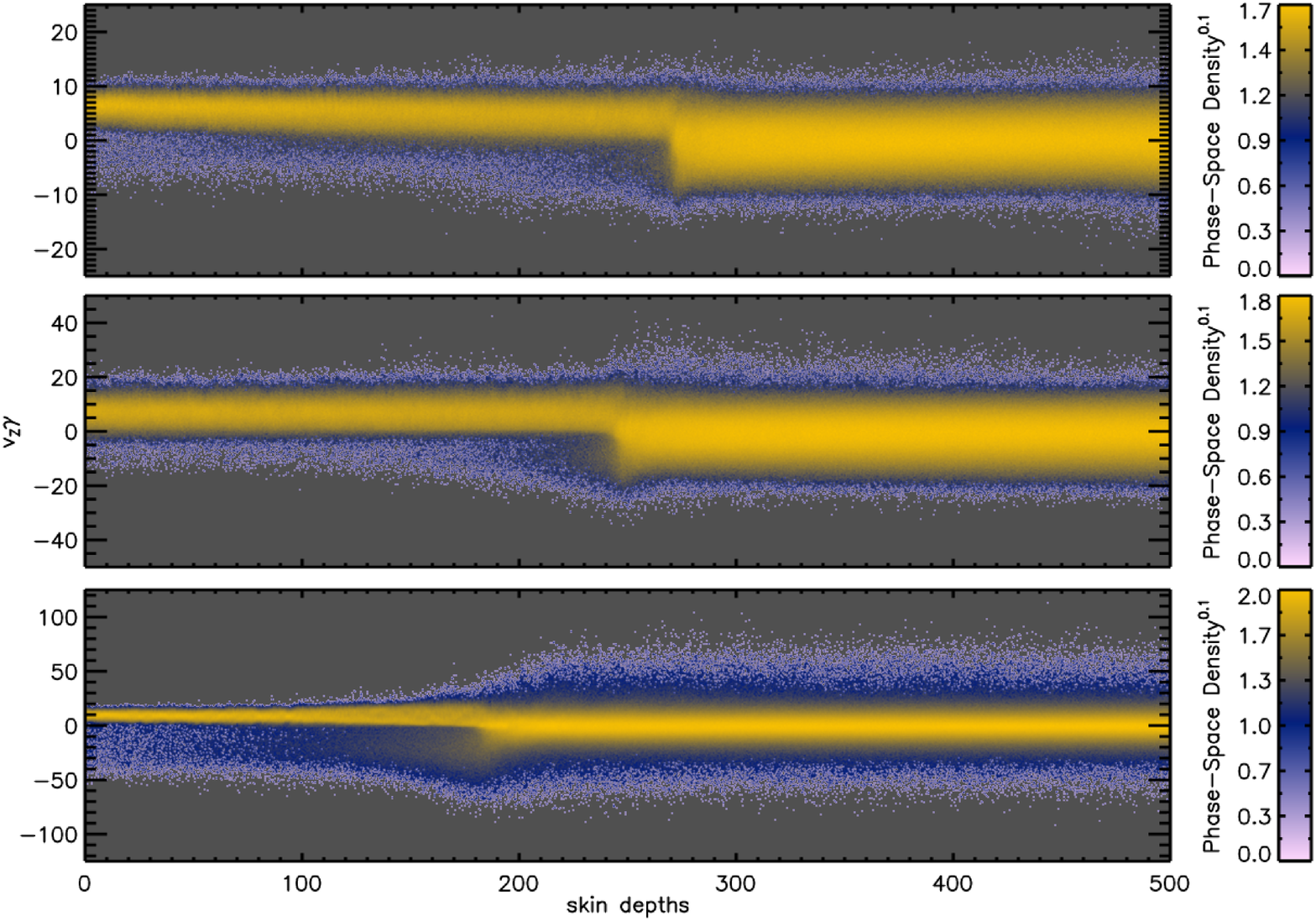}
     \end{center}
\caption{Left panel: Magnetic field density normalized to the upstream kinetic energy density $\epsilon_B$
at $\wpet = 1600$, shown in a 500 $\delta_e$ cutout near the shock interface.
To enhance the dynamic range a signed $\epsilon_B^{1/2}$ is shown. Right panel: Corresponding phase space
density. Notice how not only the phase space is quenched by the cooling, but also the magnetic field density at
the shock interface, and downstream of the shock it depends on the cooling rate. The velocity range is different
for the different cases.}\label{fig:cooling}
\end{figure*}
We have made long term 2D2V simulations of the shock, using reflecting boundaries and different cooling
times, including a run without cooling for reference, and with electron-ion and pair plasmas. The runs were
done with cubic interpolation, sixth order field pusher and a 17-point current density filter. We used the sliding
window technique to be able to follow the development of the shock up to $\wpet=5000$, and used more than
5 billion particles to model the shocks. In all cases the upstream
Lorentz factor is $\Gamma=10$. While below we present results from runs with 20 cells per skin depth, we
have used both 10, 20, and 40 cells per skin depth and between 12 and 24 particles per species per cell and
find converged results. Examples of the magnetic field and phase space density for different cooling times
can be seen in figure \ref{fig:cooling}.
The shocks can phenomenologically be categorized into three types: shocks in the strong cooling regime, radiative
shocks, and weakly radiative shocks. In the strong cooling regime, there are no traces of a power law tail of
accelerated particles, and both the evolution, shock jump conditions, and the micro structure itself are qualitatively
different from a normal non-radiating shock. In particular, the upstream is completely unperturbed by the
existence of the shock. This is found for cooling times $T_\textrm{cool} \lesssim 200$.
In a strongly radiating shock the micro structure is disturbed, and the downstream magnetic
islands only survive very close to the shock interface, but the upstream does have resemblance to a normal
collisionless shock. The power law tail of accelerated particles is completely gone, and the shock jump conditions
are altered. This is found for cooling times $200 \lesssim T_\textrm{cool} \lesssim 1000$.
The weakly radiative shocks are similar in structure
to a non-radiating shock, but with mildly perturbed shock jump conditions and propagation velocity
(see figure \ref{fig:jump}). The
power law index for the high energy population of accelerated electrons is steeper than for a non-radiating shock,
with a dependence on the cooling time.
The magnetic energy density
decreases significantly approximately 150 $\delta_e$ away from the shock interface, but the time a high energy particle
spends that close to the shock transition, where the bulk of the cooling happens, is a stochastic function of its angle to
the shock normal, and the number of scatterings on fluctuations in the electromagnetic field. In principle, if we simulate
for a long enough time, with a long cooling time and with high statistics, the powerlaw tail of accelerated particles will
grow over more than a decade in energy, and a well defined cooling break should emerge with two different slopes clearly
visible. In practice, given the tangled nature and stochastic propagation, the break will most probably be smooth, significantly
smeared out around the characteristic energy, where electrons start to be efficiently cooled. In these exploratory
simulations the emergence of a cooling break does not occur.  Instead, in the case of weakly radiative shocks, the
high-energy part of the particle distribution (PDF) is a powerlaw with an exponential cut-off, but with a steeper
powerlaw index than in the non-radiating case (see fig.~\ref{fig:pdf}). It is well known that in PIC simulations of non-radiating
shocks the upstream region affected by high energy particles produced at the shock interface only grows with time,
and it has been an open question what the long term structure looks like\cite{Keshet:2009}. This is different
for radiatively cooled shocks, where at large times the shock settles down to a quasi-steady state,
making it possible to draw conclusions about the long term behavior. The extent of the upstream is
limited, and the powerlaw part of the PDF does not grow in time.
Any collisionless shock is radiatively cooling, given long enough time, and from our simulations it
is clear that the impact of cooling for weakly radiating shocks is greater than speculated in e.g. Medvedev \&
Spitkovsky\cite{Medvedev:2009}, where analytic estimates were used.
Given the possibility of collisionless shocks to mediate secondary instabilities such as the Bell instability far upstream,
by the generation of streaming cosmic rays\cite{Niemiec:2008}, it would be interesting to understand the
impact on collisionless shock for the very large cooling times expected in e.g.~GRB afterglows.
We speculate that using progressively larger cooling times in sufficiently large simulations could give insight
in how to scale the solution to arbitrarily long cooling times.

\subsection{The linear magnetized Kelvin-Helmholtz instability}
Earth's magnetopause constitutes an important region in space. It is the boundary layer, separating the Earth's magnetosphere
from the solar wind. In this region, the Kelvin-Helmholtz instability (KHI) is driven by velocity shears between the magnetosheath
and the magnetospheric plasma at low latitude (close to the magnetic equator). We have selected the problem of the linear
magnetized KHI to compare with results using our local MHD \staggercode{}. Essentially, we use the
setup for the SWIFF Magnetopause Challenge~\cite{SWIFF_D2.1_KH} code comparison, except that the amplitude is slightly different
and there are half the number of skin depths in the kinetic case. The MHD and PIC code setups are nearly identical, the only difference
being the addition of microphysical parameters, and that in the PIC case we must ensure that the initial condition is a kinetic
equilibrium respecting constraints like Gauss law.

We compare our PIC results against those obtained with the \staggercode{}, also developed and maintained at the Niels Bohr
Institute\citet{1997LNP...489..179N}. This finite difference mesh based MHD code is a fully 3D resistive and compressible MHD
code. The MHD variables are located on staggered meshes, and the discretization is very similar to the \ppcode{}, with sixth order
spatial derivatives, and fifth order interpolation of variables. The time integration of the MHD equations is performed using an explicit
3rd order low storage Runge-Kutta method\citep{1980JCoPh..35...48W}.

The experiment is a periodic 2D3V setup, and the box size is $L_x=90\pi$, $L_y=30\pi$. The initial velocity field $\textbf{V} = V_y(x) \textbf{e}_y$
contains a periodic double shear layer, to avoid boundary effects, with a velocity amplitude $A_{eq}=1$. The sheared velocity jumps are
located at $L_x / 4$ and $3/4 \ L_x$, and the transition width is $a=3$. With these parameters, the initial
velocity profiles is defined as:
$$V_y(x) = \tanh \left( \frac{x - L_x / 4}{3} \right) - \tanh \left( \frac{x - 3/4 L_x}{3} \right)-1~.$$

Initially, $J_{eq} = \EE = 0$, $\BB = B_0 sin(\theta) \textbf{e}_y + B_0 cos(\theta)  \textbf{e}_z$, with $B_0 = 1$,
$\theta = 0.05$. The density, pressure and Alfv\'en velocity are all unity: $\rho=P=V_A=1$. For the MHD
we use an adiabatic equation of state with $\gamma_\textrm{gas} = 5/3$,
and impose a small perturbation in the velocity field to seed the KHI $\delta\textbf{V}=\textbf{e}_z\times\nabla\psi$, where
$$\psi = \epsilon f(x) \sum_{m=1}^{N_y/4} cos(2 \pi m y / L_y + \phi_m )/m ~,$$
and
$$f(x) = \exp \left[- \left(\frac{x - L_x/4} {3}\right)^2 \right] + \exp \left[ - \left(\frac{x - 3/4L_x} {3}\right)^2 \right]$$ and $\epsilon$
is such that $\max(| \delta V |) \simeq 10^{-3}$. $\phi_m$ are random phases.

\begin{figure*}[htbp]
    \centering
        \includegraphics[width=\textwidth]{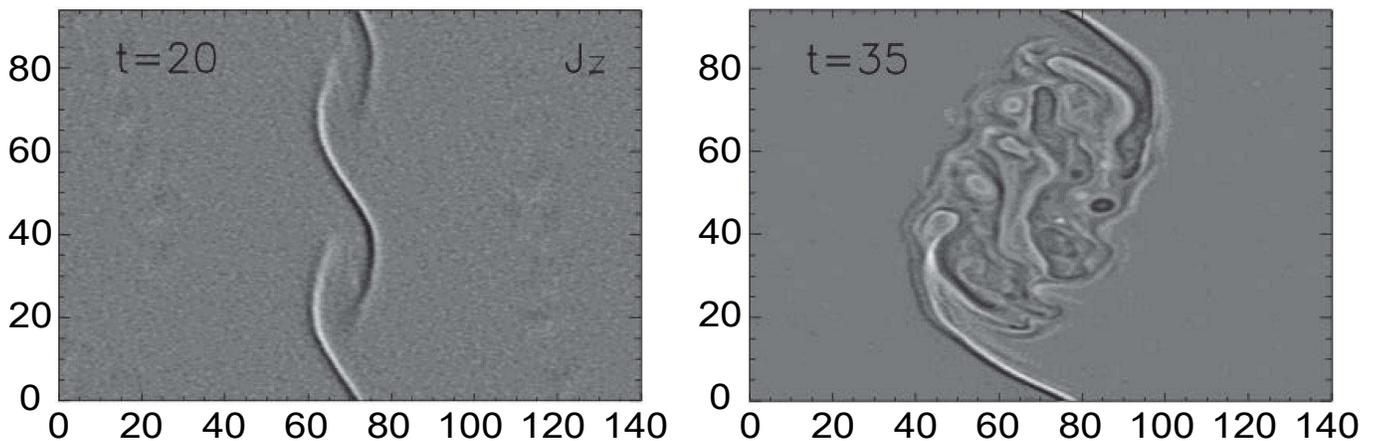}
    \caption{Early non-linear (left) and late non-linear vortex paring (right panel) stages of a PIC code model of the
    Kelvin-Helmholtz instability. The small scale electron density waves barely visible in the left panel are probably
    because of the initial condition not being a perfect kinetic equilibrium\cite{Belmont:2012.1,Belmont:2012.2}.}
    \label{fig:small_PIC_KH_fig}
\end{figure*}

In the MHD case we resolve the box with $N_x=1536$, $N_y=512$ cells. The PIC setup was prepared with physical initial
conditions essentially identical to the MHD setup, using the technique of section \ref{sec:embed} to reach an approximate
kinetic equilibrium. The objective was to produce the ion-scale KHI, while resolving the electron skin-depth
$\delta_e$ and having well separated inertial scales with $m_i/m_e =64$. We use a setup similar to what was used for a
hybrid code in the SWIFF comparison\cite{SWIFF_D2.1_KH},
but limit the number of ion skin depths to $45\pi \times 15\pi$, and the resolution to $N_x=6144$, $N_y=2048$. To have a reasonable
plasma frequency, we rescale the speed of light to $c=10$, and use $\delta_e = 6 \Delta x$, and 20 particles per cell per species
with a total of roughly 500 million particles in the box.
Our choice of the shearing jump amplitude ($V_0=\pm1.0$) and width ($a=3.0$) selects a fastest growing mode (FGM) leading to
production of two vortices, which pair up and merge during the early and late non-linear stages of the KHI.
Figure~\ref{fig:small_PIC_KH_fig} shows the vortices just prior to, and well after the vortex merging in the PIC model.

The growth rates for the two extremum cases of transverse ($\BB_0 \perp \vv_0$) and parallel ($\BB_0 \parallel
\vv_0$) configurations were investigated theoretically and further calculated
by Miura\cite{Miura:1982.1,Miura:1982.2}. We may assume that for a weak parallel component,
i.e.~$B_{0z} \ll B_{0y}$ ($B_{0z} = 0.05 B_{0y}$), the instability evolves almost as the ideal
transverse case to a good approximation. It was also determined that the stability criterion for $V_0=2$ is
$M_f\equiv{V_0}/{\sqrt{v_A^2+c_s^2}}<2$. Here, $M_f$ is the fast mode magneto-acoustic Mach number.

For our setup $M_f=\sqrt{2}$, with both $v_A^2=1$,$c_s^2=1$, using $V_0=2$ and $a=3$. From
Miura 1982\citep{Miura:1982.1}, we find a growth rate for our
specific setup to be ${2a}\gamma_{KH}/{V_0}\approx0.162$, which leads to the final result of
$$\gamma_{KH,FGM}\approx0.055\pm0.002.$$

To measure the KHI rate growth --- in both the MHD and PIC cases --- we calculate the quantity
\begin{equation}
    \left|\tilde{V_x}(x_i,k_y,t_n)\right|^2 = \left| \frac{1}{N_y} 
    \sum^{N_y-1}_{j=0} V_x(x_i,y_j,t_n) e^{-i2\pi\frac{k_yj}{N_y}} \right|^2~,
\end{equation}
i.e.~the power spectra along the $y$-axis of the $x$-component of velocity, $V_x$, at constant $x_i$.
This is then averaged in the $x$-direction
\begin{equation}
	\mathcal{Q}(t_n) = \frac{1}{x_{hi}-x_{lo}} \sum^{x_{hi}}_{x_{lo}} \left|\tilde{V_x}(x_i,k_y,t_n)\right|^2~,
\end{equation}
for two different sets of $\{x_{lo},x_{hi}\}$, namely i) centered on one shearing layer, 
5$\Delta_x$ wide, and ii) across the entire half-volume in the x-direction. We then make a fit 
to exponential growth to obtain $\gamma_{KI}(mode)$ for the FGM.

\begin{table}[!h]
 \caption{Growth rate, $\gamma_{KH,FGM}$ for the magnetized KHI; comparison between identical runs
 with the \ppcode{} and the \staggercode.}
 \label{tab:kh_growth_rates_compare}
 \begin{tabular}{ l c c c }
   \colrule\colrule
                &  PIC  & ~ &  MHD   \\
   \colrule
   Right        & 0.058 & ~ & 0.058  \\
   Total        & 0.050 & ~ & 0.056  \\
   \colrule
 \end{tabular}
\end{table}
We find that growth rates from both the MHD and PIC simulations agree well with theory, to
within about 10\% (see table \ref{tab:kh_growth_rates_compare}).
Averaging over all layers in the simulation half-plane yields a lower growth rate with the MHD in closest agreement with theory.
This is expected, since averaging over only 5 layers does not capture the entire width of the shearing layer which
is $W_{shear}\sim50\Delta_x$. For the PIC growth rate results discrepancies in physics and uncertainties in data fitting, due to mode
coupling are higher than for the MHD case; and growth rates therefore differ by as much as 10\% in the in the volume-averaged
case. The explanation for a slightly lower growth rate in the total volume averaged PIC case may be due to
enhanced dissipation and intrinsic noise properties of the PIC code, or the development of secondary instabilities.

Concluding this test, we emphasize that a PIC code has been used to run a fully MHD problem in PIC explicit mode, with
almost identical growth rates and vortex structures in the linear phase. For the vortex-merging epoch of the PIC run, results
deviate qualitatively and quantitatively from MHD, due to kinetic effects and secondary tearing-like instabilities developing during
the merging stage.

\section{Parallelization, scaling and performance}
Modern 3D Particle-in-Cell experiments in astrophysics use billions of particles
to simulate the macroscopic structure of plasmas. To run these simulations the
code has to be massively parallel. The \ppcode{} started with a simple domain decomposition
along one axis, using MPI. Later it was changed to support a 3D MPI decomposition,
and in the current version we use a hybrid parallelization with OpenMP and MPI to scale
the code effectively up to hundreds of thousands of cores (with 262.144 cores being the
largest case tried so far). The OpenMP hybrid approach
is also necessary when running the GPU version of the code, because normally the
ratio of CPU cores to GPUs in a GPU system is larger than one.

\subsection{MPI}
\begin{figure}
     \begin{center}
           \includegraphics[width=\linewidth]{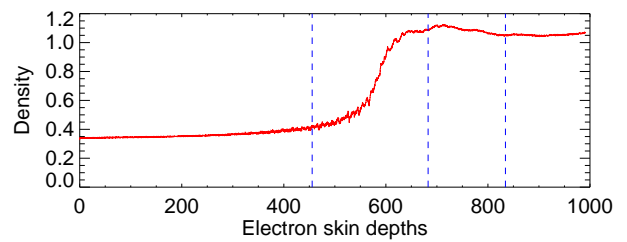}
     \end{center}
     \caption{Density profile of a 2D collisionless shock. The dashed lines indicate the $z$-boundaries of MPI domains.
     They are updated dynamically to equalize the load.} \label{fig:loadbalance}
\end{figure}

\begin{figure}
     \begin{center}
           \includegraphics[width=\linewidth]{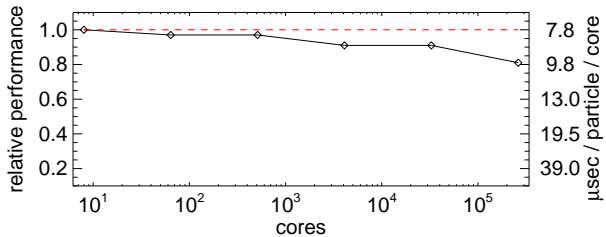}
     \end{center}
     \caption{Weak scaling in pure MPI mode when running on the BG/P machine JUGENE. The setup is a relativistic
     two-stream experiment with periodic boundaries and a $\Gamma=10$ streaming motion. There are roughly 80
     particles per cell with $16^3$ cells per MPI domain. This experiment is not using the charge conserving current
     deposition, and the scaling of the current code is significantly better.} \label{fig:mpi_scaling}
\end{figure}

The MPI parallelization has been designed to be as transparent as possible,
and consists of different modules, all collected in a single file. This makes it easy to
compile the \ppcode{} both with and without the MPI library.
\begin{itemize}
\item{Particles: After moving the particles on a node, and applying the physical boundary
conditions, on each thread we check sequentially in the $x$-, $y$-, and $z$-direction which
particles have moved to other threads, and interchange particles accordingly. A reverse transversion of
the linked list structure containing the send particles makes it efficient to store the received particles
in the same slots.}
\item{Fields: When the physical boundary conditions are applied for the fields, ghostzones are
also exchanged between threads.}
\item{I/O: The snapshot and restart file format is binary, and the code uses MPI-IO. For the particles
each attribute (ie.~the $x$-position) is stored in a single block, and the IO-types are all 64-bit, making
it possible to store billions of particles in a single snapshot with GB/s of performance. The code
stores everything in a few files, and restarts can be done on an arbitrary number of threads.}
\item{Synthetic Spectra: The spectra are sampled on sub-volumes of the data, and each thread
only allocates data for the sub-volumes that intersect with the local domain. This makes the
mechanism scalable to $\mathcal{O}(10^5)$ sub-volumes, without wasting memory. The lowest
ranked thread for a given sub-volume writes the data, making the I/O scalable too.}
\item{Particle Tracing: Particle tracing is done locally on each thread, while only the master thread
writes the I/O. This is a potential bottleneck, and on systems with relatively weak CPUs, e.g. Blue-Gene/P
we are limited to tracing $\sim$1 million particles, while for x86-based clusters with global Lustre
filesystems we can effectively trace up to $\sim$10 million particles, without significant slowdowns.}
\item{2D-slices: The code can extract (averaged) 2D slices on-the-fly while running. While this
is very convenient for movie making, for large runs thread contention becomes a problem when reducing
the data on a single thread: When thousands of threads try to communicate with a single thread, the
slow down can be significant. We use single-sided MPI to effectively circumvent the thread contention,
making slice writing scalable to at least 100.000 MPI threads on network transports that support RDMA.}
\end{itemize}
A major problem with simple domain decomposed particle-in-cell codes is that systematic local
over-densities of particles easily occur, for example in collisionless shock or laser wake-field simulations. This leads to severe
load imbalance. To work around this the \ppcode{} has a simple dynamic load-balance feature. If enabled, mesh-slices along
the $z$-direction can be exchanged in-between neighboring threads, according to a user defined cost formula.
In the current version particles and cells are assigned a certain cost, and this makes it possible to dynamically sample
local spikes in the particle density (see Figure~\ref{fig:loadbalance}). When a slice of cells is moved from a
thread to the neighbor thread the corresponding field values, particle data, synthetic spectra sub-volumes, and local
random number generators are updated. The same mechanism is used to make an optimal distribution of cells
when restarting a run.
We have tested the MPI performance in a simple, but realistic setup, with two counter streaming plasma beams on the
Blue-Gene/P machine JUGENE running in pure MPI mode. Figure \ref{fig:mpi_scaling} shows weak scaling behavior of the
code from 8 to 262.144 cores. This was done a few years ago, with a version of the code without charge conservation,
and the scaling of the current code is significantly better.
\begin{figure*}
     \begin{center}
           \includegraphics[width=0.32\linewidth]{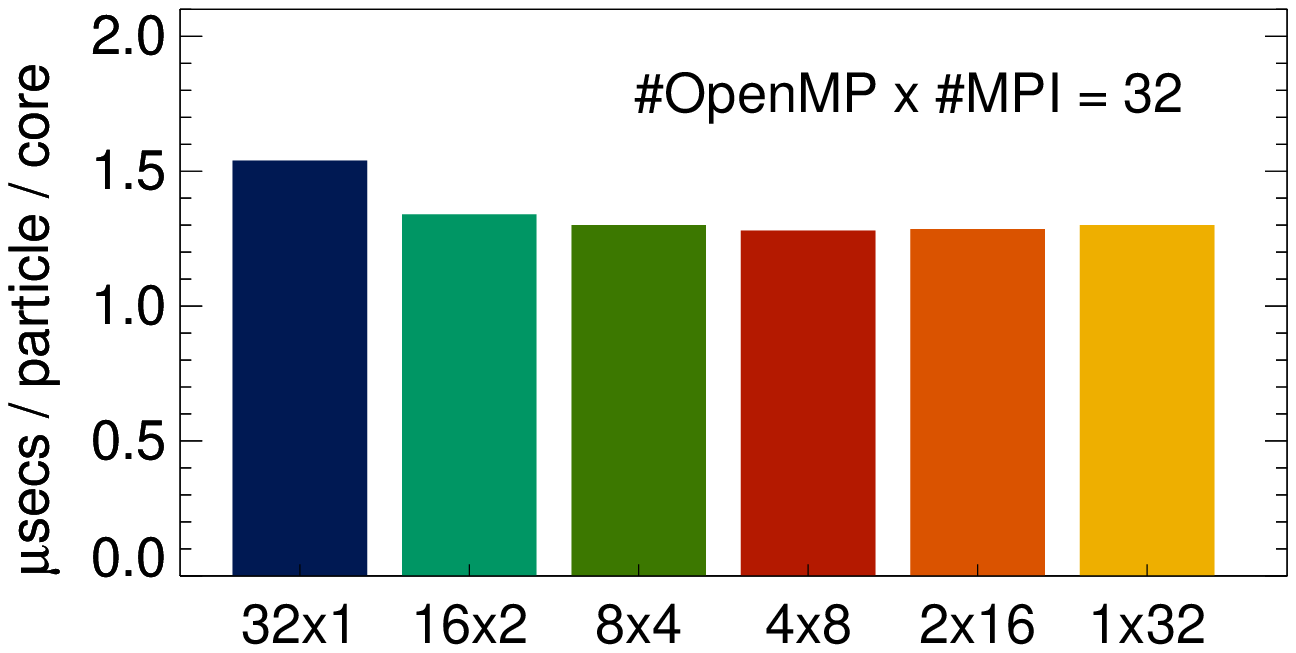}
           \includegraphics[width=0.32\linewidth]{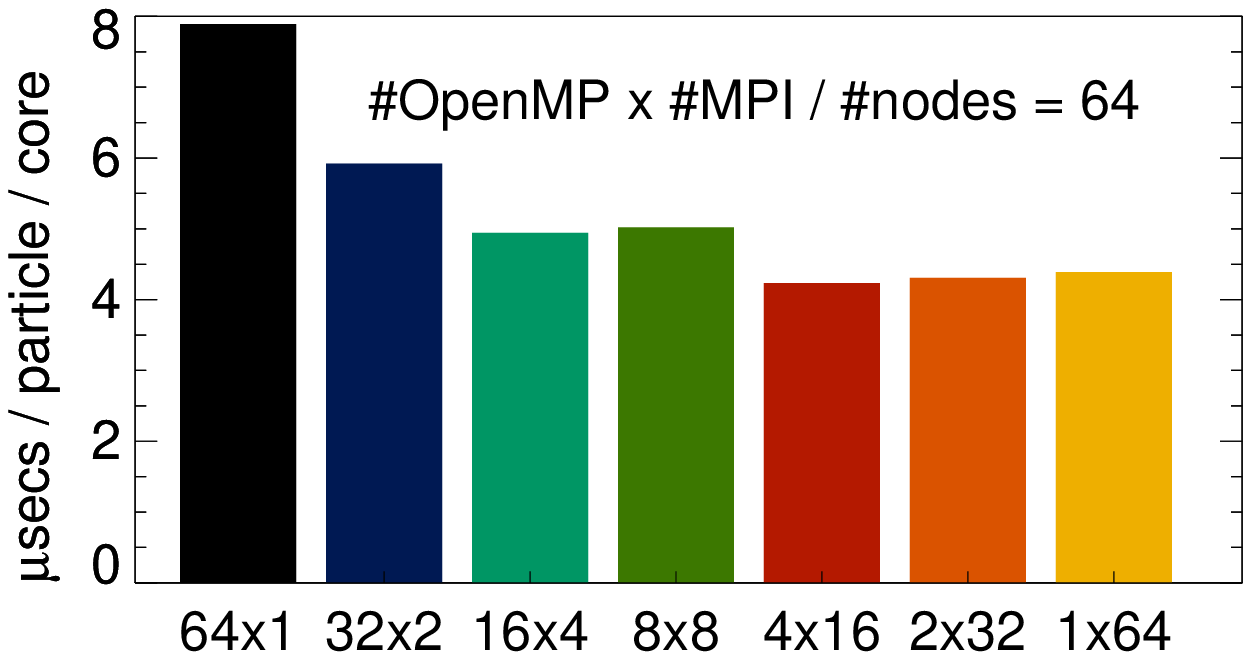}
           \includegraphics[width=0.32\linewidth]{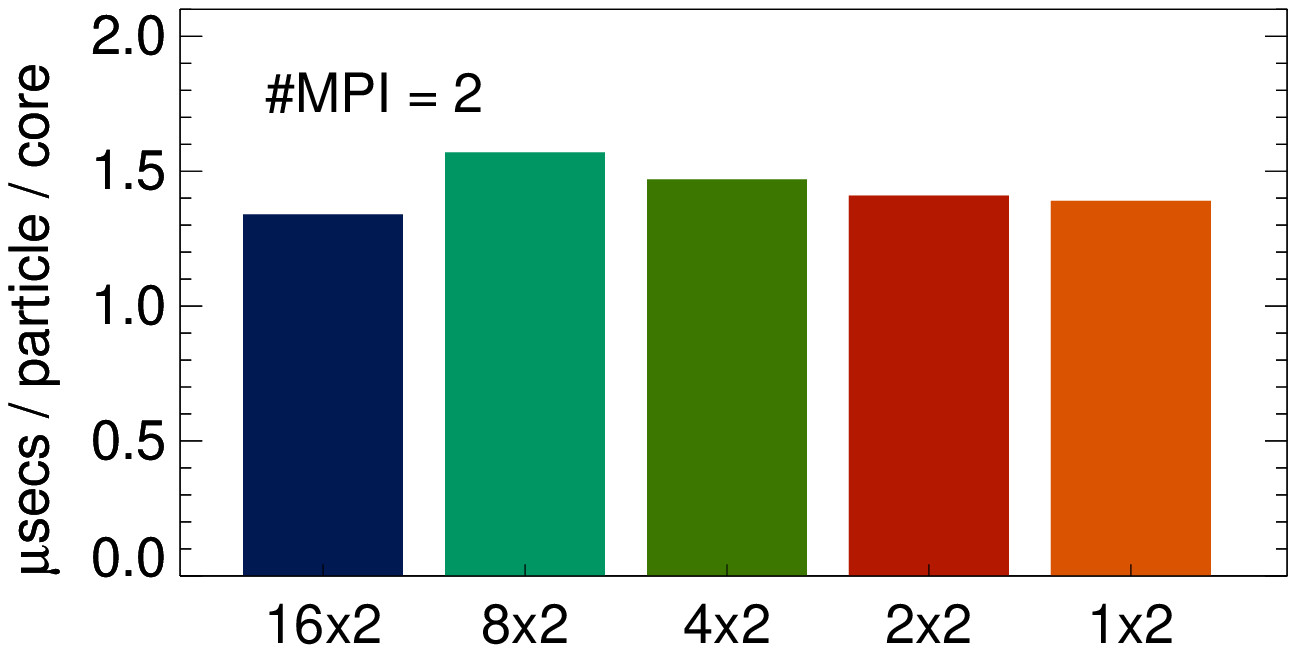}
     \end{center}
     \caption{Weak scaling in hybrid mode when running on a 16 core E5-2670 2.6GHz Sandy Bridge
     node and the BG/Q machine JUQUEEN. The setup is a relativistic two-stream experiment with periodic
     boundaries and a $\Gamma=10$ streaming motion. There are 60 particles per cell. This experiment is
     using the charge conserving current deposition. The label gives number of OpenMP threads times MPI threads,
     and on both machines we take advantage of hyperthreading, while performance is measured per physical
     core. The left panel shows weak scaling for a fixed domain size of $16^3$ cells per thread. The middle
     panel shows weak scaling on 64 nodes or 4096 threads on JUQUEEN with only $12^3$ cells per thread.
     The right panel shows strong scaling keeping the domain constant at $50^3$ cells.} \label{fig:openmp}
\end{figure*}

\subsection{OpenMP}
The 3D MPI domain decomposition has served well, scaling the code, for most applications, to 10.000 cores on
x86 clusters and more than 100.000 cores on Blue-Gene/P. To scale the code to a million threads or more on for example
Blue-Gene/Q, one must introduce a new layer of parallelization. We have used OpenMP for a number of reasons: It is
relatively easy to get almost perfect scaling for a low number of cores, it can be done incrementally, and it is a
natural thing to do for the GPU version of the \ppcode{}.
Furthermore, using a hybrid parallelization, the size of each MPI domain becomes bigger, and
the particle load-balance requirements, due to fluctuations and variations in number density, become smaller.
This is important for experiments with some level of
density fluctuations, even if the scaling per se is good in pure MPI mode. In the current version the most CPU consuming
parts of the code have been OpenMP parallelized:
\begin{itemize}
\item{The Mover / Charge deposition is trivially parallelized by allocating one set of charge density fields per thread,
and parallelizing the update of particles on a per cell basis.}
\item{The field solver consist of simple differential operators, and we use loop-based parallelization for perfect speedup.}
\item{The sorting of particles is more challenging: We first parallelize on the number of species, this is
embarrassing parallel. Then on a nested level we first partition the particle data with nested parallel sweeps in 128 sets,
and afterwards use quick-sort to sort each set of particles in parallel.}
\item{Sending particles between threads: We first parallelize on the number of species, this is
embarrassing parallel. Then on a nested level we use loop parallelization to select particles to be send, and multiplex MPI
communication with OMP sections.}
\item{Several other auxiliary routines have been parallelized and made thread safe: The random generators,
synthetic spectra sampling, initial- and boundary conditions.}
\end{itemize}
The rest of the code, mostly diagnostics and I/O, can be incrementally OpenMP parallelized as needed. We find excellent
scaling inside x86 nodes, with optimal performance at 4 MPI threads per socket / 8 per node, only loosing about 5\%
when using 1 MPI thread per socket. The 5\% is easily regained in higher MPI efficiency and better load-balance when running
on a large number of nodes (top panel in \fig{fig:openmp}).
We also find that we gain roughly 15\% in performance by running with hyperthreading (32 compared to 16 threads in bottom panel in
\fig{fig:openmp}). Finally, on Blue-Gene/Q it is absolutely crucial to use all 64 hardware threads on a 16 core node. It
improves overall performance by a factor of 2. But to effectively use the massive amount of threads exposed in the system,
we need OpenMP, since otherwise the domain size is too small. Currently we use 4 or 16 OpenMP threads per node,
depending on the scale of
the problem. Even though we loose 16\% performance by going from 4 to 16 OpenMP threads, for runs with e.g. particle tracing
enabled or other IO done only through the master thread, and for very large runs, where load imbalance can be more than
20\%, it is advantageous to use 16 OpenMP threads per MPI thread. In simpler and smaller scale runs we use 4
OpenMP threads per MPI thread.

\section{Concluding Remarks}
In this paper we present the \ppcode{} particle-in-cell code, with new numerical methods, physical extensions,
and parallelization techniques. Originally the main motivation for developing a new particle-in-cell code from
scratch was to implement a modern, modularized and extendible code, with modern numerical techniques. In the paper
we have demonstrated how the extension of the classical PIC framework to higher
order spatial and temporal derivatives, together with a novel charge conservation scheme, gives a much higher accuracy
and better stability than traditional, second order PIC codes, when using the same number of grid points.
Conversely, fewer mesh points are needed to reach a given level of fidelity in realistic three dimensional
kinetic astrophysical setups, which leads
to large savings in computational costs. The excellent scalability of the code, and the ability to do
diagnostics on-the-fly, makes it possible to achieve a high scientific turnaround, where numerical experiments with
more than 100 billion particles can be performed in a day, with dedicated use of petascale computational resources.

Among the novel physical extensions of the \ppcode{} code are the description of binary interactions by particle splitting,
the ability to include radiative cooling in a self-consistent manner, the possibility of embedding the kinetic model in an
MHD snapshot, and the application of a sliding simulation window. Other extensions are in development or are planned
for the future, among them nested grids, proper inclusion of neutral particles and dust, self gravity,
smooth transition from the kinetic to
the MHD regime, and simpler binary collision drag descriptions, relevant for studying fractionation processes.

The \ppcode{} has already been in full production mode for some years, the user base is growing
steadily, and we expect that it will be used extensively at the Niels Bohr Institute and elsewhere
for many years to come.
Hopefully the detailed description of the algorithms documented in this paper will also be of use to others,
when developing particle-in-cell techniques and applying them astrophysical plasma physics problems in the future.

\begin{acknowledgments}
We thank Juri Poutanen for funding the workshop 'Thinkshop 2005' in Stockholm 2005;
\url{http://comp.astro.ku.dk/Twiki/view/NumPlasma/ThinkShop2005}, which initialized the
development of the \ppcode{} code. Further, we wish to acknowledge Boris Stern for
ample help with Compton scattering Large Particle Monte-Carlo techniques.
TH and {\AA}N are supported by the Centre for Star and Planet Formation, which is
financed by the Danish National Science Foundation.
{\AA}N and JTF are supported by SWIFF; the research leading to these results has
received funding  from the European Commission's Seventh Framework Programme
(FP7/2007-2013) under the grant agreement SWIFF (project n$^\texttt{o}$ 263340, www.swiff.eu).
Computer time was provided by the Danish Center for Scientific Computing (DCSC),
and through the PRACE project 'Ab Initio Modeling of Solar Active
Regions' running on JUGENE and JUQUEEN at the J\"ulich Supercomputing Centre; we thank
the centre staff for their support, in particular in connection with I/O scaling issues,
and for help with execution of full machine jobs.
\end{acknowledgments}

\bibstyle{pop}
\bibliography{references}

\begin{thebibliography}{52}%
\makeatletter
\providecommand \@ifxundefined [1]{%
 \@ifx{#1\undefined}
}%
\providecommand \@ifnum [1]{%
 \ifnum #1\expandafter \@firstoftwo
 \else \expandafter \@secondoftwo
 \fi
}%
\providecommand \@ifx [1]{%
 \ifx #1\expandafter \@firstoftwo
 \else \expandafter \@secondoftwo
 \fi
}%
\providecommand \natexlab [1]{#1}%
\providecommand \enquote  [1]{``#1''}%
\providecommand \bibnamefont  [1]{#1}%
\providecommand \bibfnamefont [1]{#1}%
\providecommand \citenamefont [1]{#1}%
\providecommand \href@noop [0]{\@secondoftwo}%
\providecommand \href [0]{\begingroup \@sanitize@url \@href}%
\providecommand \@href[1]{\@@startlink{#1}\@@href}%
\providecommand \@@href[1]{\endgroup#1\@@endlink}%
\providecommand \@sanitize@url [0]{\catcode `\\12\catcode `\$12\catcode
  `\&12\catcode `\#12\catcode `\^12\catcode `\_12\catcode `\%12\relax}%
\providecommand \@@startlink[1]{}%
\providecommand \@@endlink[0]{}%
\providecommand \url  [0]{\begingroup\@sanitize@url \@url }%
\providecommand \@url [1]{\endgroup\@href {#1}{\urlprefix }}%
\providecommand \urlprefix  [0]{URL }%
\providecommand \Eprint [0]{\href }%
\providecommand \doibase [0]{http://dx.doi.org/}%
\providecommand \selectlanguage [0]{\@gobble}%
\providecommand \bibinfo  [0]{\@secondoftwo}%
\providecommand \bibfield  [0]{\@secondoftwo}%
\providecommand \translation [1]{[#1]}%
\providecommand \BibitemOpen [0]{}%
\providecommand \bibitemStop [0]{}%
\providecommand \bibitemNoStop [0]{.\EOS\space}%
\providecommand \EOS [0]{\spacefactor3000\relax}%
\providecommand \BibitemShut  [1]{\csname bibitem#1\endcsname}%
\let\auto@bib@innerbib\@empty
\bibitem [{\citenamefont {{Evans}}\ and\ \citenamefont
  {{Harlow}}(1957)}]{Harlow:1957}%
  \BibitemOpen
  \bibfield  {author} {\bibinfo {author} {\bibfnamefont {M.~W.}\ \bibnamefont
  {{Evans}}}\ and\ \bibinfo {author} {\bibfnamefont {F.~H.}\ \bibnamefont
  {{Harlow}}},\ }\href@noop {} {\enquote {\bibinfo {title} {The
  particle-in-cell method for hydrodynamic calculations},}\ }\bibinfo {type}
  {Tech. Rep.}\ \bibinfo {number} {LA-2139}\ (\bibinfo  {institution} {Los
  Alamos Scientific Laboratory},\ \bibinfo {year} {1957})\BibitemShut {NoStop}%
\bibitem [{\citenamefont {{Harlow}}(1964)}]{Harlow:1964}%
  \BibitemOpen
  \bibfield  {author} {\bibinfo {author} {\bibfnamefont {F.~H.}\ \bibnamefont
  {{Harlow}}},\ }\href@noop {} {\bibfield  {journal} {\bibinfo  {journal}
  {Methods Comput. Phys.}\ }\textbf {\bibinfo {volume} {3}},\ \bibinfo {pages}
  {319 } (\bibinfo {year} {1964})}\BibitemShut {NoStop}%
\bibitem [{\citenamefont {Birdsall}\ and\ \citenamefont
  {Langdon}(1985)}]{birdsall:1985}%
  \BibitemOpen
  \bibfield  {author} {\bibinfo {author} {\bibfnamefont {C.~K.}\ \bibnamefont
  {Birdsall}}\ and\ \bibinfo {author} {\bibfnamefont {A.~B.}\ \bibnamefont
  {Langdon}},\ }\href@noop {} {\emph {\bibinfo {title} {Plasma physics via
  computer simulation}}}\ (\bibinfo  {publisher} {McGraw-Hill, New York :},\
  \bibinfo {year} {1985})\ pp.\ \bibinfo {pages} {xxiii, 479 p. :}\BibitemShut
  {NoStop}%
\bibitem [{\citenamefont {{Hockney}}\ and\ \citenamefont
  {{Eastwood}}(1988)}]{hockney:1988}%
  \BibitemOpen
  \bibfield  {author} {\bibinfo {author} {\bibfnamefont {R.~W.}\ \bibnamefont
  {{Hockney}}}\ and\ \bibinfo {author} {\bibfnamefont {J.~W.}\ \bibnamefont
  {{Eastwood}}},\ }\href@noop {} {\emph {\bibinfo {title} {Bristol: Hilger,
  1988}}},\ edited by\ \bibinfo {editor} {\bibnamefont {{Hockney, R.~W.~\&
  Eastwood, J.~W.}}}\ (\bibinfo  {publisher} {Taylor \& Francis},\ \bibinfo
  {year} {1988})\BibitemShut {NoStop}%
\bibitem [{\citenamefont {{Lapenta}}, \citenamefont {{Brackbill}},\ and\
  \citenamefont {{Ricci}}(2006)}]{Lapenta:2006}%
  \BibitemOpen
  \bibfield  {author} {\bibinfo {author} {\bibfnamefont {G.}~\bibnamefont
  {{Lapenta}}}, \bibinfo {author} {\bibfnamefont {J.~U.}\ \bibnamefont
  {{Brackbill}}}, \ and\ \bibinfo {author} {\bibfnamefont {P.}~\bibnamefont
  {{Ricci}}},\ }\href {\doibase 10.1063/1.2173623} {\bibfield  {journal}
  {\bibinfo  {journal} {Physics of Plasmas}\ }\textbf {\bibinfo {volume}
  {13}},\ \bibinfo {pages} {055904} (\bibinfo {year} {2006})}\BibitemShut
  {NoStop}%
\bibitem [{\citenamefont {Chaniotis}\ and\ \citenamefont
  {Poulikakos}(2004)}]{Chaniotis2004}%
  \BibitemOpen
  \bibfield  {author} {\bibinfo {author} {\bibfnamefont {A.}~\bibnamefont
  {Chaniotis}}\ and\ \bibinfo {author} {\bibfnamefont {D.}~\bibnamefont
  {Poulikakos}},\ }\href {\doibase 10.1016/j.jcp.2003.11.026} {\bibfield
  {journal} {\bibinfo  {journal} {Journal of Computational Physics}\ }\textbf
  {\bibinfo {volume} {197}},\ \bibinfo {pages} {253 } (\bibinfo {year}
  {2004})}\BibitemShut {NoStop}%
\bibitem [{\citenamefont {{Yee}}(1966)}]{yee}%
  \BibitemOpen
  \bibfield  {author} {\bibinfo {author} {\bibfnamefont {K.}~\bibnamefont
  {{Yee}}},\ }\href {\doibase 10.1109/TAP.1966.1138693} {\bibfield  {journal}
  {\bibinfo  {journal} {IEEE Transactions on Antennas and Propagation}\
  }\textbf {\bibinfo {volume} {14}},\ \bibinfo {pages} {302} (\bibinfo {year}
  {1966})}\BibitemShut {NoStop}%
\bibitem [{\citenamefont {{Yoshida}}(1990)}]{Yoshida:1990}%
  \BibitemOpen
  \bibfield  {author} {\bibinfo {author} {\bibfnamefont {H.}~\bibnamefont
  {{Yoshida}}},\ }\href {\doibase 10.1016/0375-9601(90)90092-3} {\bibfield
  {journal} {\bibinfo  {journal} {Physics Letters A}\ }\textbf {\bibinfo
  {volume} {150}},\ \bibinfo {pages} {262} (\bibinfo {year}
  {1990})}\BibitemShut {NoStop}%
\bibitem [{\citenamefont {{Forest}}\ and\ \citenamefont
  {{Ruth}}(1990)}]{Forest:1990}%
  \BibitemOpen
  \bibfield  {author} {\bibinfo {author} {\bibfnamefont {E.}~\bibnamefont
  {{Forest}}}\ and\ \bibinfo {author} {\bibfnamefont {R.~D.}\ \bibnamefont
  {{Ruth}}},\ }\href {\doibase 10.1016/0167-2789(90)90019-L} {\bibfield
  {journal} {\bibinfo  {journal} {Physica D Nonlinear Phenomena}\ }\textbf
  {\bibinfo {volume} {43}},\ \bibinfo {pages} {105} (\bibinfo {year}
  {1990})}\BibitemShut {NoStop}%
\bibitem [{\citenamefont {{Candy}}\ and\ \citenamefont
  {{Rozmus}}(1991)}]{Candy:1991}%
  \BibitemOpen
  \bibfield  {author} {\bibinfo {author} {\bibfnamefont {J.}~\bibnamefont
  {{Candy}}}\ and\ \bibinfo {author} {\bibfnamefont {W.}~\bibnamefont
  {{Rozmus}}},\ }\href {\doibase 10.1016/0021-9991(91)90299-Z} {\bibfield
  {journal} {\bibinfo  {journal} {Journal of Computational Physics}\ }\textbf
  {\bibinfo {volume} {92}},\ \bibinfo {pages} {230} (\bibinfo {year}
  {1991})}\BibitemShut {NoStop}%
\bibitem [{\citenamefont {{Boris}}(1970)}]{boris}%
  \BibitemOpen
  \bibfield  {author} {\bibinfo {author} {\bibfnamefont {J.~P.}\ \bibnamefont
  {{Boris}}},\ }in\ \href@noop {} {\emph {\bibinfo {booktitle} {{the Fourth
  Conference on Numerical Simulation Plasmas (Naval Research Laboratory
  Washington, D.C.)}}}}\ (\bibinfo {year} {1970})\ pp.\ \bibinfo {pages}
  {3--67}\BibitemShut {NoStop}%
\bibitem [{\citenamefont {{Vay}}(2008)}]{vay:2008}%
  \BibitemOpen
  \bibfield  {author} {\bibinfo {author} {\bibfnamefont {J.-L.}\ \bibnamefont
  {{Vay}}},\ }\href {\doibase 10.1063/1.2837054} {\bibfield  {journal}
  {\bibinfo  {journal} {Physics of Plasmas}\ }\textbf {\bibinfo {volume}
  {15}},\ \bibinfo {pages} {056701} (\bibinfo {year} {2008})}\BibitemShut
  {NoStop}%
\bibitem [{\citenamefont {{Nordlund}}\ and\ \citenamefont
  {{Galsgaard}}(1997)}]{1997LNP...489..179N}%
  \BibitemOpen
  \bibfield  {author} {\bibinfo {author} {\bibfnamefont {A.}~\bibnamefont
  {{Nordlund}}}\ and\ \bibinfo {author} {\bibfnamefont {K.}~\bibnamefont
  {{Galsgaard}}},\ }in\ \href {\doibase 10.1007/BFb0105676} {\emph {\bibinfo
  {booktitle} {European Meeting on Solar Physics}}},\ \bibinfo {series}
  {Lecture Notes in Physics, Berlin Springer Verlag}, Vol.\ \bibinfo {volume}
  {489},\ \bibinfo {editor} {edited by\ \bibinfo {editor} {\bibfnamefont
  {G.~M.}\ \bibnamefont {{Simnett}}}, \bibinfo {editor} {\bibfnamefont {C.~E.}\
  \bibnamefont {{Alissandrakis}}}, \ and\ \bibinfo {editor} {\bibfnamefont
  {L.}~\bibnamefont {{Vlahos}}}}\ (\bibinfo {year} {1997})\ p.\ \bibinfo
  {pages} {179}\BibitemShut {NoStop}%
\bibitem [{\citenamefont {{Eastwood}}(1991)}]{eastwood:1991}%
  \BibitemOpen
  \bibfield  {author} {\bibinfo {author} {\bibfnamefont {J.~W.}\ \bibnamefont
  {{Eastwood}}},\ }\href {\doibase 10.1016/0010-4655(91)90036-K} {\bibfield
  {journal} {\bibinfo  {journal} {Computer Physics Communications}\ }\textbf
  {\bibinfo {volume} {64}},\ \bibinfo {pages} {252} (\bibinfo {year}
  {1991})}\BibitemShut {NoStop}%
\bibitem [{\citenamefont {{Villasenor}}\ and\ \citenamefont
  {{Buneman}}(1992)}]{villasenor:1992}%
  \BibitemOpen
  \bibfield  {author} {\bibinfo {author} {\bibfnamefont {J.}~\bibnamefont
  {{Villasenor}}}\ and\ \bibinfo {author} {\bibfnamefont {O.}~\bibnamefont
  {{Buneman}}},\ }\href {\doibase 10.1016/0010-4655(92)90169-Y} {\bibfield
  {journal} {\bibinfo  {journal} {Computer Physics Communications}\ }\textbf
  {\bibinfo {volume} {69}},\ \bibinfo {pages} {306} (\bibinfo {year}
  {1992})}\BibitemShut {NoStop}%
\bibitem [{\citenamefont {{Esirkepov}}(2001)}]{esirkepov:2001}%
  \BibitemOpen
  \bibfield  {author} {\bibinfo {author} {\bibfnamefont {T.~Z.}\ \bibnamefont
  {{Esirkepov}}},\ }\href {\doibase 10.1016/S0010-4655(00)00228-9} {\bibfield
  {journal} {\bibinfo  {journal} {Computer Physics Communications}\ }\textbf
  {\bibinfo {volume} {135}},\ \bibinfo {pages} {144} (\bibinfo {year}
  {2001})}\BibitemShut {NoStop}%
\bibitem [{\citenamefont {{Umeda}}\ \emph {et~al.}(2003)\citenamefont
  {{Umeda}}, \citenamefont {{Omura}}, \citenamefont {{Tominaga}},\ and\
  \citenamefont {{Matsumoto}}}]{umeda:2003}%
  \BibitemOpen
  \bibfield  {author} {\bibinfo {author} {\bibfnamefont {T.}~\bibnamefont
  {{Umeda}}}, \bibinfo {author} {\bibfnamefont {Y.}~\bibnamefont {{Omura}}},
  \bibinfo {author} {\bibfnamefont {T.}~\bibnamefont {{Tominaga}}}, \ and\
  \bibinfo {author} {\bibfnamefont {H.}~\bibnamefont {{Matsumoto}}},\ }\href
  {\doibase 10.1016/S0010-4655(03)00437-5} {\bibfield  {journal} {\bibinfo
  {journal} {Computer Physics Communications}\ }\textbf {\bibinfo {volume}
  {156}},\ \bibinfo {pages} {73} (\bibinfo {year} {2003})}\BibitemShut
  {NoStop}%
\bibitem [{\citenamefont {{Londrillo}}\ \emph {et~al.}(2010)\citenamefont
  {{Londrillo}}, \citenamefont {{Benedetti}}, \citenamefont {{Sgattoni}},\ and\
  \citenamefont {{Turchetti}}}]{londrillo:2010}%
  \BibitemOpen
  \bibfield  {author} {\bibinfo {author} {\bibfnamefont {P.}~\bibnamefont
  {{Londrillo}}}, \bibinfo {author} {\bibfnamefont {C.}~\bibnamefont
  {{Benedetti}}}, \bibinfo {author} {\bibfnamefont {A.}~\bibnamefont
  {{Sgattoni}}}, \ and\ \bibinfo {author} {\bibfnamefont {G.}~\bibnamefont
  {{Turchetti}}},\ }\href {\doibase 10.1016/j.nima.2010.01.055} {\bibfield
  {journal} {\bibinfo  {journal} {Nuclear Instruments and Methods in Physics
  Research A}\ }\textbf {\bibinfo {volume} {620}},\ \bibinfo {pages} {28}
  (\bibinfo {year} {2010})}\BibitemShut {NoStop}%
\bibitem [{\citenamefont {Engeln-M\"{u}llges}\ and\ \citenamefont
  {Uhlig}(1996)}]{penta:1996}%
  \BibitemOpen
  \bibfield  {author} {\bibinfo {author} {\bibfnamefont {G.}~\bibnamefont
  {Engeln-M\"{u}llges}}\ and\ \bibinfo {author} {\bibfnamefont
  {F.}~\bibnamefont {Uhlig}},\ }\href@noop {} {\emph {\bibinfo {title}
  {Numerical algorithms with C}}}\ (\bibinfo  {publisher} {Springer-Verlag New
  York, Inc.},\ \bibinfo {address} {New York, NY, USA},\ \bibinfo {year}
  {1996})\BibitemShut {NoStop}%
\bibitem [{\citenamefont {{Hededal}}(2005)}]{Hededal:2005b}%
  \BibitemOpen
  \bibfield  {author} {\bibinfo {author} {\bibfnamefont {C.}~\bibnamefont
  {{Hededal}}},\ }\emph {\bibinfo {title} {{Gamma-Ray Bursts, Collisionless
  Shocks and Synthetic Spectra}}},\ \href@noop {} {Ph.D. thesis},\ \bibinfo
  {school} {Niels Bohr Institute [astro-ph/0506559]} (\bibinfo {year} {2005}),\
  \Eprint {http://arxiv.org/abs/arXiv:astro-ph/0506559}
  {arXiv:astro-ph/0506559} \BibitemShut {NoStop}%
\bibitem [{\citenamefont {Matsumoto}\ and\ \citenamefont
  {Nishimura}(1998)}]{mersenne}%
  \BibitemOpen
  \bibfield  {author} {\bibinfo {author} {\bibfnamefont {M.}~\bibnamefont
  {Matsumoto}}\ and\ \bibinfo {author} {\bibfnamefont {T.}~\bibnamefont
  {Nishimura}},\ }\href {\doibase 10.1145/272991.272995} {\bibfield  {journal}
  {\bibinfo  {journal} {ACM Trans. Model. Comput. Simul.}\ }\textbf {\bibinfo
  {volume} {8}},\ \bibinfo {pages} {3} (\bibinfo {year} {1998})}\BibitemShut
  {NoStop}%
\bibitem [{\citenamefont {{Dunkel}}\ and\ \citenamefont
  {{H{\"a}nggi}}(2009)}]{Dunkel:2009}%
  \BibitemOpen
  \bibfield  {author} {\bibinfo {author} {\bibfnamefont {J.}~\bibnamefont
  {{Dunkel}}}\ and\ \bibinfo {author} {\bibfnamefont {P.}~\bibnamefont
  {{H{\"a}nggi}}},\ }\href {\doibase 10.1016/j.physrep.2008.12.001} {\bibfield
  {journal} {\bibinfo  {journal} {Physics Reports}\ }\textbf {\bibinfo {volume}
  {471}},\ \bibinfo {pages} {1} (\bibinfo {year} {2009})},\ \Eprint
  {http://arxiv.org/abs/0812.1996} {arXiv:0812.1996 [cond-mat.stat-mech]}
  \BibitemShut {NoStop}%
\bibitem [{\citenamefont {{Berenger}}(1994)}]{PML2D}%
  \BibitemOpen
  \bibfield  {author} {\bibinfo {author} {\bibfnamefont {J.-P.}\ \bibnamefont
  {{Berenger}}},\ }\href {\doibase 10.1006/jcph.1994.1159} {\bibfield
  {journal} {\bibinfo  {journal} {Journal of Computational Physics}\ }\textbf
  {\bibinfo {volume} {114}},\ \bibinfo {pages} {185} (\bibinfo {year}
  {1994})}\BibitemShut {NoStop}%
\bibitem [{\citenamefont {{Berenger}}(1996)}]{PML3D}%
  \BibitemOpen
  \bibfield  {author} {\bibinfo {author} {\bibfnamefont {J.}~\bibnamefont
  {{Berenger}}},\ }\href {\doibase 10.1006/jcph.1996.0181} {\bibfield
  {journal} {\bibinfo  {journal} {Journal of Computational Physics}\ }\textbf
  {\bibinfo {volume} {127}},\ \bibinfo {pages} {363} (\bibinfo {year}
  {1996})}\BibitemShut {NoStop}%
\bibitem [{\citenamefont {{Umeda}}, \citenamefont {{Omura}},\ and\
  \citenamefont {{Matsumoto}}(2001)}]{umeda:2001}%
  \BibitemOpen
  \bibfield  {author} {\bibinfo {author} {\bibfnamefont {T.}~\bibnamefont
  {{Umeda}}}, \bibinfo {author} {\bibfnamefont {Y.}~\bibnamefont {{Omura}}}, \
  and\ \bibinfo {author} {\bibfnamefont {H.}~\bibnamefont {{Matsumoto}}},\
  }\href {\doibase 10.1016/S0010-4655(01)00182-5} {\bibfield  {journal}
  {\bibinfo  {journal} {Computer Physics Communications}\ }\textbf {\bibinfo
  {volume} {137}},\ \bibinfo {pages} {286} (\bibinfo {year}
  {2001})}\BibitemShut {NoStop}%
\bibitem [{\citenamefont {Tzeng}, \citenamefont {Mori},\ and\ \citenamefont
  {Decker}(1996)}]{movingframe}%
  \BibitemOpen
  \bibfield  {author} {\bibinfo {author} {\bibfnamefont {K.-C.}\ \bibnamefont
  {Tzeng}}, \bibinfo {author} {\bibfnamefont {W.~B.}\ \bibnamefont {Mori}}, \
  and\ \bibinfo {author} {\bibfnamefont {C.~D.}\ \bibnamefont {Decker}},\
  }\href {\doibase 10.1103/PhysRevLett.76.3332} {\bibfield  {journal} {\bibinfo
   {journal} {Phys. Rev. Lett.}\ }\textbf {\bibinfo {volume} {76}},\ \bibinfo
  {pages} {3332} (\bibinfo {year} {1996})}\BibitemShut {NoStop}%
\bibitem [{\citenamefont {{Haugb{\o}lle}}(2011)}]{haugboelle:2011}%
  \BibitemOpen
  \bibfield  {author} {\bibinfo {author} {\bibfnamefont {T.}~\bibnamefont
  {{Haugb{\o}lle}}},\ }\href {\doibase 10.1088/2041-8205/739/2/L42} {\bibfield
  {journal} {\bibinfo  {journal} {ApjL}\ }\textbf {\bibinfo {volume} {739}},\
  \bibinfo {eid} {L42} (\bibinfo {year} {2011})}\BibitemShut {NoStop}%
\bibitem [{\citenamefont {{Baumann}}, \citenamefont {{Haugb{\o}lle}},\ and\
  \citenamefont {{Nordlund}}(2012)}]{baumann:2012a}%
  \BibitemOpen
  \bibfield  {author} {\bibinfo {author} {\bibfnamefont {G.}~\bibnamefont
  {{Baumann}}}, \bibinfo {author} {\bibfnamefont {T.}~\bibnamefont
  {{Haugb{\o}lle}}}, \ and\ \bibinfo {author} {\bibfnamefont
  {{\AA}.}~\bibnamefont {{Nordlund}}},\ }\href@noop {} {\bibfield  {journal}
  {\bibinfo  {journal} {ArXiv e-prints}\ } (\bibinfo {year} {2012})},\ \Eprint
  {http://arxiv.org/abs/1204.4947} {arXiv:1204.4947 [astro-ph.SR]} \BibitemShut
  {NoStop}%
\bibitem [{\citenamefont {{Baumann}}\ and\ \citenamefont
  {{Nordlund}}(2012)}]{baumann:2012b}%
  \BibitemOpen
  \bibfield  {author} {\bibinfo {author} {\bibfnamefont {G.}~\bibnamefont
  {{Baumann}}}\ and\ \bibinfo {author} {\bibfnamefont {{\AA}.}~\bibnamefont
  {{Nordlund}}},\ }\href@noop {} {\bibfield  {journal} {\bibinfo  {journal}
  {ArXiv e-prints}\ } (\bibinfo {year} {2012})},\ \Eprint
  {http://arxiv.org/abs/1205.3486} {arXiv:1205.3486 [astro-ph.SR]} \BibitemShut
  {NoStop}%
\bibitem [{\citenamefont {{Trier Frederiksen}}\ \emph
  {et~al.}(2010)\citenamefont {{Trier Frederiksen}}, \citenamefont
  {{Haugb{\o}lle}}, \citenamefont {{Medvedev}},\ and\ \citenamefont
  {{Nordlund}}}]{Trier:2010}%
  \BibitemOpen
  \bibfield  {author} {\bibinfo {author} {\bibfnamefont {J.}~\bibnamefont
  {{Trier Frederiksen}}}, \bibinfo {author} {\bibfnamefont {T.}~\bibnamefont
  {{Haugb{\o}lle}}}, \bibinfo {author} {\bibfnamefont {M.~V.}\ \bibnamefont
  {{Medvedev}}}, \ and\ \bibinfo {author} {\bibfnamefont {{\AA}.}~\bibnamefont
  {{Nordlund}}},\ }\href {\doibase 10.1088/2041-8205/722/1/L114} {\bibfield
  {journal} {\bibinfo  {journal} {\apjl}\ }\textbf {\bibinfo {volume} {722}},\
  \bibinfo {pages} {L114} (\bibinfo {year} {2010})},\ \Eprint
  {http://arxiv.org/abs/1003.1140} {arXiv:1003.1140 [astro-ph.HE]} \BibitemShut
  {NoStop}%
\bibitem [{\citenamefont {{Medvedev}}\ \emph {et~al.}(2011)\citenamefont
  {{Medvedev}}, \citenamefont {{Trier Frederiksen}}, \citenamefont
  {{Haugb{\o}lle}},\ and\ \citenamefont {{Nordlund}}}]{Medvedev:2011}%
  \BibitemOpen
  \bibfield  {author} {\bibinfo {author} {\bibfnamefont {M.~V.}\ \bibnamefont
  {{Medvedev}}}, \bibinfo {author} {\bibfnamefont {J.}~\bibnamefont {{Trier
  Frederiksen}}}, \bibinfo {author} {\bibfnamefont {T.}~\bibnamefont
  {{Haugb{\o}lle}}}, \ and\ \bibinfo {author} {\bibfnamefont
  {{\AA}.}~\bibnamefont {{Nordlund}}},\ }\href {\doibase
  10.1088/0004-637X/737/2/55} {\bibfield  {journal} {\bibinfo  {journal}
  {\apj}\ }\textbf {\bibinfo {volume} {737}},\ \bibinfo {eid} {55} (\bibinfo
  {year} {2011})},\ \Eprint {http://arxiv.org/abs/1003.0063} {arXiv:1003.0063
  [astro-ph.HE]} \BibitemShut {NoStop}%
\bibitem [{\citenamefont {{Haugb\o lle}}(2005)}]{Haugboelle:2005}%
  \BibitemOpen
  \bibfield  {author} {\bibinfo {author} {\bibfnamefont {T.}~\bibnamefont
  {{Haugb\o lle}}},\ }\emph {\bibinfo {title} {{Modelling Relativistic
  Astrophysics at the Large and Small Scale}}},\ \href@noop {} {Ph.D. thesis},\
  \bibinfo  {school} {Niels Bohr Institute [astro-ph/0510292]} (\bibinfo {year}
  {2005}),\ \Eprint {http://arxiv.org/abs/arXiv:astro-ph/0510292}
  {arXiv:astro-ph/0510292} \BibitemShut {NoStop}%
\bibitem [{\citenamefont {{Frederiksen}}(2008{\natexlab{a}})}]{Trier:2008.2}%
  \BibitemOpen
  \bibfield  {author} {\bibinfo {author} {\bibfnamefont {J.~T.}\ \bibnamefont
  {{Frederiksen}}},\ }\href {\doibase 10.1086/589648} {\bibfield  {journal}
  {\bibinfo  {journal} {\apjl}\ }\textbf {\bibinfo {volume} {680}},\ \bibinfo
  {pages} {L5} (\bibinfo {year} {2008}{\natexlab{a}})},\ \Eprint
  {http://arxiv.org/abs/0803.3618} {arXiv:0803.3618} \BibitemShut {NoStop}%
\bibitem [{\citenamefont {{Frederiksen}}, \citenamefont {{Haugb{\o}lle}},\ and\
  \citenamefont {{Nordlund}}(2008)}]{Trier:2008.3}%
  \BibitemOpen
  \bibfield  {author} {\bibinfo {author} {\bibfnamefont {J.~T.}\ \bibnamefont
  {{Frederiksen}}}, \bibinfo {author} {\bibfnamefont {T.}~\bibnamefont
  {{Haugb{\o}lle}}}, \ and\ \bibinfo {author} {\bibfnamefont {A.}~\bibnamefont
  {{Nordlund}}},\ }in\ \href {\doibase 10.1063/1.3002512} {\emph {\bibinfo
  {booktitle} {American Institute of Physics Conference Series}}},\ \bibinfo
  {series} {American Institute of Physics Conference Series}, Vol.\ \bibinfo
  {volume} {1054},\ \bibinfo {editor} {edited by\ \bibinfo {editor}
  {\bibfnamefont {M.}~\bibnamefont {{Axelsson}}}}\ (\bibinfo {year} {2008})\
  pp.\ \bibinfo {pages} {87--97},\ \Eprint {http://arxiv.org/abs/0808.0710}
  {arXiv:0808.0710} \BibitemShut {NoStop}%
\bibitem [{\citenamefont {{Frederiksen}}(2008{\natexlab{b}})}]{Trier:2008.1}%
  \BibitemOpen
  \bibfield  {author} {\bibinfo {author} {\bibfnamefont {J.~T.}\ \bibnamefont
  {{Frederiksen}}},\ }\emph {\bibinfo {title} {{Microphysical Conditioning of
  Gamma-Ray Burst Shocks}}},\ \href@noop {} {Ph.D. thesis},\ \bibinfo  {school}
  {Department of Astronomy Stockholm University 106 91 Stockholm, Sweden}
  (\bibinfo {year} {2008}{\natexlab{b}})\BibitemShut {NoStop}%
\bibitem [{\citenamefont {Baumann}(2012)}]{Baumann:2012c}%
  \BibitemOpen
  \bibfield  {author} {\bibinfo {author} {\bibfnamefont {G.}~\bibnamefont
  {Baumann}},\ }\emph {\bibinfo {title} {{Particle Dynamics in Solar Coronal
  Magnetic Reconnection Regions}}},\ \href@noop {} {Ph.D. thesis},\ \bibinfo
  {school} {Niels Bohr Institute, University of Copenhagen} (\bibinfo {year}
  {2012})\BibitemShut {NoStop}%
\bibitem [{\citenamefont {{Bret}}, \citenamefont {{Firpo}},\ and\ \citenamefont
  {{Deutsch}}(2004)}]{Bret:2004}%
  \BibitemOpen
  \bibfield  {author} {\bibinfo {author} {\bibfnamefont {A.}~\bibnamefont
  {{Bret}}}, \bibinfo {author} {\bibfnamefont {M.}~\bibnamefont {{Firpo}}}, \
  and\ \bibinfo {author} {\bibfnamefont {C.}~\bibnamefont {{Deutsch}}},\ }\href
  {\doibase 10.1103/PhysRevE.70.046401} {\bibfield  {journal} {\bibinfo
  {journal} {\pre}\ }\textbf {\bibinfo {volume} {70}},\ \bibinfo {pages}
  {046401} (\bibinfo {year} {2004})}\BibitemShut {NoStop}%
\bibitem [{\citenamefont {{Bret}}, \citenamefont {{Gremillet}},\ and\
  \citenamefont {{Bellido}}(2007)}]{Bret:2007}%
  \BibitemOpen
  \bibfield  {author} {\bibinfo {author} {\bibfnamefont {A.}~\bibnamefont
  {{Bret}}}, \bibinfo {author} {\bibfnamefont {L.}~\bibnamefont {{Gremillet}}},
  \ and\ \bibinfo {author} {\bibfnamefont {J.~C.}\ \bibnamefont {{Bellido}}},\
  }\href {\doibase 10.1063/1.2710810} {\bibfield  {journal} {\bibinfo
  {journal} {Physics of Plasmas}\ }\textbf {\bibinfo {volume} {14}},\ \bibinfo
  {pages} {032103} (\bibinfo {year} {2007})}\BibitemShut {NoStop}%
\bibitem [{\citenamefont {Tzoufras}\ \emph {et~al.}(2006)\citenamefont
  {Tzoufras}, \citenamefont {Ren}, \citenamefont {Tsung}, \citenamefont
  {Tonge}, \citenamefont {Mori}, \citenamefont {Fiore}, \citenamefont
  {Fonseca},\ and\ \citenamefont {Silva}}]{Tzoufras:2006}%
  \BibitemOpen
  \bibfield  {author} {\bibinfo {author} {\bibfnamefont {M.}~\bibnamefont
  {Tzoufras}}, \bibinfo {author} {\bibfnamefont {C.}~\bibnamefont {Ren}},
  \bibinfo {author} {\bibfnamefont {F.~S.}\ \bibnamefont {Tsung}}, \bibinfo
  {author} {\bibfnamefont {J.~W.}\ \bibnamefont {Tonge}}, \bibinfo {author}
  {\bibfnamefont {W.~B.}\ \bibnamefont {Mori}}, \bibinfo {author}
  {\bibfnamefont {M.}~\bibnamefont {Fiore}}, \bibinfo {author} {\bibfnamefont
  {R.~A.}\ \bibnamefont {Fonseca}}, \ and\ \bibinfo {author} {\bibfnamefont
  {L.~O.}\ \bibnamefont {Silva}},\ }\href {\doibase
  10.1103/PhysRevLett.96.105002} {\bibfield  {journal} {\bibinfo  {journal}
  {Phys. Rev. Lett.}\ }\textbf {\bibinfo {volume} {96}},\ \bibinfo {pages}
  {105002} (\bibinfo {year} {2006})}\BibitemShut {NoStop}%
\bibitem [{\citenamefont {{Dieckmann}}\ \emph {et~al.}(2006)\citenamefont
  {{Dieckmann}}, \citenamefont {{Frederiksen}}, \citenamefont {{Bret}},\ and\
  \citenamefont {{Shukla}}}]{Dieckmann:2006}%
  \BibitemOpen
  \bibfield  {author} {\bibinfo {author} {\bibfnamefont {M.~E.}\ \bibnamefont
  {{Dieckmann}}}, \bibinfo {author} {\bibfnamefont {J.~T.}\ \bibnamefont
  {{Frederiksen}}}, \bibinfo {author} {\bibfnamefont {A.}~\bibnamefont
  {{Bret}}}, \ and\ \bibinfo {author} {\bibfnamefont {P.~K.}\ \bibnamefont
  {{Shukla}}},\ }\href {\doibase 10.1063/1.2390687} {\bibfield  {journal}
  {\bibinfo  {journal} {Physics of Plasmas}\ }\textbf {\bibinfo {volume}
  {13}},\ \bibinfo {pages} {112110} (\bibinfo {year} {2006})}\BibitemShut
  {NoStop}%
\bibitem [{\citenamefont {{Frederiksen}}\ and\ \citenamefont
  {{Dieckmann}}(2008)}]{Frederiksen:2008}%
  \BibitemOpen
  \bibfield  {author} {\bibinfo {author} {\bibfnamefont {J.~T.}\ \bibnamefont
  {{Frederiksen}}}\ and\ \bibinfo {author} {\bibfnamefont {M.~E.}\ \bibnamefont
  {{Dieckmann}}},\ }\href {\doibase 10.1063/1.2985776} {\bibfield  {journal}
  {\bibinfo  {journal} {Physics of Plasmas}\ }\textbf {\bibinfo {volume}
  {15}},\ \bibinfo {pages} {094503} (\bibinfo {year} {2008})},\ \Eprint
  {http://arxiv.org/abs/0808.3786} {arXiv:0808.3786 [physics.plasm-ph]}
  \BibitemShut {NoStop}%
\bibitem [{\citenamefont {{Medvedev}}\ and\ \citenamefont
  {{Loeb}}(1999)}]{bib:Medvedev1999}%
  \BibitemOpen
  \bibfield  {author} {\bibinfo {author} {\bibfnamefont {M.~V.}\ \bibnamefont
  {{Medvedev}}}\ and\ \bibinfo {author} {\bibfnamefont {A.}~\bibnamefont
  {{Loeb}}},\ }\href {\doibase 10.1086/308038} {\bibfield  {journal} {\bibinfo
  {journal} {\apj}\ }\textbf {\bibinfo {volume} {526}},\ \bibinfo {pages} {697}
  (\bibinfo {year} {1999})},\ \Eprint
  {http://arxiv.org/abs/arXiv:astro-ph/9904363} {arXiv:astro-ph/9904363}
  \BibitemShut {NoStop}%
\bibitem [{\citenamefont {{Frederiksen}}\ \emph {et~al.}(2004)\citenamefont
  {{Frederiksen}}, \citenamefont {{Hededal}}, \citenamefont {{Haugb{\o}lle}},\
  and\ \citenamefont {{Nordlund}}}]{Frederiksen:2004}%
  \BibitemOpen
  \bibfield  {author} {\bibinfo {author} {\bibfnamefont {J.~T.}\ \bibnamefont
  {{Frederiksen}}}, \bibinfo {author} {\bibfnamefont {C.~B.}\ \bibnamefont
  {{Hededal}}}, \bibinfo {author} {\bibfnamefont {T.}~\bibnamefont
  {{Haugb{\o}lle}}}, \ and\ \bibinfo {author} {\bibfnamefont
  {{\AA}.}~\bibnamefont {{Nordlund}}},\ }\href {\doibase 10.1086/421262}
  {\bibfield  {journal} {\bibinfo  {journal} {\apjl}\ }\textbf {\bibinfo
  {volume} {608}},\ \bibinfo {pages} {L13} (\bibinfo {year} {2004})},\ \Eprint
  {http://arxiv.org/abs/arXiv:astro-ph/0308104} {arXiv:astro-ph/0308104}
  \BibitemShut {NoStop}%
\bibitem [{\citenamefont {{Medvedev}}\ and\ \citenamefont
  {{Spitkovsky}}(2009)}]{Medvedev:2009}%
  \BibitemOpen
  \bibfield  {author} {\bibinfo {author} {\bibfnamefont {M.~V.}\ \bibnamefont
  {{Medvedev}}}\ and\ \bibinfo {author} {\bibfnamefont {A.}~\bibnamefont
  {{Spitkovsky}}},\ }\href {\doibase 10.1088/0004-637X/700/2/956} {\bibfield
  {journal} {\bibinfo  {journal} {\apj}\ }\textbf {\bibinfo {volume} {700}},\
  \bibinfo {pages} {956} (\bibinfo {year} {2009})},\ \Eprint
  {http://arxiv.org/abs/0810.4014} {arXiv:0810.4014} \BibitemShut {NoStop}%
\bibitem [{\citenamefont {{Keshet}}\ \emph {et~al.}(2009)\citenamefont
  {{Keshet}}, \citenamefont {{Katz}}, \citenamefont {{Spitkovsky}},\ and\
  \citenamefont {{Waxman}}}]{Keshet:2009}%
  \BibitemOpen
  \bibfield  {author} {\bibinfo {author} {\bibfnamefont {U.}~\bibnamefont
  {{Keshet}}}, \bibinfo {author} {\bibfnamefont {B.}~\bibnamefont {{Katz}}},
  \bibinfo {author} {\bibfnamefont {A.}~\bibnamefont {{Spitkovsky}}}, \ and\
  \bibinfo {author} {\bibfnamefont {E.}~\bibnamefont {{Waxman}}},\ }\href
  {\doibase 10.1088/0004-637X/693/2/L127} {\bibfield  {journal} {\bibinfo
  {journal} {\apjl}\ }\textbf {\bibinfo {volume} {693}},\ \bibinfo {pages}
  {L127} (\bibinfo {year} {2009})},\ \Eprint {http://arxiv.org/abs/0802.3217}
  {arXiv:0802.3217} \BibitemShut {NoStop}%
\bibitem [{\citenamefont {{Niemiec}}\ \emph {et~al.}(2008)\citenamefont
  {{Niemiec}}, \citenamefont {{Pohl}}, \citenamefont {{Stroman}},\ and\
  \citenamefont {{Nishikawa}}}]{Niemiec:2008}%
  \BibitemOpen
  \bibfield  {author} {\bibinfo {author} {\bibfnamefont {J.}~\bibnamefont
  {{Niemiec}}}, \bibinfo {author} {\bibfnamefont {M.}~\bibnamefont {{Pohl}}},
  \bibinfo {author} {\bibfnamefont {T.}~\bibnamefont {{Stroman}}}, \ and\
  \bibinfo {author} {\bibfnamefont {K.-I.}\ \bibnamefont {{Nishikawa}}},\
  }\href {\doibase 10.1086/590054} {\bibfield  {journal} {\bibinfo  {journal}
  {\apj}\ }\textbf {\bibinfo {volume} {684}},\ \bibinfo {pages} {1174}
  (\bibinfo {year} {2008})},\ \Eprint {http://arxiv.org/abs/0802.2185}
  {arXiv:0802.2185} \BibitemShut {NoStop}%
\bibitem [{\citenamefont {{Henri}}\ \emph {et~al.}(2012)\citenamefont
  {{Henri}}, \citenamefont {{Califano}}, \citenamefont {{Faganello}},
  \citenamefont {{{\v S}ebek}}, \citenamefont {{Frederiksen}}, \citenamefont
  {{Markidis}}, \citenamefont {{Pegoraro}}, \citenamefont {{Hellinger}},
  \citenamefont {{Travnicek}}, \citenamefont {{Nordlund}},\ and\ \citenamefont
  {{Lapenta}}}]{SWIFF_D2.1_KH}%
  \BibitemOpen
  \bibfield  {author} {\bibinfo {author} {\bibfnamefont {P.}~\bibnamefont
  {{Henri}}}, \bibinfo {author} {\bibfnamefont {F.}~\bibnamefont {{Califano}}},
  \bibinfo {author} {\bibfnamefont {M.}~\bibnamefont {{Faganello}}}, \bibinfo
  {author} {\bibfnamefont {O.}~\bibnamefont {{{\v S}ebek}}}, \bibinfo {author}
  {\bibfnamefont {J.~T.}\ \bibnamefont {{Frederiksen}}}, \bibinfo {author}
  {\bibfnamefont {S.}~\bibnamefont {{Markidis}}}, \bibinfo {author}
  {\bibfnamefont {F.}~\bibnamefont {{Pegoraro}}}, \bibinfo {author}
  {\bibfnamefont {P.}~\bibnamefont {{Hellinger}}}, \bibinfo {author}
  {\bibfnamefont {P.~M.}\ \bibnamefont {{Travnicek}}}, \bibinfo {author}
  {\bibfnamefont {A.}~\bibnamefont {{Nordlund}}}, \ and\ \bibinfo {author}
  {\bibfnamefont {G.}~\bibnamefont {{Lapenta}}},\ }in\ \href@noop {} {\emph
  {\bibinfo {booktitle} {EGU General Assembly Conference Abstracts}}},\
  \bibinfo {series} {EGU General Assembly Conference Abstracts}, Vol.~\bibinfo
  {volume} {14},\ \bibinfo {editor} {edited by\ \bibinfo {editor}
  {\bibfnamefont {A.}~\bibnamefont {{Abbasi}}}\ and\ \bibinfo {editor}
  {\bibfnamefont {N.}~\bibnamefont {{Giesen}}}}\ (\bibinfo {year} {2012})\ p.\
  \bibinfo {pages} {11879}\BibitemShut {NoStop}%
\bibitem [{\citenamefont {{Williamson}}(1980)}]{1980JCoPh..35...48W}%
  \BibitemOpen
  \bibfield  {author} {\bibinfo {author} {\bibfnamefont {J.~H.}\ \bibnamefont
  {{Williamson}}},\ }\href {\doibase 10.1016/0021-9991(80)90033-9} {\bibfield
  {journal} {\bibinfo  {journal} {Journal of Computational Physics}\ }\textbf
  {\bibinfo {volume} {35}},\ \bibinfo {pages} {48} (\bibinfo {year}
  {1980})}\BibitemShut {NoStop}%
\bibitem [{\citenamefont {{Belmont}}, \citenamefont {{Aunai}},\ and\
  \citenamefont {{Smets}}(2012{\natexlab{a}})}]{Belmont:2012.1}%
  \BibitemOpen
  \bibfield  {author} {\bibinfo {author} {\bibfnamefont {G.}~\bibnamefont
  {{Belmont}}}, \bibinfo {author} {\bibfnamefont {N.}~\bibnamefont {{Aunai}}},
  \ and\ \bibinfo {author} {\bibfnamefont {R.}~\bibnamefont {{Smets}}},\ }\href
  {\doibase 10.1063/1.3685707} {\bibfield  {journal} {\bibinfo  {journal}
  {Physics of Plasmas}\ }\textbf {\bibinfo {volume} {19}},\ \bibinfo {pages}
  {022108} (\bibinfo {year} {2012}{\natexlab{a}})}\BibitemShut {NoStop}%
\bibitem [{\citenamefont {{Belmont}}, \citenamefont {{Aunai}},\ and\
  \citenamefont {{Smets}}(2012{\natexlab{b}})}]{Belmont:2012.2}%
  \BibitemOpen
  \bibfield  {author} {\bibinfo {author} {\bibfnamefont {G.}~\bibnamefont
  {{Belmont}}}, \bibinfo {author} {\bibfnamefont {N.}~\bibnamefont {{Aunai}}},
  \ and\ \bibinfo {author} {\bibfnamefont {R.}~\bibnamefont {{Smets}}},\ }in\
  \href@noop {} {\emph {\bibinfo {booktitle} {EGU General Assembly Conference
  Abstracts}}},\ \bibinfo {series} {EGU General Assembly Conference Abstracts},
  Vol.~\bibinfo {volume} {14},\ \bibinfo {editor} {edited by\ \bibinfo {editor}
  {\bibfnamefont {A.}~\bibnamefont {{Abbasi}}}\ and\ \bibinfo {editor}
  {\bibfnamefont {N.}~\bibnamefont {{Giesen}}}}\ (\bibinfo {year} {2012})\ p.\
  \bibinfo {pages} {2360}\BibitemShut {NoStop}%
\bibitem [{\citenamefont {{Miura}}\ and\ \citenamefont
  {{Pritchett}}(1982)}]{Miura:1982.1}%
  \BibitemOpen
  \bibfield  {author} {\bibinfo {author} {\bibfnamefont {A.}~\bibnamefont
  {{Miura}}}\ and\ \bibinfo {author} {\bibfnamefont {P.~L.}\ \bibnamefont
  {{Pritchett}}},\ }\href {\doibase 10.1029/JA087iA09p07431} {\bibfield
  {journal} {\bibinfo  {journal} {\jgr}\ }\textbf {\bibinfo {volume} {87}},\
  \bibinfo {pages} {7431} (\bibinfo {year} {1982})}\BibitemShut {NoStop}%
\bibitem [{\citenamefont {Miura}(1982)}]{Miura:1982.2}%
  \BibitemOpen
  \bibfield  {author} {\bibinfo {author} {\bibfnamefont {A.}~\bibnamefont
  {Miura}},\ }\href {\doibase 10.1103/PhysRevLett.49.779} {\bibfield  {journal}
  {\bibinfo  {journal} {Phys. Rev. Lett.}\ }\textbf {\bibinfo {volume} {49}},\
  \bibinfo {pages} {779} (\bibinfo {year} {1982})}\BibitemShut {NoStop}%
\end{thebibliography}%

\end{document}